\documentclass[showpacs,preprintnumbers,amsmath,amssymb,APSl,prd,nofootinbib,superscriptaddress, twocolumn]{revtex4-2}

\usepackage{dcolumn}% Align table columns on decimal point
\usepackage{bm}% bold math
\usepackage{ifpdf}
\usepackage{hyperref}
\usepackage{bm}
\usepackage{xcolor,color,graphicx,graphics}
\usepackage[spanish,english]{babel}%, portuguese
\usepackage[latin1]{inputenc}
\usepackage[OT1]{fontenc}
\usepackage{latexsym,amssymb,amsmath,amsfonts}
\usepackage{makeidx}
\usepackage{epsfig,subfigure}
\usepackage{natbib}
\usepackage{epstopdf}
\usepackage{mathrsfs}
\hypersetup{colorlinks=true, linkcolor=blue, citecolor=green}
\usepackage{enumerate}
\usepackage{xcolor}
 \usepackage{multirow}
\usepackage{enumitem,kantlipsum}

\definecolor{red}{rgb}{1,0,0}

\def\+{^\dagger}

\def\<{\leftarrow}
\def\>{\rightarrow}

\def\({\left(}
\def\){\right)}

\newcommand{\bi}{\begin{itemize}} 				\newcommand{\ei}{\end{itemize}}
\newcommand{\benu}{\begin{enumerate}} 		\newcommand{\enu}{\end{enumerate}}
\newcommand{\bd}{\begin{dinglist}{0}}     \newcommand{\ed}{\end{dinglist}}
\newcommand{\bfig}{\begin{figure}[htbp]}  \newcommand{\efig}{\end{figure}}
        			
\newcommand{\bc}{\begin{center}} 				  \newcommand{\ec}{\end{center}}
\newcommand{\be}{\begin{equation}} 				\newcommand{\ee}{\end{equation}}
\newcommand{\bsub}{\begin{subequations}}  \newcommand{\esub}{\end{subequations}}
\newcommand{\ben}{\begin{eqnarray}} 			\newcommand{\een}{\end{eqnarray}}
\newcommand{\ba}[1]{\begin{array}{#1}} 		\newcommand{\ea}{\end{array}}
\newcommand{\bea}{\begin{equation}\begin{array}{rcl}}
\newcommand{\eea}{\end{array}\end{equation}}

\begin{document}
\title{Photon rings as tests for alternative spherically symmetric geometries with thin accretion disks}

	\author{Lu\'is F. Dias da Silva} \email{fc53497@alunos.fc.ul.pt}
\affiliation{Instituto de Astrof\'{i}sica e Ci\^{e}ncias do Espa\c{c}o, Faculdade de Ci\^{e}ncias da Universidade de Lisboa, Edif\'{i}cio C8, Campo Grande, P-1749-016 Lisbon, Portugal}	
	\author{Francisco S. N. Lobo} \email{fslobo@ciencias.ulisboa.pt}
\affiliation{Instituto de Astrof\'{i}sica e Ci\^{e}ncias do Espa\c{c}o, Faculdade de Ci\^{e}ncias da Universidade de Lisboa, Edif\'{i}cio C8, Campo Grande, P-1749-016 Lisbon, Portugal}
\affiliation{Departamento de F\'{i}sica, Faculdade de Ci\^{e}ncias da Universidade de Lisboa, Edif\'{i}cio C8, Campo Grande, P-1749-016 Lisbon, Portugal}
	\author{Gonzalo J. Olmo} \email{gonzalo.olmo@uv.es}
\affiliation{Departamento de F\'isica Te\'orica and IFIC, 
Centro Mixto Universidad de Valencia - CSIC. 
Universidad de Valencia, Burjassot-46100, 
Valencia, Spain}
\affiliation{Universidade Federal do Cear\'a (UFC), Departamento de F\'isica,\\ Campus do Pici, Fortaleza - CE, C.P. 6030, 60455-760 - Brazil.}
\author{Diego Rubiera-Garcia} \email{drubiera@ucm.es}
\affiliation{Departamento de F\'isica Te\'orica and IPARCOS,
	Universidad Complutense de Madrid, E-28040 Madrid, Spain}

\date{\today}
\begin{abstract}

The imaging by the Event Horizon Telescope (EHT) of the supermassive central objects at the heart of the M87 and Milky Way (Sgr A$^\star$) galaxies, has marked the first step into peering at the photon rings and central brightness depression that characterize the optical appearance of black holes surrounded by an accretion disk. Recently, Vagnozzi et. al. [S.~Vagnozzi, \textit{et al.} arXiv:2205.07787 [gr-qc]] used the claim by the EHT that the size of the {\it shadow} of Sgr A$^\star$ can be inferred by calibrated measurements of the bright ring enclosing it, to constrain a large number of spherically symmetric space-time geometries. In this work we use this result to study some features of the first and second photon rings of a restricted pool of such geometries in thin accretion disk settings. The emission profile of the latter is described by calling upon three analytic samples belonging to the family introduced by Gralla, Lupsasca and Marrone, in order to characterize such photon rings using the Lyapunov exponent of nearly bound orbits and discuss its correlation with the luminosity extinction rate between the first and second photon rings. We finally elaborate on the chances of using such photon rings as observational discriminators of alternative black hole geometries using very long baseline interferometry.

\end{abstract}

\maketitle

\section{Introduction}

One of the core results of the theory of black holes within General Relativity (GR) is the universality of the Kerr hypothesis, namely, that every black hole of the universe is described by two parameters: mass and angular momentum (since the electric charge is typically neglected in astrophysical environments) \cite{Kerr:1963ud}. This hypothesis is deeply anchored in the uniqueness theorems, though the addition of matter fields allows to find hairy black holes under certain circumstances \cite{Herdeiro:2015waa}. Since testing the validity of the Kerr hypothesis is nearly impossible (see however \cite{Bambi:2016sac}), one typically performs instead null-tests, i.e., tests with electromagnetic or gravitational waves on the compatibility of the Kerr black hole with current observations, and the feasibility of every alternative to it (be a modified black hole or a horizonless compact object) to also match such observations \cite{Cardoso:2019rvt}.

Recently, the progress in the development of very long baseline interferometry (VLBI) has paid off via the imaging by the Event Horizon Telescope (EHT) Collaboration of the central supermassive objects at the heart of the M87 \cite{EventHorizonTelescope:2019dse} and Milky Way (Sgr A$^\star$) \cite{EventHorizonTelescope:2022wkp} galaxies. Such observations report the presence of a bright ring of radiation enclosing a central brightness depression, which are the two most salient features of images found using General Relativistic Magneto-HydroDynamic (GRMHD) simulations of the accretion flow surrounding a Kerr black hole. The first such feature comes from the presence of a region of bound unstable orbits in the effective potentials seen by photons (the photon shell \cite{Hadar:2022xag}), allowing for strongly lensed trajectories that orbit the black hole $n$ (half-)times. If the disk is optically thin (i.e. transparent to its own radiation), such trajectories create a thin {\it photon ring} whose features interpolate between two extreme scenarios. On one end,  the photon ring converges to a {\it critical} curve in the image plane of the observer, while the central brightness depression entirely fills it: this is the {\it shadow} of Falcke's view \cite{Falcke:1999pj,Narayan:2019imo}. This arises, in particular in accretion disks which have a spherically symmetric inflow. On the opposite end,  the photon ring is decomposed into an infinite sequence of rings \cite{Gralla:2019xty}, each of them being a gravitationally lensed image of the direct emission region outside the event horizon  but exponentially dimmed in luminosity, the latter captured by the Lyapunov exponent of nearly bound geodesics in a given geometry, while the size of the central brightness depression can be strongly reduced \cite{Chael:2021rjo}. The latter scenario happens not only in infinitesimally thin-disk equatorial flows but also as long as there are gaps in the emission region.

This field of imaging compact objects illuminated by their accretion disk is thus entering a golden era in which it represents a promising opportunity to both test the reliability of the Kerr solution and to explore the plausibility of any of its alternatives to describe observed images. However, such an opportunity can be spoiled by the large uncertainties in the modelling of the disk together with its entanglement with the background geometry in the generation of such images, rendering the quest for reliable observational discriminators between the Kerr solution and its many alternatives a main object of interest in the scientific community. This can be pursued via the two main features of such images - photon ring and central brightness depression -, since they carry a wealth of information about the underlying space-time geometry and, consequently, on the case to test GR itself \cite{Gralla:2020srx,Staelens:2023jgr}.

For the former feature, as larger values of $n$ are considered, the theoretical properties of the corresponding photon rings grow less dependent on the features of the disk and more on the background geometry, thus offering us a way out of the ``contamination" enacted by the disk \cite{Eichhorn:2022oma}. Indeed, despite the exponentially-suppressed luminosity of photon rings, their sharp features make them die off slowly in the Fourier domain and, as a consequence, tend to dominate the interferometric signal in very high-frequencies, leading to the VLBI field. The most promising target in this sense is the $n=2$ ring.  While its detection hinges on observational capabilities that surpass those currently available, either because they require observations at higher frequencies or longer baselines, these are expected to be achievable with the next generation EHT (ngEHT) observations \cite{Tiede:2022grp} and through space-based interferometry. In this regard, prospects have been recently reviewed in the literature, see e.g. \cite{Johnson:2020,Gralla:2020nwp,Gralla:2020yvo,Vincent:2022,Gurvits:2022wgm}. Given the fact that one expects significant deviations in the shape, diameter, width, and relative luminosity of the $n=2$ ring \cite{Paugnat:2022qzy} (and of the $n=1$ one to a lesser extent \cite{Cardenas-Avendano:2023dzo}) for alternative (non-Kerr) geometries, precise observations of this ring could be potentially used to constrain them \cite{Wielgus:2021peu}.

For the latter feature, while the precise size of the outer edge of the central brightness depression cannot be directly determined by the EHT Collaboration given the fact that it cannot measure luminosity contrasts below $\sim 10\%$ of its peak, it has been recently reported that it can be {\it indirectly} inferred (after proper calibration accounting for theoretical and observational uncertainties and subject to several caveats) by a correlation between the observed size of the bright $n=0$ ring (which is caused by the disk's direct emission) and the shadow's size itself \cite{EventHorizonTelescope:2022xqj}. Assuming the validity of this correlation and the assumptions upon which it holds, a collective effort was made by Vagnozzi et. al. (hereafter Vea) in \cite{Vagnozzi:2022moj} to constrain the parameter space of a plethora of alternative spherically symmetric geometries motivated by fundamental or phenomenological considerations. One should note, however, that this observation alone does not single out specific metrics to represent current images but rather their compatibility with them, since the shadow is known to be degenerate between different background geometries \cite{Lima:2021las}.

The main aim of this paper is to combine the two ingredients discussed above, taking a restricted pool of the alternative spherically symmetric geometries considered by Vea, and generate their images when illuminated by an equatorial orbital infinitesimally-thin accretion disk (i.e. the object is seen face-on). Such an assumption on the geometry of the disk is motivated in order to enhance the opportunity to clearly visualize the photon rings. Indeed, the accretion disk features are the weakest thread in the generation of black hole images due to the not so well understood physics of the magnetized plasma, so different pools of assumptions upon its optical, geometrical, and emission properties (among others \cite{Gold:2020iql}) are needed in order to optimize our chances to seek any putative deviation from the Kerr metric under different physical conditions. In our case, the emission properties are set via the consideration of a bunch of analytic models introduced by Gralla, Lupsasca and Marrone (hereafter GLM models) in \cite{Gralla:2020srx}, and whose usefulness (of some of them) in matching the results of some scenarios of GMRHD simulations for the accretion flow have been explored in several works \cite{Vincent:2022,Cardenas-Avendano:2023dzo}. We shall use three picks of such models: one truncated at a certain distance from the event horizon in order to clearly isolate the $n=1$ and $n=2$ photon rings, and two extending to the event horizon with different peaks and decays. In the former model we provide captions of the photon rings, while in the latter models we supply the full images for all these geometries. In all cases we compute the Lyapunov exponent of nearly-bound orbits (a sensitive quantity to deviations from Kerrness) and seek for the presence of any correlation with the actual suppression of luminosity between the $n=2$ and $n=1$ rings, a potential observable of VLBI projects.

This paper is organized as follows: in Sec. \ref{C:II} we set the theoretical framework, build the (null) geodesic motion in spherically symmetric backgrounds, upgrade the formalism to account for those cases in which the matter source is a magnetic monopole from non-linear models of electrodynamics, discuss the notion of critical curves, briefly describe the EHT calibrated measurement of Sgr A$^\star$ shadow, and set the emission (GLM) models used in this work. In Sec. \ref{C:III} we provide an explanation of the choice of spherically symmetric geometries from Vea and the refinements made upon the space of parameters in each case. The generation of images and discussion of the physical results obtained is provided in Sec. \ref{C:IV}, and we conclude in Sec. \ref{C:V} with further thoughts and prospects.

\section{Theoretical framework} \label{C:II}

\subsection{Null geodesics in spherically symmetric space-times}\label{subsec:nullgeo}

We consider the motion of null particles in a spherically symmetric space-time suitably written as
\begin{equation} \label{eq:SSS}
ds^2=-A(r)dt^2+B(r)dr^2 +C(r)d\Omega^2 \ .
\end{equation}
Note that these three functions can always be reduced to just two via a change of coordinates; however, the radial function $C(r)$ cannot always be trivialized to $C(r)=r^2$, so this shape grants us a larger freedom to work with. We assume such particles to follow geodesics of the background metric as $g_{\mu\nu}k^{\mu} k^{\nu}=0$, where $k^{\mu}=\dot{x}^{\mu}$ is the photon's wave number, and a dot represents a derivative with respect to the affine parameter. Using the freedom granted by the spherical symmetry of the system, we can assume the motion to take place along $\theta=\pi/2$ without any loss of generality, so that the above equation reads
\begin{equation} \label{eq:geo1}
-A\dot{t}^2+B\dot{r}^2+C\dot{\phi}^2=0 \ .
\end{equation}
Using the conserved quantities of the system, namely, the energy per unit mass, $E=A\dot{t}$, and the angular momentum per unit mass, $L=C\dot{\phi}$, the above equation can be suitably rewritten (after re-absorbing a factor $L^2$ in the definition of the affine parameter) as
\begin{equation} \label{eq:geoeq}
AB \dot{r}^2=\frac{1}{b^2}-V_{eff}(r) \ ,
\end{equation}
where $b \equiv \frac{L}{E}$ is the impact parameter, and the effective potential reads as
\begin{equation}\label{eq:effpot}
V_{eff}=\frac{A(r)}{C(r)} \ .
\end{equation}
Unstable bound orbits correspond to critical (maxima) points of the effective potential and are an essential theoretical concept for the characterization of black hole images. They are found as the solutions of the equations
\begin{equation} \label{eq:cip}
b_c^2=V_{eff}^{-1}(r_{ps}) \hspace{0.1cm} , \hspace{0.1cm} V_{eff}'(r_{ps})=0 \hspace{0.1cm}, \hspace{0.1cm} V_{eff}''(r_{ps})<0 \ ,
\end{equation}
where primes denote derivatives with respect to the radial coordinate $r$. Here $r_{ps}$ is dubbed as photon sphere and $b_c$ as the critical impact parameter. In the image plane of the observer the photon sphere is mapped into a {\it critical curve}, which splits the light rays issued from the observer's screen backwards towards the black hole into two well-distinguished regions: those with $b>b_c$ find a turning point at some radius $r>r_{ps}$, while those with $b<b_c$ eventually intersect the event horizon of the black hole. Those that have $b \gtrsim b_c$ approach asymptotically (in the observer's image plane) the critical curve and may linger there indefinitely, turning the black hole an arbitrarily large number of times before being released to asymptotic infinity. The angle turned by every photon upon deflection by the black hole is found by re-writing Eq. (\ref{eq:geoeq}) into the more convenient form
\begin{equation}
\frac{d\phi}{dr} = - \frac{b}{C(r)} \frac{\sqrt{AB}}{\sqrt{1-b^2\frac{A(r)}{C(r)}}} \ .
\end{equation}

This equation is the main tool we shall be using for the ray-tracing behind the generation of black hole images. This is done by integrating a set of light trajectories (for a range of $b$) backwards from the observer's screen and classify them according to the number $n$ of times it has (half-)circled the black hole.  This is of interest since, provided that there are gaps in the emission region of the disk (i.e. as long as the disk is not completely spherical, see however \cite{Narayan:2019imo})., and assuming that the disk is transparent to its own radiation (i.e. optically thin) at the emission frequencies, every trajectory turning $n$-half times will boost its luminosity by picking additional photons from the disk on its winding around the black hole. This is the reason behind the existence, in this scenario, of a nested sequence of photon rings on top of the direct emission of the disk, the latter corresponding to those photons that travel from the disk to the observer without undergoing additional turns around the black hole. The characterization of such photon rings is the main object of interest in this work.

\subsection{Effective null geodesics from non-linear electrodynamics}

The above formalism needs to be upgraded when the matter fields threading the geometry belong to non-linear electrodynamics (NED): generalizations of Maxwell electrodynamics via new contributions in the field invariants. NEDs have been longed employed in the literature due their connections to singularity resolution \cite{Dymnikova:2015hka} and effective approaches to quantum electrodynamics \cite{MarotoBook}, enhancing the geometrical and thermodynamical features of charged black holes \cite{Rasheed:1997ns}. Furthermore, they are very flexible in supporting (perhaps with scalar fields added as well) many ad hoc solutions of interest \cite{Bronnikov:2021uta}. Due to the latter feature, many of the alternative geometries ever proposed in the literature are supported by NEDs, and in such a case it has been long recognized in the literature that photons propagate along null geodesics of an effective metric induced by the non-linearity of the matter fields \cite{Novello:1999pg,DeLorenci:2000yh}. Since three of the sixteen geometries considered in this work have been shown to be supported by specific NED models, we are driven to generalize the equations of geodesic motion in NED-supported geometries for the purpose of casting images of the corresponding objects.

NEDs are generically defined by two field invariants constructed from the field strength tensor $F_{\mu\nu}$ and its dual $F_{\mu\nu}^*$ as
\begin{equation}\label{eq:effgeo}
F=\frac{1}{4}F_{\mu\nu}F^{\mu\nu} \quad ; \quad G=\frac{1}{4}F_{\mu\nu}F^{*\mu\nu}
\end{equation}
and which can be written in terms of the electric and magnetic fields. However, for purely electric or magnetic fields (the latter being the case of interest for our purposes in this work) the field invariant $G$ can be made to vanish, and only $F$ remains, so that NEDs correspond to choices of a function $\mathcal{L}(F)$. It was proven in \cite{Novello:1999pg} (see also \cite{DeLorenci:2000yh}) that in geometries threaded by such NEDs, the effective geometry such photons propagate on, $g^e_{\mu\nu}k^{\mu} k^{\nu}=0$, is related to the background geometry via the relation
\begin{equation} \label{eq:eff}
g^{\mu\nu}_e=\mathcal{L}_F g^{\mu\nu} - \mathcal{L}_{FF} {F^\mu}_{\alpha} F^{\alpha \nu} \ ,
\end{equation}
where $\mathcal{L}_F  \equiv d\mathcal{L}/dF$. For purely magnetic configurations $A_{\mu}=q_m \cos \theta \delta_{\mu}^{\phi}$, where $q_m$ is the magnetic charge, the NED field equations provide a single solution for the invariant $F$ for every NED as given by
\begin{equation}
F=\frac{q_m^2}{2r^4} \ .
\end{equation}
With these definitions we can repeat the derivation performed in Sec. \ref{subsec:nullgeo}. According to the relation (\ref{eq:eff}) between the effective and background metrics, and the properties of the NED field in this magnetically charged, spherically symmetric scenario, a suitable line element for the effective geometry can be written as
\begin{equation}\label{eq:effds}
ds_{e}^2=H(r)(-A(r)dt^2+B(r)dr^2) +h(r)C(r)d\Omega^2 \ ,
\end{equation}
so that the two functions $H(r)$ and $h(r)$ encode the deviations between the effective and background metrics. By working out the relation (\ref{eq:eff}) with the parametrization above, such functions are explicitly given in the present framework by
\begin{equation}
H(r)=\mathcal{L}_F +2F\mathcal{L}_{FF} \,, \qquad h(r)=\mathcal{L}_F \ .
\end{equation}
This way Eq.(\ref{eq:geo1}) gets replaced by
\begin{equation}
-H(A\dot t^2+B\dot r^2)+hCd\dot\phi^2 = 0 \ ,
\end{equation}
where now the energy reads as $E=HA\dot{t}$ and the angular momentum as $L=hC\dot{\phi}$, allowing to rewrite the previous equation as
\begin{equation} \label{eq:effgeoeq}
AB \left( \frac{dr}{d\phi} \right)^2 = \frac{C^2h^2}{H^2} \left( \frac{1}{b^2}-V^e_{eff}(r) \right) \ ,
\end{equation}
and once again we can identify a potential $V^e_{eff}(r)$ under these effective geodesics as
\begin{equation} \label{eq:effVef}
V^e_{eff}(r)\equiv \frac{A(r)}{C(r)}\frac{H(r)}{h(r)} \ .
\end{equation}
Unstable bound photon orbits must thus satisfy conditions (\ref{eq:cip}), but now with respect to the new effective potential (\ref{eq:effVef}). Obviously, this means that there will be differences in the quantitative values of the critical curve and its associated impact parameter and, consequently, in the features of the corresponding optical appearances. To work out the latter, we just need to (once again) find an equation for the deflection angle as a function of the radial coordinate, which is just Eq. (\ref{eq:effgeoeq}) rewritten as
\begin{equation}\label{eq:geo}
\frac{d\phi}{dr} = \pm \frac{b}{C(r)}\frac{H(r)}{h(r)} \frac{\sqrt{AB}}{\sqrt{1-b^2\frac{A(r)}{C(r)}\frac{H(r)}{h(r)} }} \ ,
\end{equation}
and thus we are done regarding this aspect.

\subsection{EHT shadow boundary constraints}

On the observer's screen, the critical curve (the image of the photon sphere (\ref{eq:cip})) separates the scattered orbits from the captured ones and thus marks the boundary of the shadow. We can explicitly write the critical curve as the solution of the equation
\begin{equation}
C'(r_{ps}) A(r_{ps})-C(r_{ps})A'(r_{ps})=0 \ .
\end{equation}
The shadow's radius in this view corresponds to Falcke's idea of a region entirely filling the critical curve, that is
\begin{equation}
r_{sh}=\sqrt{\frac{C(r)}{A(r)}} \Bigg \vert_{r=r_{ps}} \ ,
\end{equation}
and obviously it coincides with $b_c$. In most spherically symmetric space-times the radial function trivializes to $C(r)=r^2$ and one recovers a more well-known expression, $r_{sh}=r/\sqrt{A(r)}\vert_{r=r_{ps}}$. However, the shadow's radius defined this way is actually directly unobservable given the lack of photon sensitivity below a certain threshold of the peak intensity. The EHT collaboration copes with this by appealing to the radius of the bright ring of radiation created by the direct emission, which is measurable provided that two main conditions are met \cite{EventHorizonTelescope:2022xqj}:

\begin{enumerate}

\item A sufficiently bright source and strongly lensed supply of photons near the horizon is present and;

\item The accretion flow is geometrically thick and furthermore optically thin (i.e. transparent to its own radiation) at the observing wavelengths,

\end{enumerate}
and then use the observations of the properties of such photon rings as a proxy for the size of the shadow. In addition to these two conditions, a calibration factor must be introduced, which accounts for both theoretical and observational sources of uncertainty in how reliable such a proxy between the bright ring's radius and the shadow's size is. This inference is possible for Sgr A$^\star$ thanks to the fact that its mass-to-distance ratio $M/D$ is known via the tracking of the orbits of the so-called $S$-stars. In particular, the $S0-2$ star  \cite{GRAVITY:2018ofz} has been tracked by two instruments (Keck and VLTI), whose combined (and uncorrelated, since they are obtained from two independent instruments) data allow to quantify the fractional deviation $\delta$ between the inferred radius of a Schwarzschild black hole of angular size (dimensionless form) $\theta_{sh,Sch}=6\sqrt{3}\theta_g$, where $\theta_g=M/D$ is its angular gravitational radius, as \cite{EventHorizonTelescope:2022xqj}
\begin{equation} \label{eq:delta}
\delta  \equiv  \frac{r_{sh}}{r_{sh,Sch}}-1  \approx -0.060 \pm 0.065 \ .
\end{equation}
In turn, this constraint can be transformed into the shadow's size as
\begin{equation} \label{eq:sh1s}
4.54 \lesssim r_{sh}/M \lesssim 5.22 \ ,
\end{equation}
at $1\sigma$ and
\begin{equation} \label{eq:sh2s}
4.21 \lesssim r_{sh}/M \lesssim 5.66 \ ,
\end{equation}
at $2\sigma$. As one can see from these inferred constraints, bounds on the shadow's size are much more generous in alternative spherically symmetric geometries that reduce it, which is actually the majority of the models considered in this work, as we shall see later. For the sake of this work, and to enhance any potential differences in their cast images, we shall take as our reference value the $2\sigma$ bound of Eq. (\ref{eq:sh2s}) in order to constrain the parameter space of each geometry.  Note also that the addition of rotation would modify the shadow's size (and the photon sphere towards the photon shell), though this is assumed to be small as happens in the Kerr solution \cite{Psaltis:2018xkc}: a more complete analysis of this problem should however take this ingredient into account.

Before continuing, there are some caveats in the inference above worth discussing.  The EHT analysis of Sgr A$^*$, as explained on its paper VI  \cite{EventHorizonTelescope:2022xqj}, introduces a ``calibration factor" $\alpha_c \equiv \hat{d}_m/d_{sh}$, which quantifies the degree to which the ``observed" bright ring diameter $\hat{d}_m$ tracks the diameter of the shadow $d_{sh}$. However, such an observed diameter will differ from the true ring diameter $d_m$ due to both theoretical $\alpha_1$ and measurement $\alpha_2$ uncertainties as $\alpha_c=\alpha_1 \times \alpha_2$. The former are the ones we are interested here: these are related to the physics of image formation near the event horizon of the black hole, and come from the theoretical uncertainties on the value of $\alpha_1  \equiv d_m/d_{sh}$ that results from analyzing a given background space-time geometry in combination with a modelling for the accretion flow. In practical terms, the EHT assumes Kerr (Schwarzschild) as the default geometry for the computation of the $\alpha_1$ part of the calibration factor; hence by assuming here it at face value we introduce a bias in the extrapolation of these results to alternative geometries since the latter might modify such a factor. This way, our analysis (and the one of Vea) implicitly assumes the EHT analysis based on the Schwarzschild geometry for the connection provided by the fractional deviation $\delta$ between the measured ring diameter and the critical curve. Should (any of) the alternative geometries modify such a correspondence between observed ring and critical curve, then the domain of validity of (the space of parameters of) such geometries could be extended beyond the bounds imposed by (\ref{eq:delta}) that will appear in our Section \ref{C:III} below. This weakness could be overcome by carrying out a similar analysis as the one of the EHT on the distributions of deviations for the alternative geometries as compared to the Schwarzschild on a case-by-case basis, something far beyond the scope of the present work.

\subsection{Photon rings and central brightness depression in a thin-disk emission model}

Under the hypothesis that the universality of the Kerr (Schwarzschild) solution is replaced by the universality of an alternative metric, the image cast from any such object should be compatible with any scenario for the accretion flow. We shall thus use this idea to employ the bounds on the shadow's size above to constrain the space of parameters of alternative spherically symmetric geometries to subsequently enact their predictions in the opposite end of the geometry of the accretion flow, namely, that in which the disk is infinitesimally thin. In such a case, the outer edge of the central brightness depression does not coincide with the critical curve \cite{Gralla:2019xty} since the dark region is strongly reduced\footnote{Note, however, that there is a lower limit for the size of the central brightness depression assuming an arbitrary equatorial emission model, in such a case depending only on the background geometry. This is dubbed in  \cite{Chael:2021rjo} as the {\it inner shadow}.}. Therefore, for the sake of our analysis we shall reserve the word {\it shadow} for the region completely filling the critical curve and associated to Falcke's view and the EHT bounds above, and use {\it central brightness depression} for the reduced-size dark region when the disk is infinitesimally thin. In any case, our main concern here is to characterize the properties of the photon rings (which are directly observable) in order to compare the pool of alternative compact objects.

The proper treatment of the imaging of a black hole surrounded by its accretion disk requires the use of GRMHD simulations of the plasma making up the disk under a pool of assumptions for the particles' velocities and temperature, the opacity and geometrical shape of the disk, its magnetic properties, and so on, see e.g. \cite{EventHorizonTelescope:2022urf}. However, it is possible to develop semi-analytic approximations to this problem capable to capture the most influential features of the disk contributing to the image. For the sake of this work we focus on the Gralla-Lupsasca-Marrone (GLM) models, which are based on the Standard Unbound (SU) Johnson distribution, and read as \cite{Gralla:2020srx}
\begin{equation} \label{eq:GLM}
I(r;\gamma,\mu,\sigma)=\frac{\exp \left(-\frac{1}{2} (\gamma + \text{arcsinh} \left(\frac{r-\mu}{\sigma} ) \right)^2 \right)}{\sqrt{(r-\mu)^2 + \sigma^2 }} \ .
\end{equation}
These GLM models assume a monochromatic emission (in the frame of disk), and contain three freely-adjustable parameters which control the features of the disk's intensity: $\gamma$ is related to its rate of growth from asymptotic infinity to the peak, $\mu$ shifts the profile to a desired location, while $\sigma$ sets its dilation. By exploring a range of parameters of such profiles one can be able to mimic the results of certain scenarios for the accretion flow within GRMHD simulations \cite{Vincent:2022,Cardenas-Avendano:2023dzo}.  This yields a simplified analytical and numerical treatment of the photon ring features of the image, this way allowing for a more efficient comparison of different geometries of the shadow caster.

 For an optically thin flow with a purely equatorial emission, every emitted photon suffers gravitational redshift on its run-away from the black hole. In the absence of absorption, this can be computed according to Liouville's theorem, which demands the conservation of the flux $I_{\nu_0}/\nu_0^3= I_{\nu_e}/\nu_e^3$, where $\nu_0$ and $\nu_e$ refer to the frequency in the observer's and emitter's frames, respectively. Using the monochromatic character of $I_{\nu_e} \equiv I(r)$, and the fact that in the spherically symmetric geometry (\ref{eq:SSS}) one has the relation $\nu_o=A^{3/2}(r) \nu_e$, so we can integrate the above relation between fluxes to all frequencies as $I_{ob}= \int d\nu_{0} I(r)$ to find the result\footnote{Note that in the case in which we are using effective geodesics, we need to add a factor $H^2(r)$ to this expression in order to account for the transformation of line element (\ref{eq:effds}). Absorption and other potential transport effects emerging from nonlinearities in the electromagnetic Lagrangian will be studied elsewhere.}
\begin{equation} \label{eq:Io}
I_{ob}= \sum_{n=0}^{2} A^2(r)I(r) \ ,
\end{equation}
where in this expression we have introduced the contributions up to the second photon ring, $n=2$, our target in this work. 

\begin{figure}[t!]
\includegraphics[width=8.2cm,height=5cm]{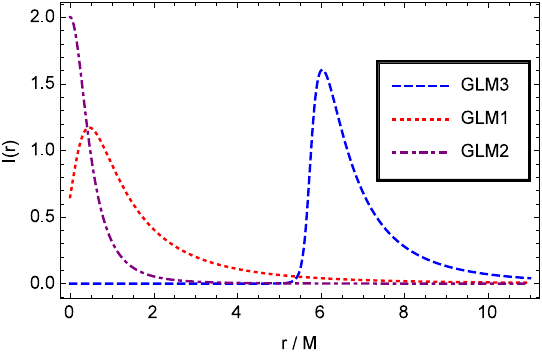}
\caption{The GLM intensity profiles (\ref{eq:GLM}) for the choices (\ref{eq:GLM3}), (\ref{eq:GLM1}) and (\ref{eq:GLM2}), respectively.}
\label{fig:GLM}
\end{figure}

For the sake of this work we shall suitably adapt to the non-rotating case the three original models included in \cite{Gralla:2020srx}, which correspond to the following choices
\begin{eqnarray}
\text{GLM3}&:& \gamma=-2 \,, \qquad \mu=\frac{17M}{3} \, , \qquad \sigma=\frac{M}{4} \label{eq:GLM3}  \ , \\
\text{GLM1}&:&   \gamma=-\frac{3}{2} \, , \quad \mu=0 \, , \qquad \sigma=\frac{M}{2} \label{eq:GLM1} \ , \\
\text{GLM2}&:&   \gamma=0 \, , \qquad \mu=0 \, , \qquad \sigma=\frac{M}{2} \ .   \label{eq:GLM2}
\end{eqnarray}
These profiles are depicted in Fig. \ref{fig:GLM}. GLM3 has a peak brightness located slightly above the corresponding innermost stable circular orbit of a Schwarzschild black hole (i.e. $r \gtrsim 6M$), while GLM1/GLM2 go all the way down to $r=0$ (note that the black hole horizon will appear well before getting there) with different shapes. While the latter two models are thus more suitable to describe the overflow of the plasma in orbit around the heart of M87 and Sgr A$^\star$, the inner edge of the direct emission of the disk in such cases will be smaller than the one of the $n=2$ ring: consequently, photon rings will be stacked on top of the direct emission, troubling their direct visualization. We thus employ GLM3 to complement the analysis, since in such a case the inner edge of the direct emission is truncated at a larger distance to allow for such a visualization. In order to perform our simulations we use our own Geodesic Rays and Visualization of IntensiTY profiles (GRAVITYp) ray-tracing code.

\section{Choice of spherically symmetric space-times} \label{C:III}

In this work we consider 16 alternative spherically symmetric space-times extracted out of the work \cite{Vagnozzi:2022moj}. In what follows we explain the motivation behind such a pick of space-times and the choice of parameters for the sake of the generation of images. Regarding such parameters, they are pushed as far as possible to be compatible with the shadow's radius (defined according to Falcke's view, as discussed in the previous section) at $2\sigma$, as given by Eq. (\ref{eq:sh1s}). In doing that, not every metric proposal saturates the EHT bound(s). This is due to the fact that either a) the model's parameter must be bounded below the EHT constraints in order for an event horizon to be present, b) there is no limit the parameter can be pushed to before incurring in incompatibilities with the EHT bound. Furthermore, in some cases the EHT bound is weaker than other bounds found via analysis of several astrophysical phenomena. For the sake of our analysis, Vea constraints on the viable parameter space will be refined to more finely match the shadow's size limits: this is so because at such large deviations from the Schwarzschild's prediction the features of the photon rings become more sensitive to small modifications in the model's parameters. On the other hand, the spherical symmetry of the system strongly simplifies the problem as compared to the realistic rotating case, requiring less sophisticated treatment of the geodesic curves and, by extension, less computing power.

Since we are dealing with spherically symmetric space-times we start our considerations from the Schwarzschild black hole
\begin{equation}
A(r)=1-\frac{2M}{r} \ .
\end{equation}
Having a single parameter, the Schwarzschild black hole (BH) predicts a unique event horizon, $r_h=2M$, a unique critical impact parameter, $b_c=3\sqrt{3}M$ (hence a single shadow's radius), and a unique photon sphere radius, $r_{ps}=3M$. This way, it is the benchmark every other metric is tested against. Furthermore, in order to interpret the parameter $M$ as the mass as seen from an asymptotic observer, our analysis of spherically symmetric space-times will only consider those metrics whose behavior at large distances (assuming asymptotic flatness) is dominated by the (Schwarzschild) mass term. This will allow us to compare the predictions of all alternative models on as an equal-footing as possible.

It is important to stress that we consider this pool of geometries in a (mostly) theory-agnostic approach, namely, disregarding the theory combining gravitational (i.e. either GR or modified gravity) plus matter fields they come from, and some potential drawbacks such theories and their corresponding geometries may have\footnote{Nonetheless, we shall not be oblivious to the fact that three of the geometries considered here have been identified to be derived from reasonable enough NED theories; hence the development of the framework of effective geodesics in the previous section.}. The latter comes mostly from the violation of the energy conditions and potential instabilities which may render the configurations non-viable, but for the sake of this work we are only interested in the comparison between their cast images. Note, however, that since some of these geometries can be framed within a modified gravity perspective, some of these drawbacks of their GR-formulation (most notably the violation of the energy conditions for ``regular" geometries) may be potentially lifted.

\subsection{Geometries and shadow constraints}

\begin{enumerate}[leftmargin=*]
    \item {\bf Reissner-Nordstr\"om (RN) BH}. The canonical modification of the Schwarzschild geometry is to add a charge term to form the Reissner-Nordstr\"om solution
\begin{equation}
A(r)=1-\frac{2M}{r} + \frac{q_e^2}{r^2} \ .
\end{equation}
A critical curve is present in this model if the electric charge fulfils the bound $q_e^2 \leq (9/8)M^2$. Compatibility with $2\sigma$ shadow's radius (\ref{eq:sh2s}) allows us to push the electric charge to the value $q_e =0.939 M$. Since this is below the bound $q_e^2 \leq M^2$ marking the transition from charged black holes to over-charged (naked singularity) solutions, an event horizon will be present in this case. Note, however, that such a value is well above reasonable estimates on how much charged a black hole may be from astrophysical considerations \cite{Blandford:1977ds,Zajacek:2018ycb}, though we shall disregard such a fact in order to have an overall view on how images from charged space-times look like before engaging in other samples.

  \item {\bf Euler-Heisenberg (EH) NED BH}. Our first example of a NED-supported geometry is a natural generalization of the RN geometry via the function
  \begin{equation}
  \mathcal{L}(F)=-F + 4 \mu F^2 \ ,
  \end{equation}
and its spherically symmetric geometry (interpreted as supported by a magnetic monopole with charge $q_m$) is characterized by the function \cite{Yajima:2000kw}
\begin{equation}
A=1-\frac{2M}{r} + \frac{q_m^2}{r^2} - \frac{2\mu q_m^4}{5r^6} \ .
\end{equation}
Images of these configurations were discussed in \cite{Allahyari:2019jqz,Wen:2022hkv}. Note that here $\mu$ is a constant which can be related to the effective series expansions of Quantum Electrodynamics the EH model is derived from \cite{DobadoBook}, but  Vea take it as a free parameter and fix it to $\mu=0.3$ on grounds of this value to approximately correspond to the maximum coupling in which the perturbative QED expansion to be meaningful; furthermore the shadow's size is much more dependent on the value of $q_m$ than the one of $\mu$. For such a value they report the constraint $q_m \lesssim 0.8M$ though we find we can push it a bit harder up to $q_m=0.88M$ for our generation of images.

\item  {\bf Bardeen's regular BH}. Bardeen's proposal \cite{Bardeen} is to remove curvature singularities at the center of black holes by replacing the  point-like region of the Schwarzschild/RN black hole by a de Sitter core \cite{Ansoldi:2008jw}; this is achieved via a magnetically charged solution defined in terms of the metric function
\begin{equation}
A(r)=1-\frac{2Mr^2}{(r^2+q_m^2)^{3/2}} \ ,
\end{equation}
with $q_m \leq \sqrt{16/27} M \approx 0.77M$ in order to describe a black hole. Vagnozzi et al. report that all values within this range are compatible with $2\sigma$ shadow's radius. It is known that Bardeen's space-time can be obtained as a solution of the Einstein field equations coupled to an NED \cite{Beato:2000yh}. However, such a function has a bizarre shape that does not lead to functions $H$ and $h$ which smoothly recover the background geodesics in the $q_m \to 0 $ limit, and hence we follow the same route as Vea and consider images generated within background geodesics.

\item {\bf Hayward's regular BH}. Hayward's model is based on similar premises as that of Bardeen's one, and furthermore it has been widely studied as a toy-model to simulate gravitational collapse and development of de Sitter cores. Its metric function reads \cite{Hayward:2005gi}
\begin{equation}
A(r)=1-\frac{2Mr^2}{r^3+2l^2M} \ ,
\end{equation}
with the same bound as Bardeen, $l \lesssim \sqrt{16/27} M $, to describe a black hole. This is another instance of a space-time geometry that can also be obtained as a solution of the Einstein field equations coupled with NED, but whose $H$ and $h$ functions do not smoothly recover the background geodesics in the $l \to 0 $ limit. Likewise in the Bardeen model, Hayward's solution is compatible with the $2\sigma$ bounds at every $l$, so we again push the parameter $l$ of the model until nearly saturating the critical bound to describe a black hole.

\item {\bf Frolov BH}.  Frolov's choice \cite{Frolov:2016pav} is similar in spirit to both the Bardeen and Hayward models, but it contains two parameters. The metric function is given by
\begin{equation}
A(r)=1-\frac{(2Mr-q_e^2)r^2}{r^4+(2Mr+q_e^2)l^2} \ ,
\end{equation}
where $0 < q_e \leq 1$ is seen as an electric charge, and again $l \lesssim \sqrt{16/27} M $. In Vea  \cite{Vagnozzi:2022moj} they propose to fix $l=0.3$, which results in a constraint $q_e \lesssim 0.9M$. However, at the value saturating this bound Frolov's solution does not describe a black hole, but instead a naked object by a small margin; for instance, a value of $q_e=0.875M$ describes a black hole instead, but only a slightly larger shadow radius. Configurations without event horizons may produce additional photon ring contributions due to light rays that travel above (but near) the critical curve, and are reflected back due to the presence of local maxima or an infinite potential slope; several such examples have been worked out recently in the literature, see e.g. \cite{Wielgus:2020uqz,Guerrero:2022msp,Tsukamoto:2022vkt}. For the sake of our work here, their analysis would muddy the comparison of alternative spherically symmetric geometries on equal-footing since we are only interested on the $n=2$ ring and not in higher-order rings, so we opt for considering Frolov black holes with $q_e=0.875$.

\item {\bf Kazakov-Solodukhin (KS) regular BH}. It arises in a string-inspired model and is given by \cite{Kazakov:1993ha}
\begin{equation}
A(r)=-\frac{2M}{r} + \frac{\sqrt{r^2-l^2}}{r} \ ,
\end{equation}
Despite its shape it actually reduces to the Schwarzschild metric at large distances, $r \gg l$, so it belongs to our acceptable class of models. The single parameter of the model is required to be positive $l>0$ in order to avoid the central singularity. Vea report that  $2\sigma$ observations require that $l \lesssim M$, but we need to decrease it down to $l= 0.942M$ to saturate the shadow's bound.

\item {\bf Sen BH}. This proposal \cite{Sen:1992ua} belongs to dilaton gravity and also includes a magnetic charge contribution, now within the mass term as (in the non-rotating limit)
\begin{equation}
A(r)=1-\frac{2M}{r+q_m^2/M} \ ,
\end{equation}
where $q_m \lesssim 0.75M$ at $2\sigma$ but X-ray reflection spectroscopy yield a slightly stronger constraint $q_m \lesssim 0.6M$ \cite{Tripathi:2021rwb}, so we take here the latter bound.

\item {\bf Einstein-Maxwell-Dilation (EMD) BH}.  A model in which an additional scalar field is included - the dilaton - within GR coupled to a Maxwell field (EMD gravity) yields a line element given by \cite{Gibbons:1987ps}
\begin{equation}
A(r)=1-\frac{2M}{r} \left( \sqrt{1+\frac{q_e^4}{4M^2 r^2}} - \frac{q_e^2}{2Mr} \right) \ ,
\end{equation}
with $q_e \lesssim M$ in Vea which we slightly refine as $q_e=0.995M$.

\item {\bf Dark matter (DM)-surrounded BH}. A model incorporating a surrounding dark matter fluid via a correcting term to the Schwarzschild solution was proposed in \cite{Li:2012zx} as given by the line element
\begin{equation}
A(r)=1-\frac{2M}{r} +\frac{k}{r} \log \left(\frac{r}{\vert k \vert} \right) \ ,
\end{equation}
Vea report $k  \lesssim 0.15M$ but we find we just need that constant to take the value $k=0.128M$ to saturate the bound on the shadow's size.

\item {\bf Simpson-Visser (SV) black bounce BH}. Our first example of a non-trivial $C(r)$ function is provided by the so-called black bounce, which denotes a metric originally introduced by Simpson and Visser \cite{Simpson:2018tsi}, and whose philosophy is to replace the radial coordinate of the Schwarzschild solution by a radial function implementing a bounce, the latter interpreted as the throat of a wormhole. In order to do it so, Simpson and Visser follow the prescription of Ellis \cite{Ellis:1973yv} from the shift of the radial coordinate, so that the metric functions read as
\begin{equation}
A(r)=1-\frac{2M}{(r^2+a^2)^{1/2}} \, , \qquad C(r)=r^2+a^2 \ .
\end{equation}
Because of the way it is built, this model has the same critical impact parameter and photon sphere radius as its seed metric - the Schwarzschild black hole - for every $a$, so it is not constrained by the EHT results at all. For the sake of our images (a detailed analysis was made by some of us in \cite{Guerrero:2021ues})  we choose to remain within the sub-class of these configurations that have an horizon (corresponding to $0<a \leq 1$, so we take the value  $a=0.5$).

\item {\bf Loop Quantum Gravity (LQG) BH.} A solution found within the context of Loop Quantum Gravity takes the form \cite{Modesto:2008im}
\begin{equation}
A(r)=\frac{(r-r_-)(r-r_+)(r+r_{\star})^2}{r^4} \ ,
\end{equation}
with the definitions $r_+=r_S(1+P)^2, r_-=r_S P^2/(1+P)^2, r_{\star}=\sqrt{r_+ r_-}$ and $P$ is a parameter of the theory. Vagnozzi et. al report the constraint $P \lesssim 0.08M$ for compatibility with $2\sigma$ shadow; we refine such a constraint as $P=0.082M$.

\item {\bf Conformal scalar model (ConfSca) BH}. This is an example of a family of configurations which look like the RN one but with a minus sign in front of the charge term, i.e. \cite{Astorino:2013sfa}
\begin{equation}
A(r)=1 - \frac{r_S}{r} -\frac{s}{r^2} \ ,
\end{equation}
so we can take it as a benchmark for this kind of metrics. While Vea (note that we have reversed the sign for $s$ as compared to them) report that $s \lesssim 0.4M$, we find we can push it up to $s=0.45M$ for compatibility with $2\sigma$ shadow's size.

\item {\bf Janis-Newman-Winicour (JNW) naked singularity}. For completeness, and for the sake of comparison with black hole images given its historic relevance, we consider the naked singularity of the Janis-Newman-Winicour solution, supported by a massless scalar field, and given by the function \cite{Janis:1968zz}
\begin{equation}
A(r)= \left[1-\frac{2M}{r(1-\nu)} \right]^{1-\nu} ; C(r)=r^2 \left[1-\frac{2M}{r(1-\nu)}\right]^{\nu}
\end{equation}
where $\nu$ is a parameter related to the scalar charge of the field supporting it. Vea report the constraint $\nu \lesssim 0.45M$, though we find we can push it up to $\nu=0.4835M$. The absence of a horizon means that light rays above the maximum of the potential will find no obstacle to reach the center of the solution, thus posing a different scenario than black hole space-times.

\item {\bf Bronnikov's regular NED BH}. Bronnikov's model is another example of a regular magnetic black hole solution given by the metric function \cite{Bronnikov:2000vy}
\begin{equation}
A(r)=1-\frac{2M}{r} \left(1-\tanh \left[\frac{q_m^2}{2Mr}\right] \right) \ ,
\end{equation}
and supported by a well-behaved NED of the form
\begin{equation}
\mathcal{L}(F)= 4F \cosh^{-2} \left[a (2F)^{1/4} \right] \ ,
\end{equation}
where the constant $a$ is related to the magnetic charge via the relation $a=q_{m}^{3/2}/(2M)$ in order to remove the central singularity.  Vea report the constraint $q_m \lesssim M$ at $2\sigma$; however we find a much more restricted range, saturated at $q_m=0.905M$, which is the value we take here.

\item {\bf The Ghosh-Kumar (GK) BH}. This is a simple modification of the Schwarzschild geometry via the function \cite{Ghosh:2021clx,Ghosh:2014pba,Culetu:2014lca,Simpson:2019mud}
\begin{equation}
A(r)=1-\frac{r_S}{\sqrt{r^2+q_m^2}} \ ,
\end{equation}
and thus, close in spirit to the black bounce geometries such as the SV one above. As in the Bardeen and Hayward solutions, despite the fact that it can be generated within a NED, the corresponding function is of bizarre shape and so they are the corresponding effective geodesics functions. This way, we opt for considering the usual background geodesics as in Vea, and push a little bit their bound of $q_m \lesssim 1.6M$ up to $q_m = 1.63M$ for the generation of our images.

\item {\bf The Ghosh-Culetu-Simpson-Visser (GCSV) NED regular BH}. It corresponds to the function \cite{Ghosh:2014pba,Culetu:2014lca,Simpson:2019mud}
\begin{equation}
A(r)=1-\frac{r_S}{r} e^{-q_m^2/r_S} \ .
\end{equation}
In Vea they report that $q_m$ can be pushed up to $\vert q_m \vert \lesssim  M$ using the background geodesics. We instead opt for considering the effective ones given the fact that the model is  supported by a NED with Lagrangian density \cite{Kumar:2020ltt}
\begin{equation}
L(F)= F \exp \left[-\frac{q_m}{r_S} (2q_m^2 F)^{1/4} \right] \ ,
\end{equation}
whose associated effective functions $H$ and $h$ turn out to be well behaved. Furthermore, this analysis complements the one carried out in Ref. \cite{Guerrero:2022msp} about the multi-ring structure of the sub-family of configurations without event horizons.

\end{enumerate}

\begin{figure}[t!]
\includegraphics[width=8.5cm,height=7cm]{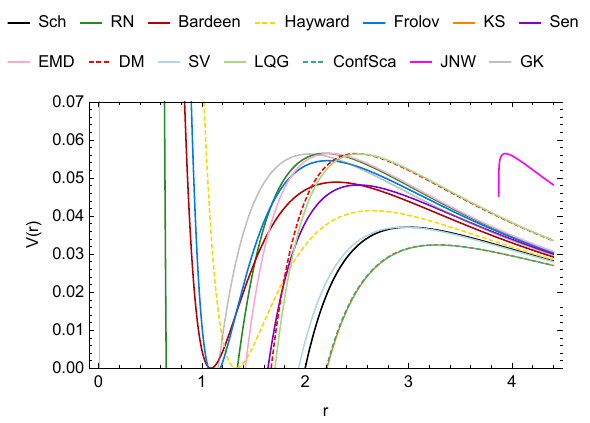}
\includegraphics[width=8.5cm,height=7cm]{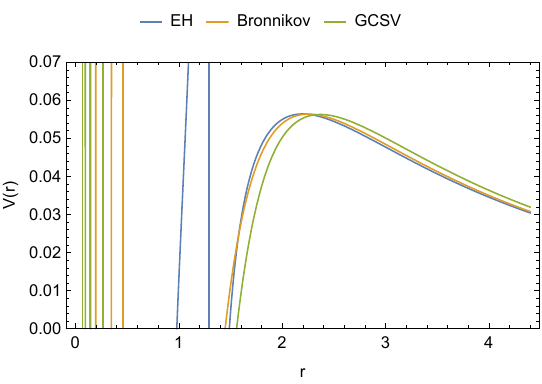}
\caption{The effective potential for spherically symmetric geometries with background geodesics (top) and for effective ones (bottom). Only the outermost part of the potential with $V>0$ is relevant for generation of images, since zeros in $V(r)$ mean presence of horizons.}
\label{fig:potentials}
\end{figure}

The effective potential for this set of sixteen spherically symmetric geometries (plus Schwarzschild) is depicted in Fig. \ref{fig:potentials} for models with background geodesics and effective ones, respectively. Some comments are in order. The fact that the JNW geometry lacks horizons makes its potential qualitatively deviate from the others, not being defined everywhere. As for the potentials of the effective geodesics, they show weird behaviors in the innermost region; however, being covered by a horizon, such a part of the potential plays no role in the generation of images. This would not be so in those cases in which the EHT bound is not saturated and further pushing the space of parameters of the geometries would make the horizon go away. In such a case the internal shape of the potential {\it does} matter in the generation of a multi-ring structure provided that it has additional minima/maxima or an infinite slope at the center; however such a feature will not be present in our images.

\subsection{Some comments on the discarded models}

There are many other models whose constraints from the shadow's radius are reported within Vea \cite{Vagnozzi:2022moj} and which are not considered in this work. Here we briefly provide the reasons why (besides practical reasons of keeping the length of the paper within reasonable limits). First, as mentioned before, we do not consider models which are not asymptotically flat, which would prevent the identification of the constant $M$ as the asymptotic mass of the space-time and thus the generation of images of the corresponding objects on an equal-footing. This leaves outside of our analysis models such as $f(R)$ [R], the DS wormhole [K], Rindler [AH], or the topological defect [AJ]. Second, we do not consider space-times that can be rewritten (via e.g. a simple redefinition of constants) in an usual RN-like form, since the same constraints placed upon the RN solution can be converted into constraints on each theory's parameters and this way the photon rings features are the same. This includes as examples BHCSH [Q1] if $s>0$, Horneski [S1] if $p<0$, MOG [T], braneworlds [U], GUPa [AL A] and GUPb [AL B], or the second non-commutative gravity model [AM B]. Third, we exclude those models which have too tight constraints on their space of parameters to significantly alter photon ring features, or are directly ruled out: this includes the SV WH [I2], the Morris-Thorne WH [J], the null NS [O], Aether models [Z], 4D Gauss-Bonnet gravity [AA], or asymptotically-safe gravity [AB]. We have also avoided consideration of glued solutions containing potential discontinuities, or others demanding excessive computational times.

\section{Results and physical discussion} \label{C:IV}

\subsection{Lyapunov exponents and extinction rates}

\begin{table*}[t]
\begin{tabular}{|c|c||c||c|c|c|c|c|}
\hline
Space-time & $r_h$ & $b_{ps}$ & $r_{ps}$ & {\bf Lyapunov} [$I_1/I_2$]   & $I_{GLM3}^{\frac{n=1}{n=2}}$   & $I_{GLM1}^{\frac{n=1}{n=2}}$   & $I_{GLM2}^{\frac{n=1}{n=2}}$   \\ \hline
LQG   & 1.708 &  4.216 &2.521 & 3.372 [29.150] & 35.59   & 31.01  & 29.53 \\ \hline
KS  & 2.214 & 5.559 & 3.279  & 3.288 [26.809]    & 30.95   & 28.24  & 26.93  \\ \hline
ConfSca    & 2.204 &  5.556 & 3.274  & 3.278 [26.530]   & 30.66  & 27.96  & 26.65  \\ \hline
DM   & 1.671 & 4.212 & 2.493 & 3.268 [26.259]   & 32.28  & 27.95  & 26.56   \\ \hline \hline
Schwarzschild & 2 & $3\sqrt{3}$ & $3$ & 3.150 [23.352]      & 27.83    & 24.74   & 23.45  \\ \hline \hline
SV  & 2 &  $3\sqrt{3}$ & 3 & 3.107 [22.367]    & 26.79  & 23.69  & 22.41  \\ \hline
JNW NS  & 0 & 4.213   &  1.453    & 3.096 [22.128]   & 21.02   & 20.52  & 19.95  \\ \hline
Sen  & 1.64 & 4.558 & 2.514 & 2.887 [17.946]    &  22.96   & 19.27   & 18.02     \\ \hline
GCSV (e)   & 1.560 &  4.217  & 2.370   & 2.693 [14.782]   & 19.88  & 15.73 & 14.16            \\ \hline
EMD   & 1.421 & 4.211 & 2.234 & 2.665 [14.381]   & 19.47  & 15.62  & 14.43  \\ \hline
Bronnikov (e)  & 1.449 &  4.213 & 2.253 & 2.587 [13.292]  & 18.34  & 14.51  & 13.15 \\ \hline
Hayward  & 1.337  & 4.916 & 2.652 & 2.542 [12.708]    & 17.45   & 14.02  & 12.77   \\ \hline
RN   & 1.343 & 4.209 & 2.197 & 2.527 [12.524]    & 17.26  & 13.78  & 12.61   \\ \hline
EH (e)   & 1.490  & 4.212 & 2.197 & 2.418 [11.229]    & 18.41   & 12.20  & 10.88     \\ \hline
Frolov  & 1.179  & 4.283 & 2.216 & 2.403 [11.066]   & 15.58  & 12.41 & 11.17          \\ \hline
Bardeen  & 1.093 & 4.524 & 2.301 & 2.253 [9.516]    & 13.27  & 10.75  & 9.56   \\ \hline
GK   & 1.158 &  4.216 & 2.038 & 2.100 [8.166]  & 11.74 & 9.21  & 8.21  \\ \hline
\end{tabular}
\caption{The alternative spherically symmetric geometries considered in this work (see the main text for abbreviations and model's parameters chosen) ordered in decreasing values of their Lyapunov exponent, the latter computed for the $n=2$ trajectory (see the corresponding discussion in the text). Here we list those quantities relevant for the generation of images (in units of $M$; (e) denotes quantities computed in the effective propagation geometry) as well as those relevant to characterize them; $r_h$: horizon radius; $b_{ps}$: critical impact parameter; $r_{ps}$: photon sphere radius; Lyapunov exponent of nearly bound orbits and its associated (theoretical) luminosity extinction rate $I_1/I_2$ [in brackets]; $I^{\frac{n=1}{n=2}}$: the (observable) extinction rate (sub-labels for GLM type of emission profile). Digit precision limited to three decimals for theoretical quantities and to two for observational ones. We single out using double rows/columns the Schwarzschild solution as the benchmark metric, and the critical impact parameter (the shadow's radius in the EHT interpretation) as the inferred quantity by the EHT, acting as the constraint the parameter space of all these geometries is subjected to.}
\label{table}
\end{table*}

In Table \ref{table} we report our findings on the main geometrical and image features of the alternative spherically symmetric space-times considered in the previous section, organized according to decreasing values of the Lyapunov exponent. The latter is a measure of the instability scale of nearly bound orbits, namely, those which hover very close to the critical curve, $r \approx r_m + \delta r_0$ where $\delta r_0 \ll r_m$. This way, after a number of half-orbits $n$, the particle will be located at
\begin{equation}
\delta r_n \approx e^{\gamma n} \delta  r_0 \ ,
\end{equation}
(for a detailed account of such orbits, see \cite{Cardoso:2008bp}). The Lyapunov exponent $\gamma$ is the number we are interested in here, since it controls the flux of intensity among successive images of the disk, that is \cite{Johnson:2020}
\begin{equation}
\frac{I_{n+1}}{I_{n}} \sim  e^{-\gamma} \hspace{0.2cm} \text{for} \hspace{0.2cm} n \gg 1 \ .
\end{equation}
It turns out that such a number is an universal quantifier of a given geometry in the limit $n \to \infty$, in which it loses its entire dependence on the accretion disk modelling. Since in this work we are interested in the $n=2$ ring, which offers a good compromise between a weak enough dependence on the disk's emission modelling and realistic/optimistic interferometric detection in the future, we shall approximate it by the $I_2/I_1$ flux. In this regard, we are taking advantage of the fact that the sequence of photon rings quickly approximate the (gravitationally lensed) critical curve, the latter corresponding to the limit $n \to \infty$. Indeed, in the Schwarzschild geometry, for $n=2$ the corresponding Lyapunov exponent approximates the exact value $\gamma =-\pi$ by an error of $\sim 0.3\%$, far below other observational uncertainties in this problem. We find its value for every geometry by tracking the relative locations of the $n=1$ and $n=2$ trajectories of the light rays in their winding around the black hole.

Following this approach we report the values of such a Lyapunov exponent in Table \ref{table}, where we observe that eleven geometries decrease its value, four increase it, and one leaves it unchanged. By inspection of this Table we see that any correlation between such an index and the compactness (i.e. the mass-horizon radius ratio) is weak. Indeed, while most alternative geometries are more compact than its Schwarzschild counterpart (and in such a case its shadow's radius is also smaller), and  lower horizon radius tend to decrease the corresponding value of the Lyapunov exponent, this trend is weak and furthermore contaminated by the three geometries supported by effective geodesics (note that models that do not saturate the EHT bound, and the JNW by its lack of a horizon, must also be left aside from this comparison by obvious reasons). As for the photon sphere radius, no correlation with the Lyapunov exponent is found. In any case and, as already pointed out by Vea, most geometries (indeed in our case all except three) decrease the critical impact parameter (the shadow's radius in the EHT interpretation), where the corresponding constraints leave a wider margin for modifications with respect to the predictions of the Schwarzschild geometry.

\begin{figure*}[t]
\includegraphics[width=4.25cm,height=3.75cm]{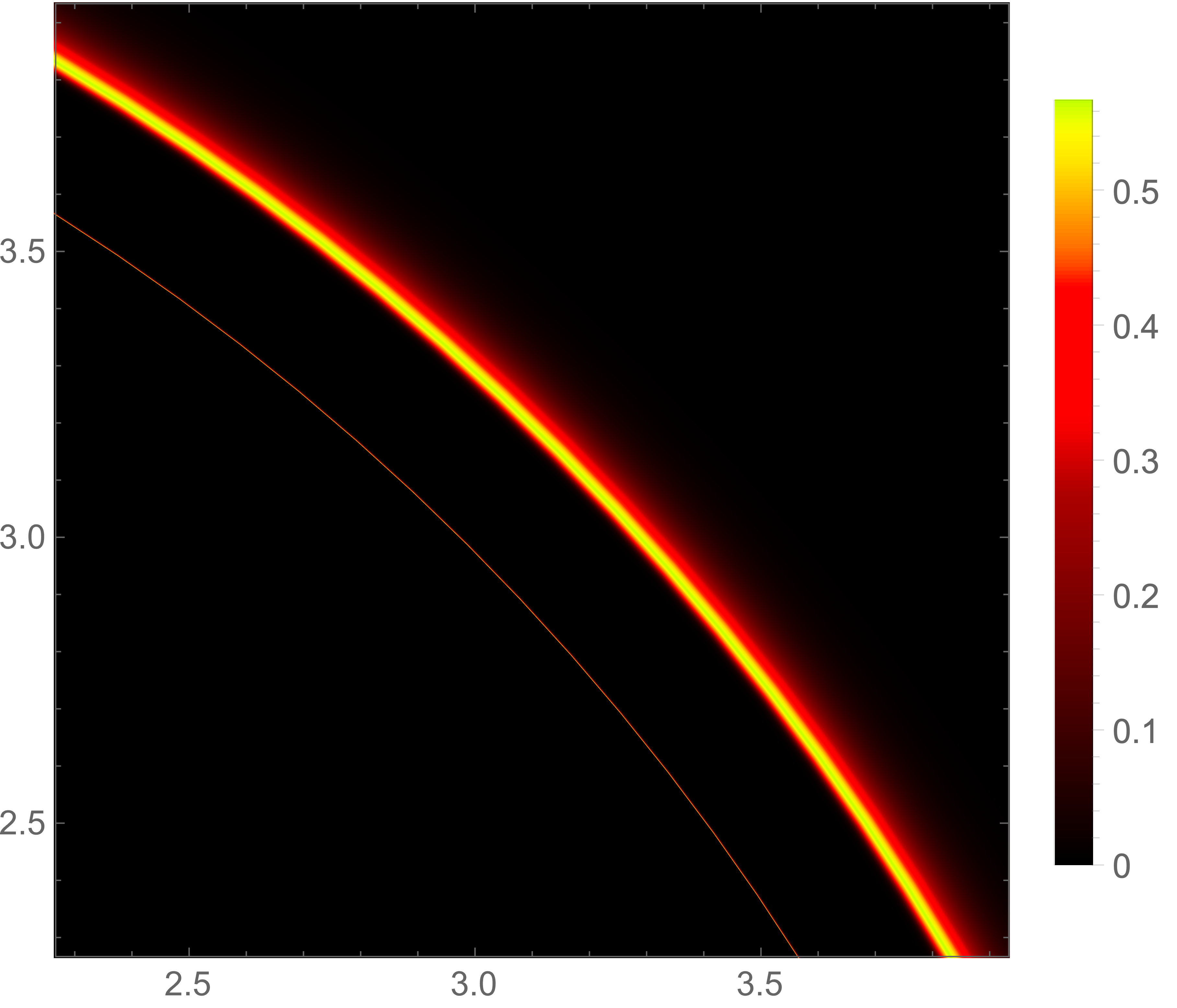}
\includegraphics[width=4.25cm,height=3.75cm]{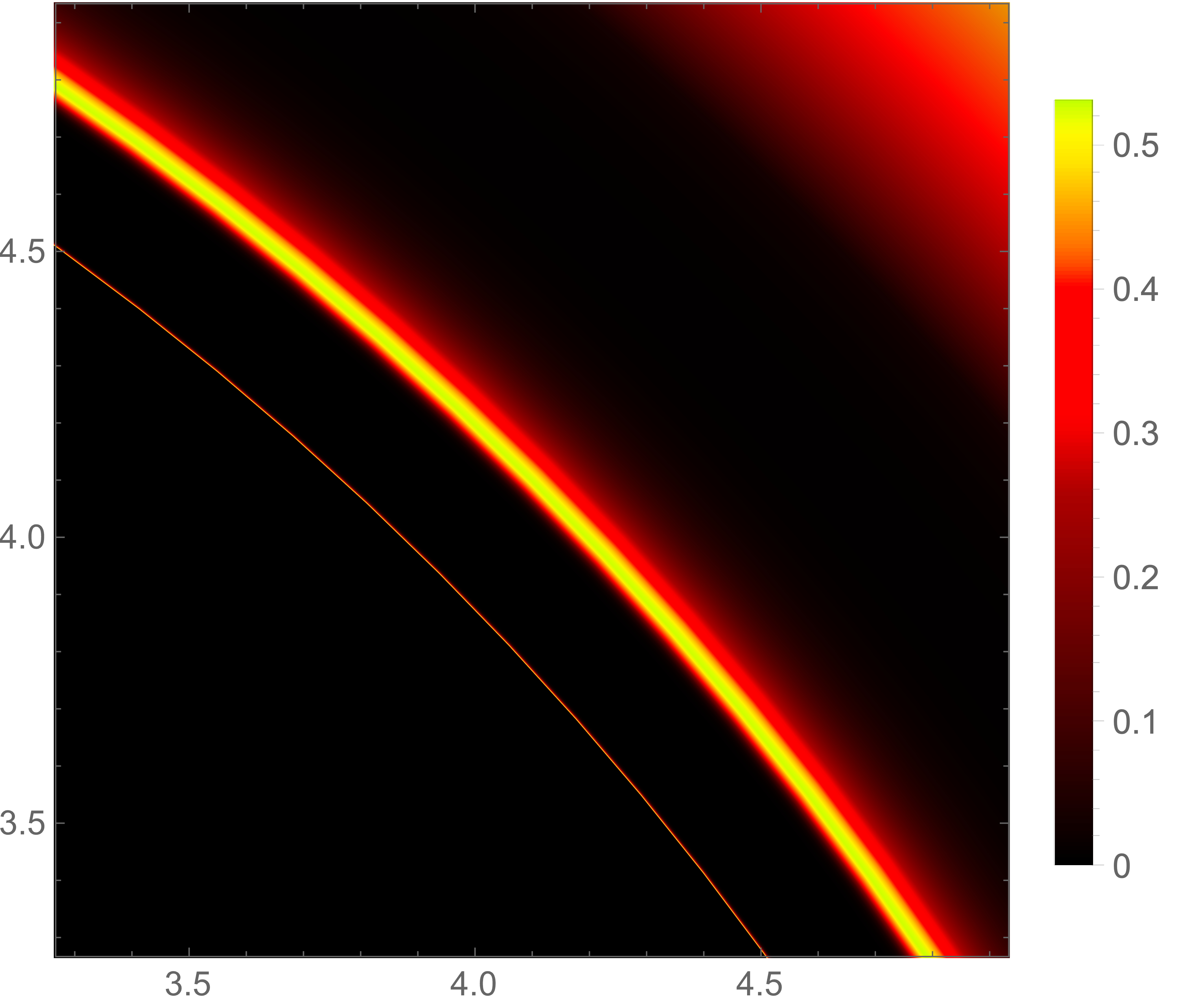}
\includegraphics[width=4.25cm,height=3.75cm]{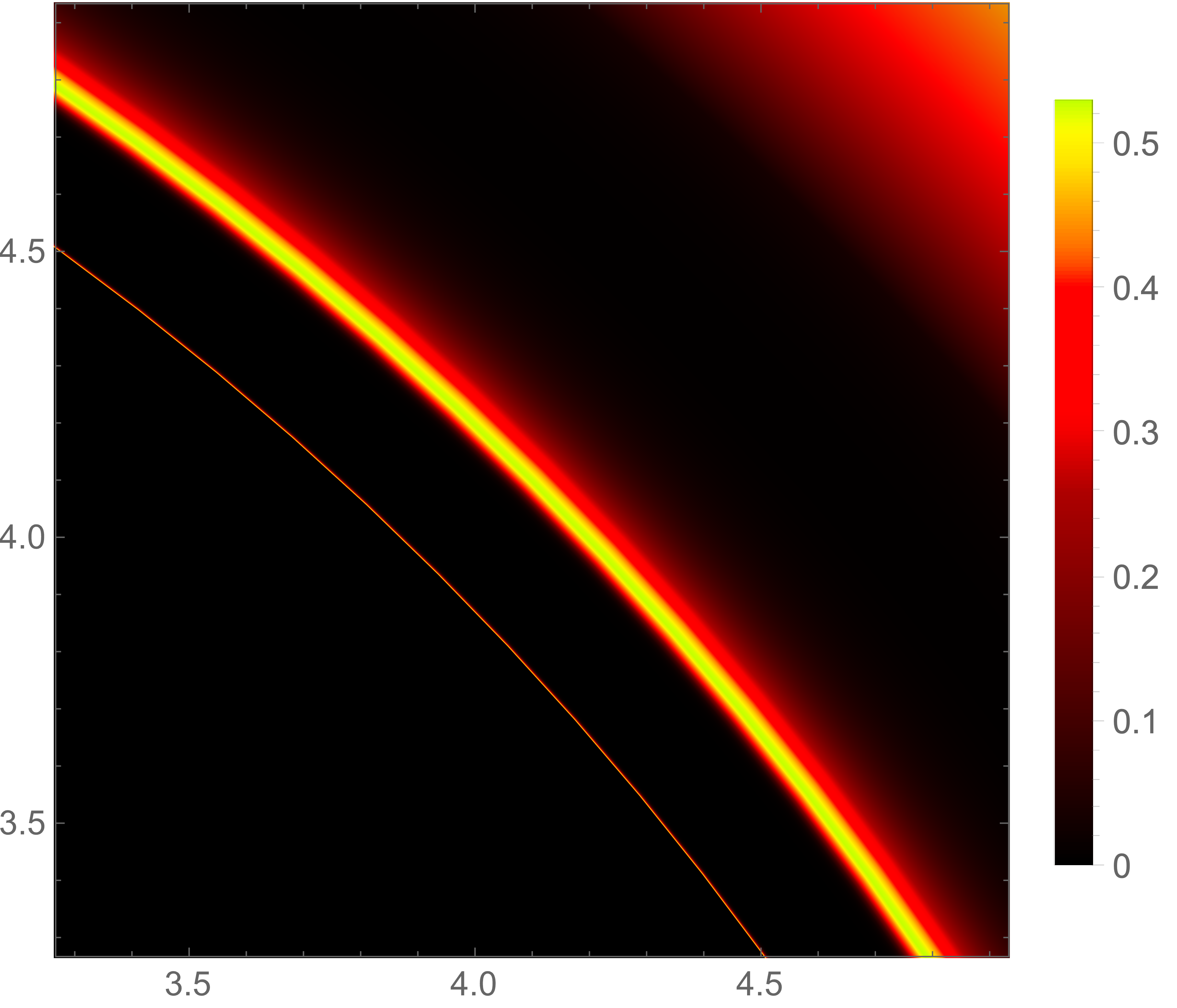}
\includegraphics[width=4.25cm,height=3.75cm]{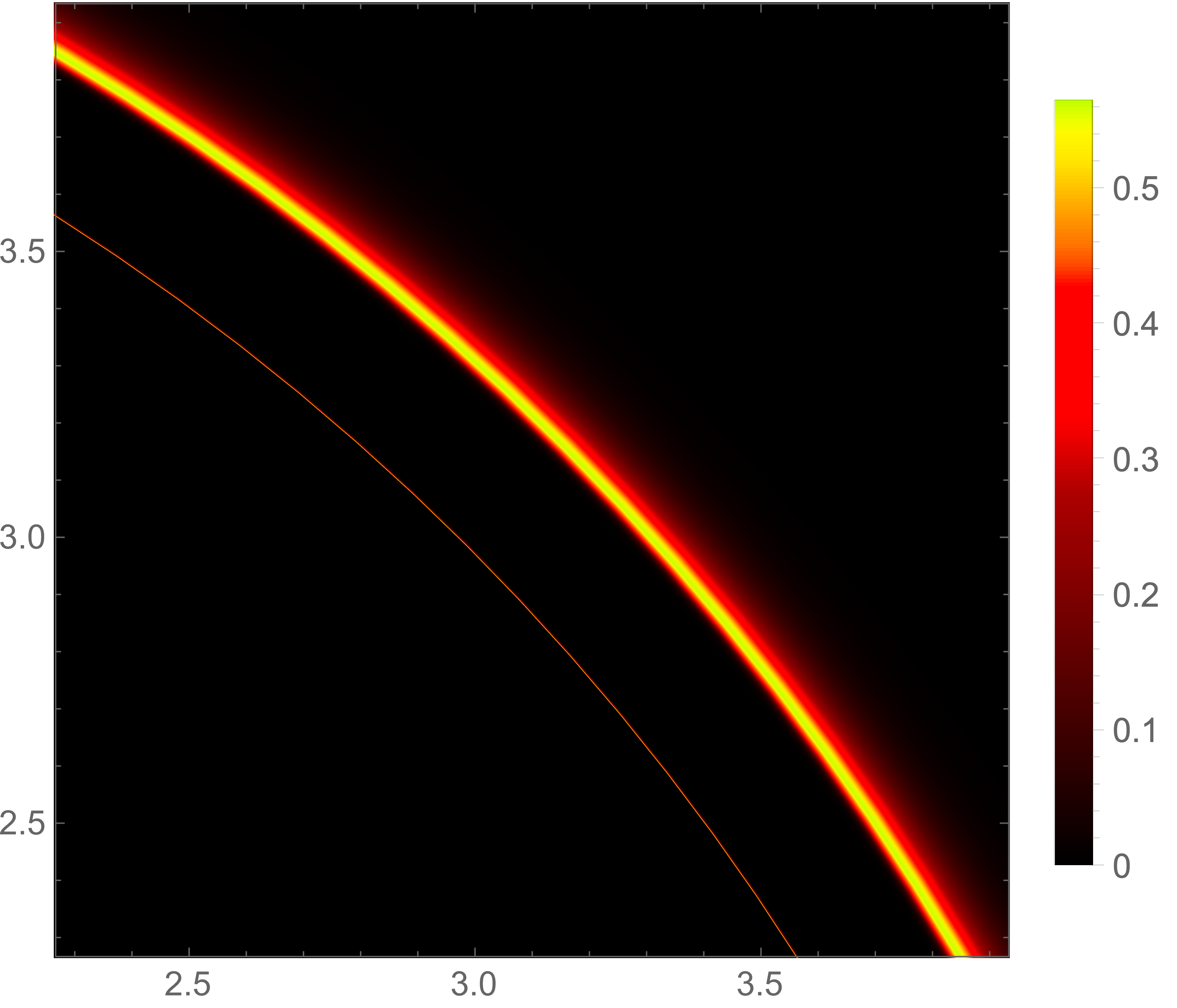}
\includegraphics[width=4.25cm,height=3.75cm]{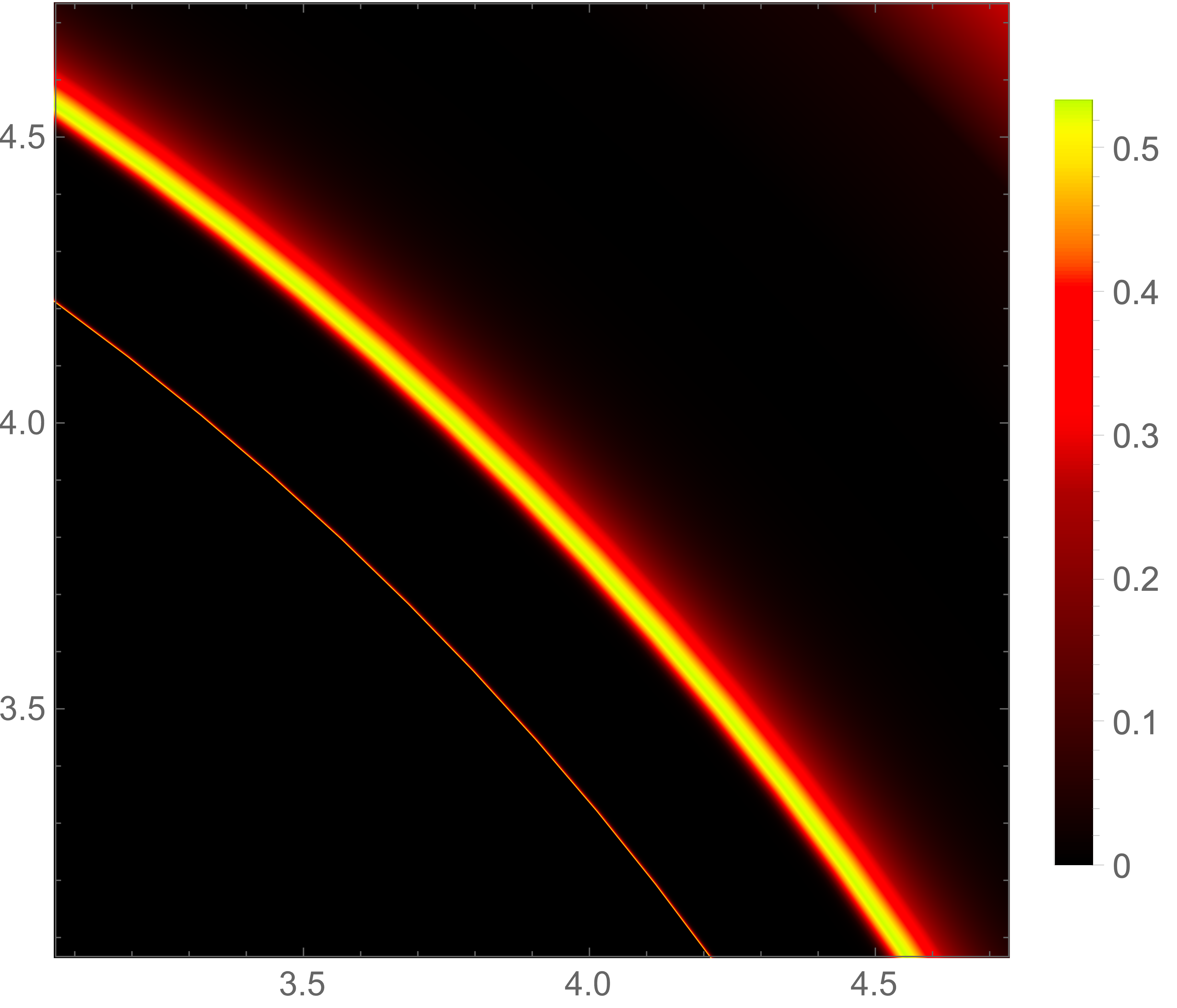}
\includegraphics[width=4.25cm,height=3.75cm]{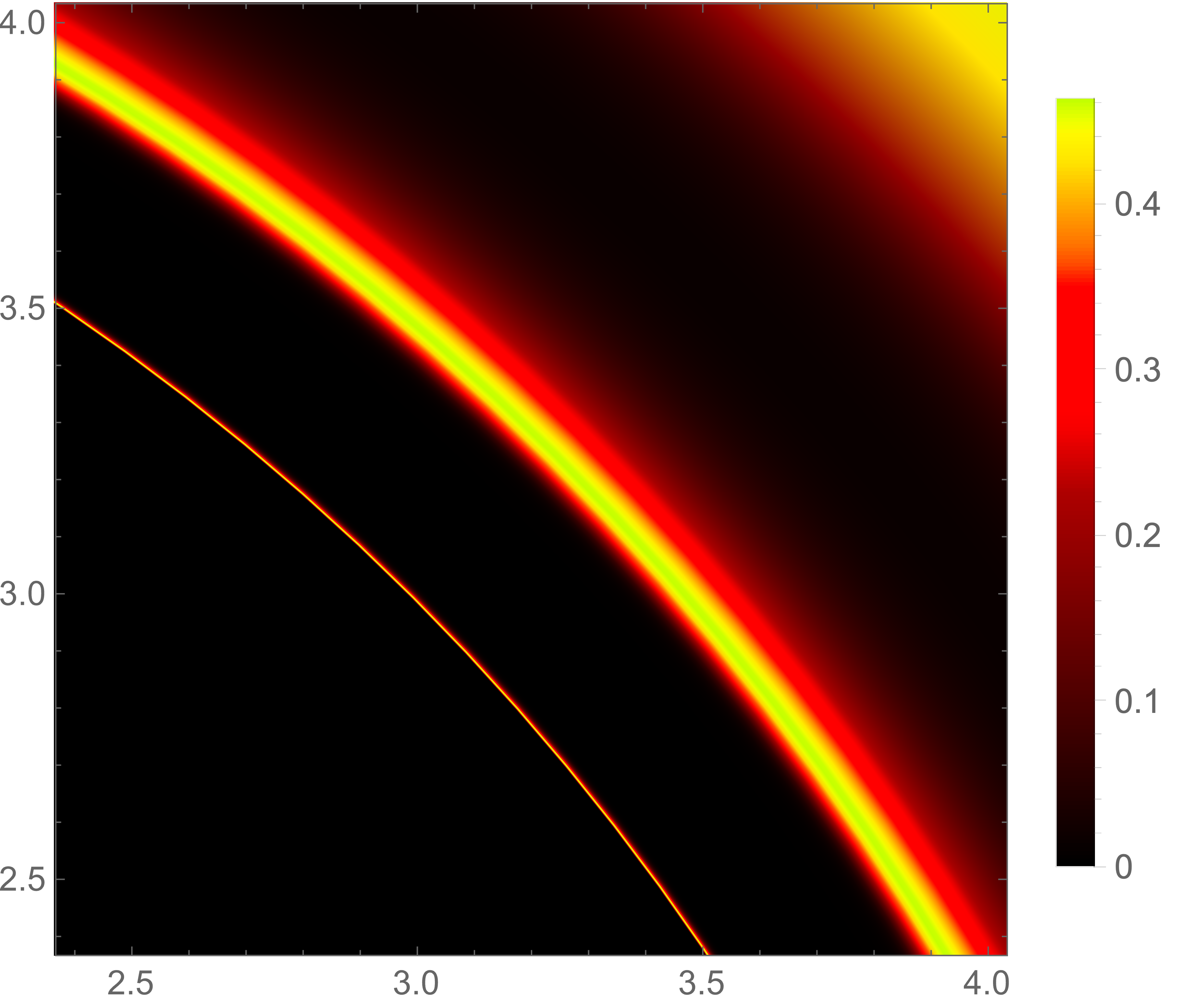}
\includegraphics[width=4.25cm,height=3.75cm]{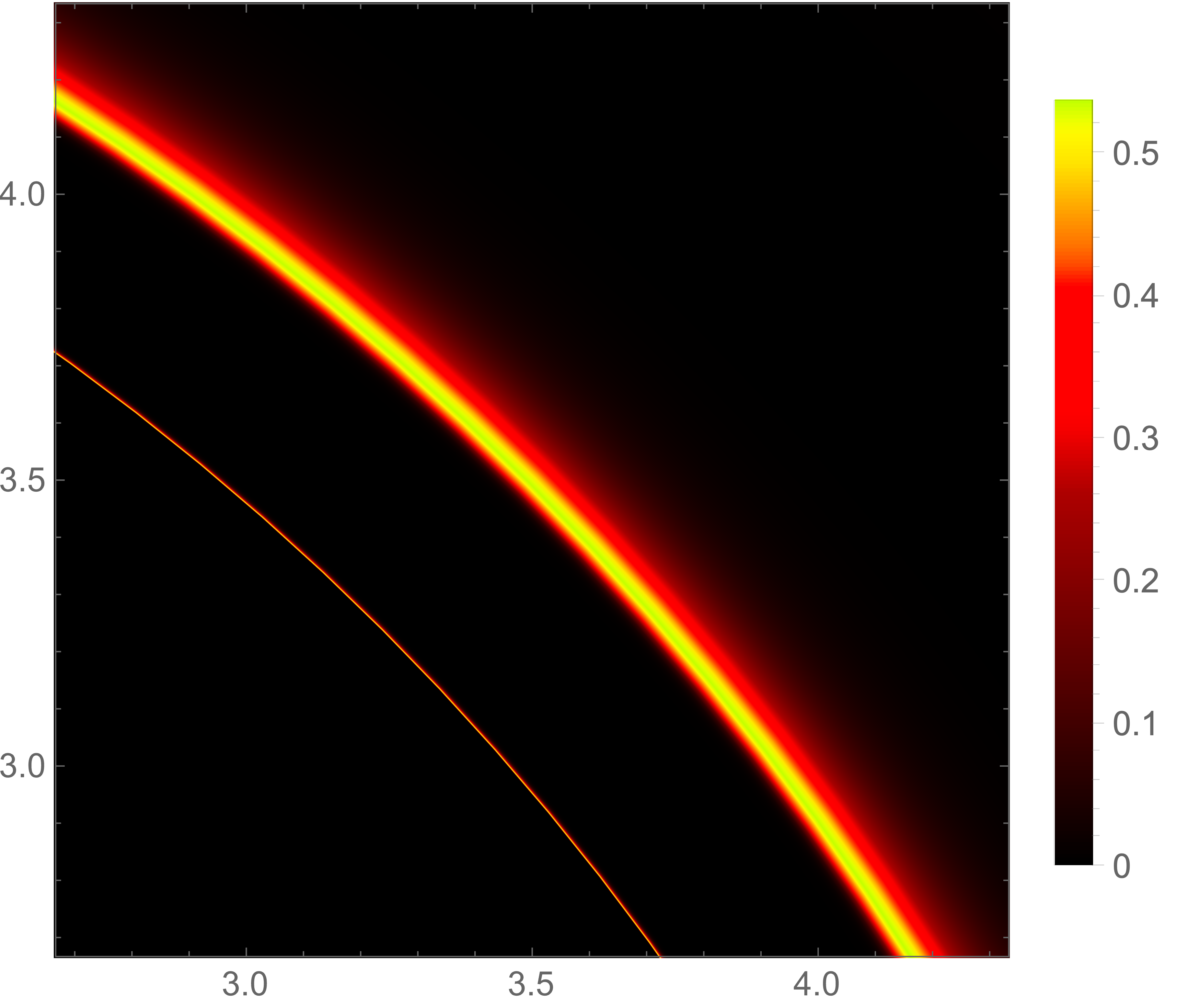}
\includegraphics[width=4.25cm,height=3.75cm]{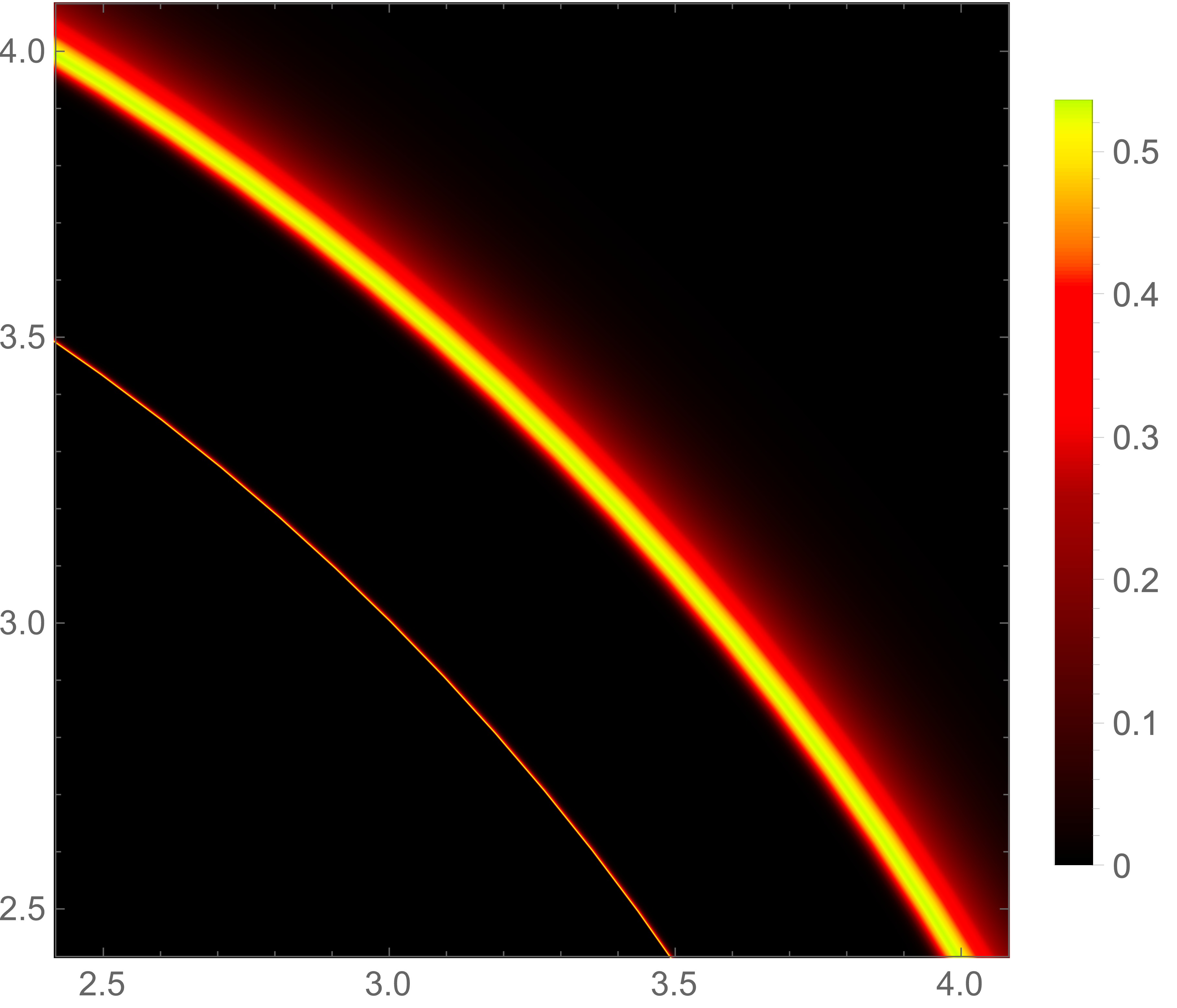}
\includegraphics[width=4.25cm,height=3.75cm]{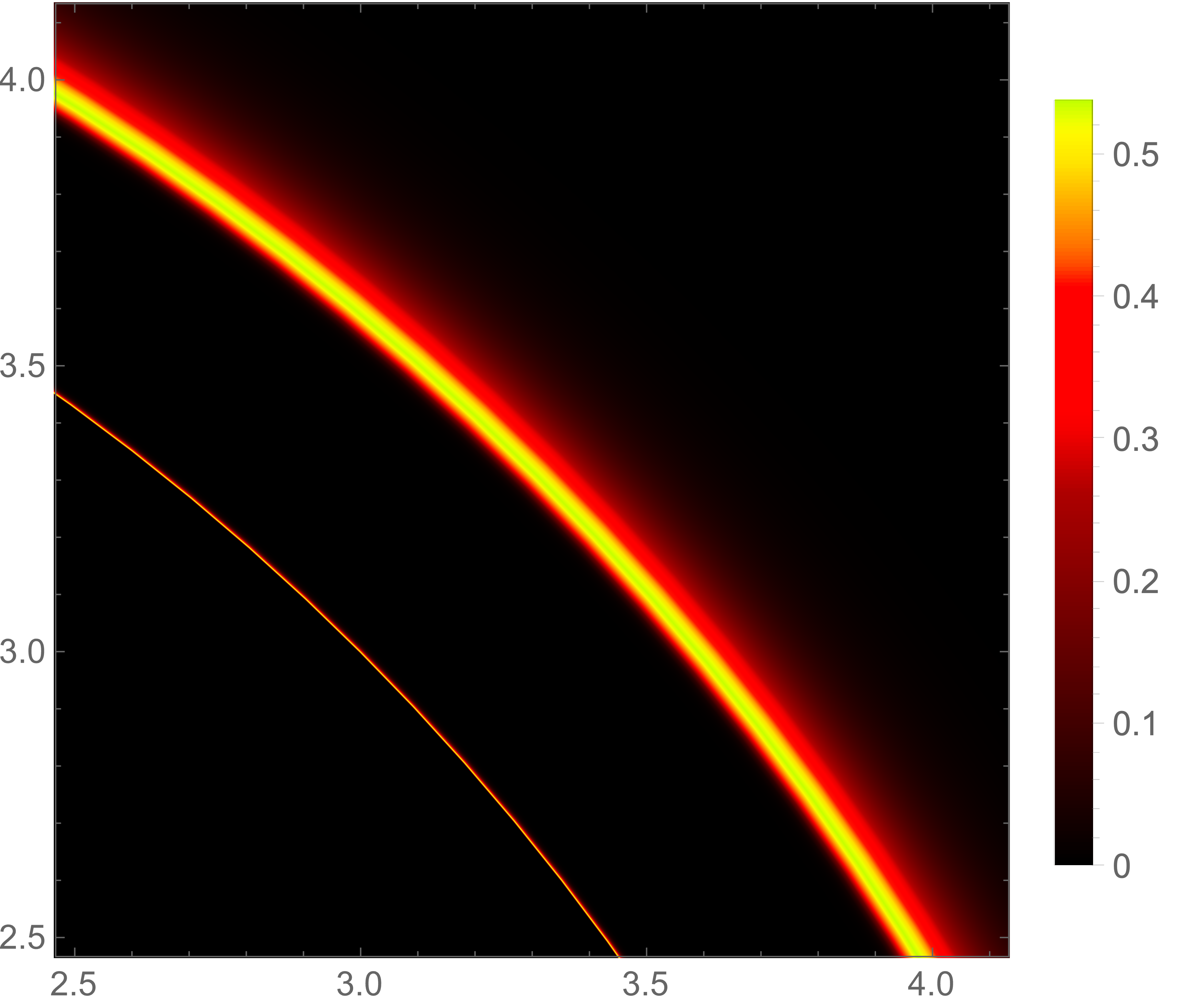}
\includegraphics[width=4.25cm,height=3.75cm]{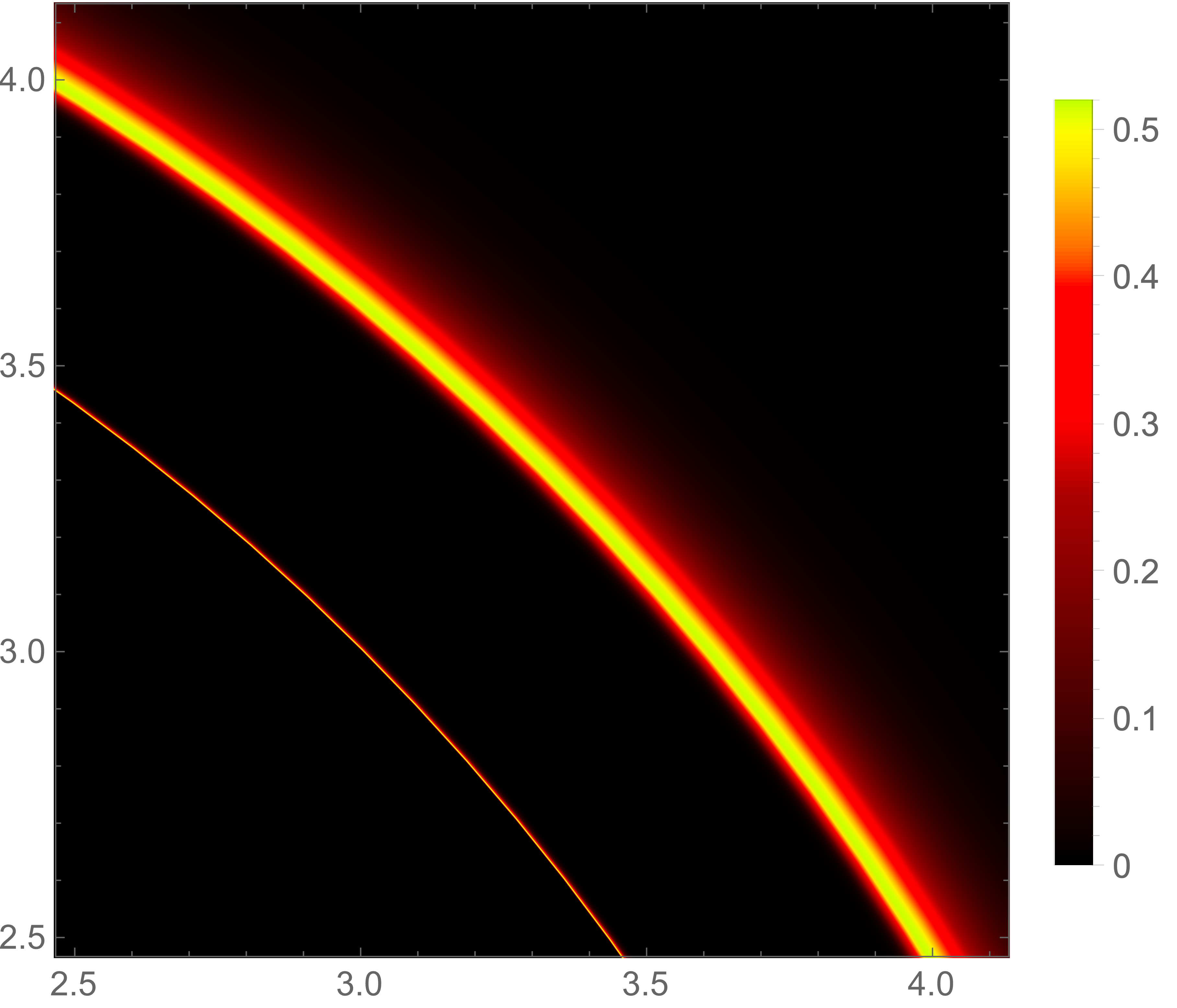}
\includegraphics[width=4.25cm,height=3.75cm]{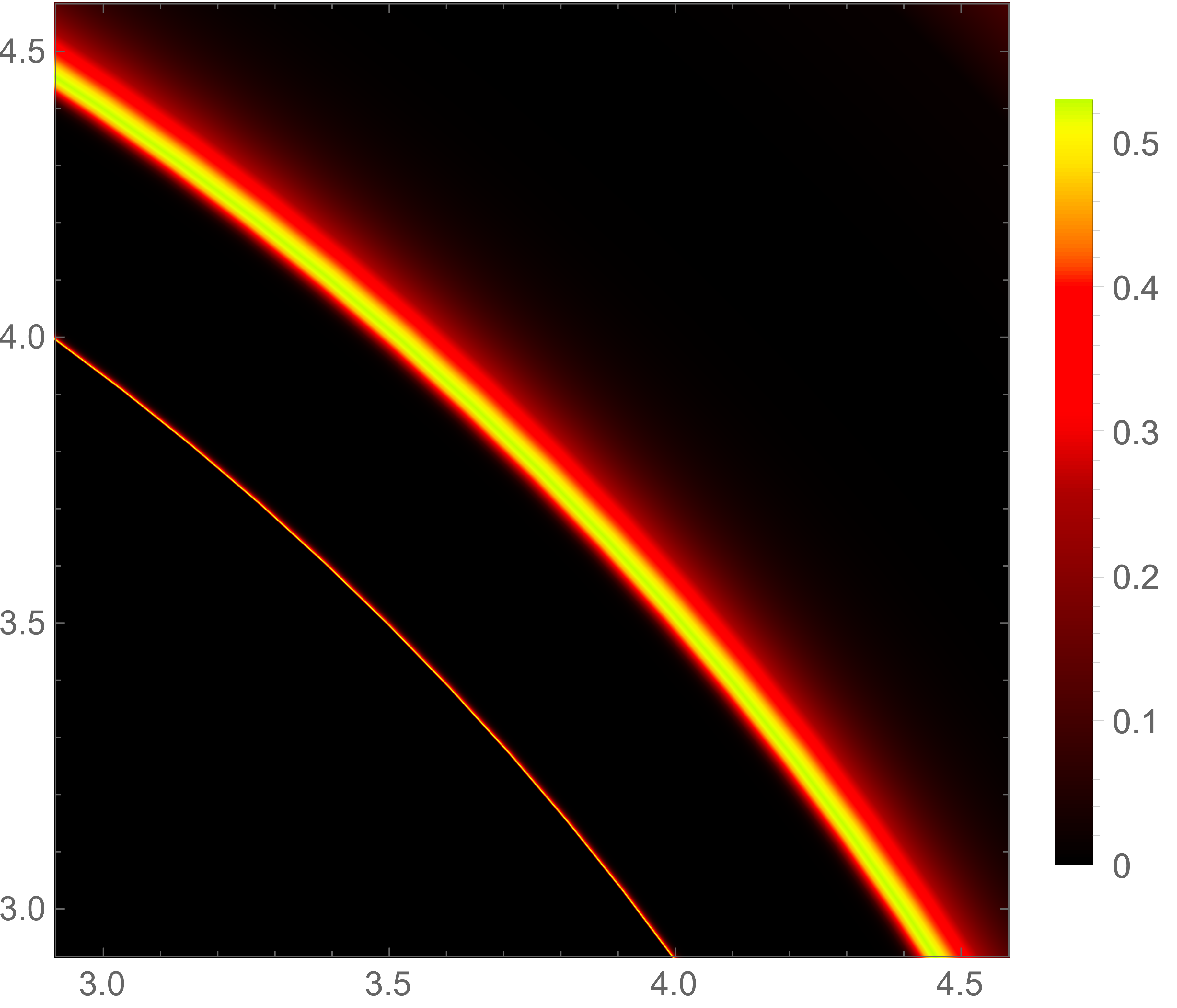}
\includegraphics[width=4.25cm,height=3.75cm]{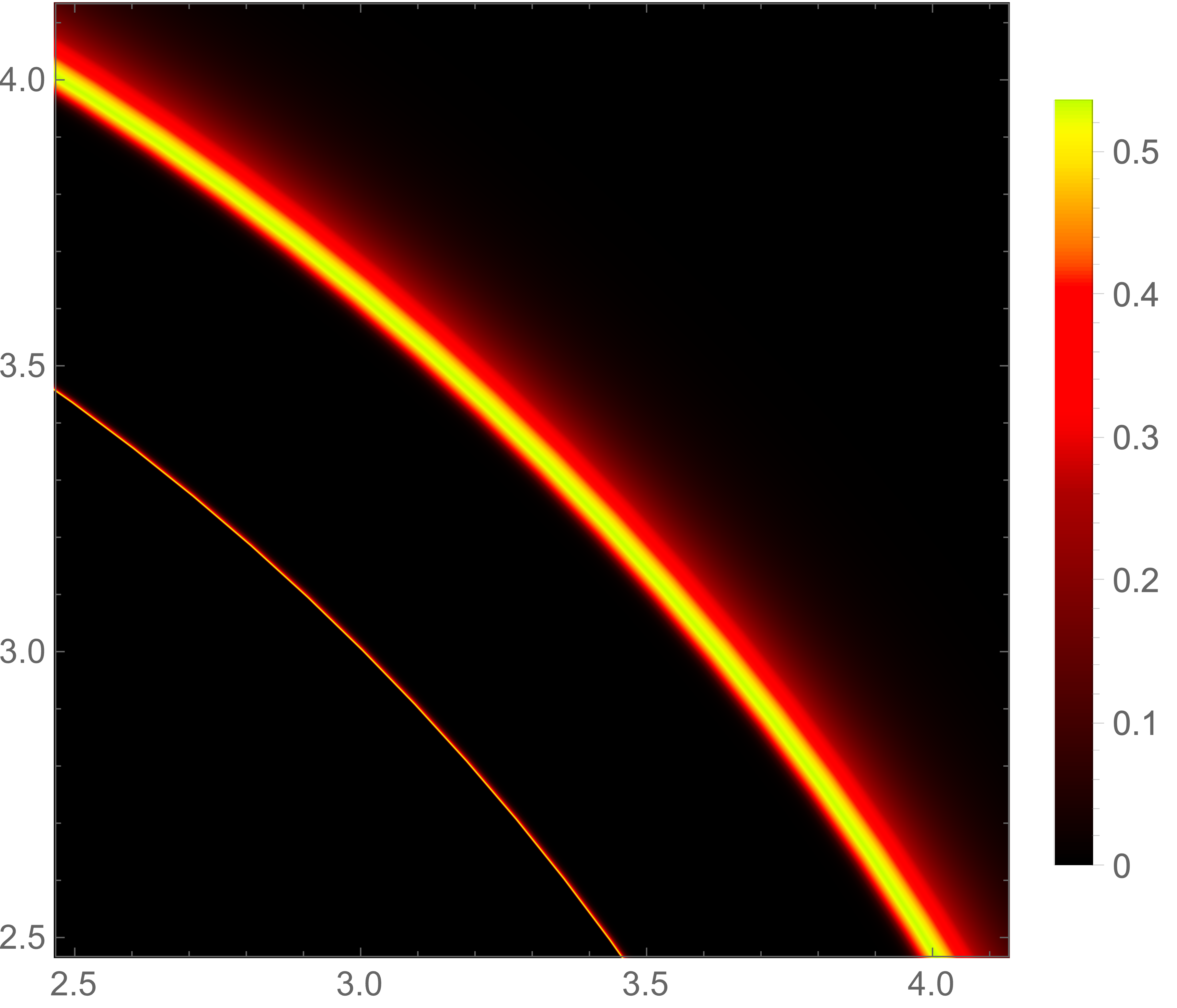}
\includegraphics[width=4.25cm,height=3.75cm]{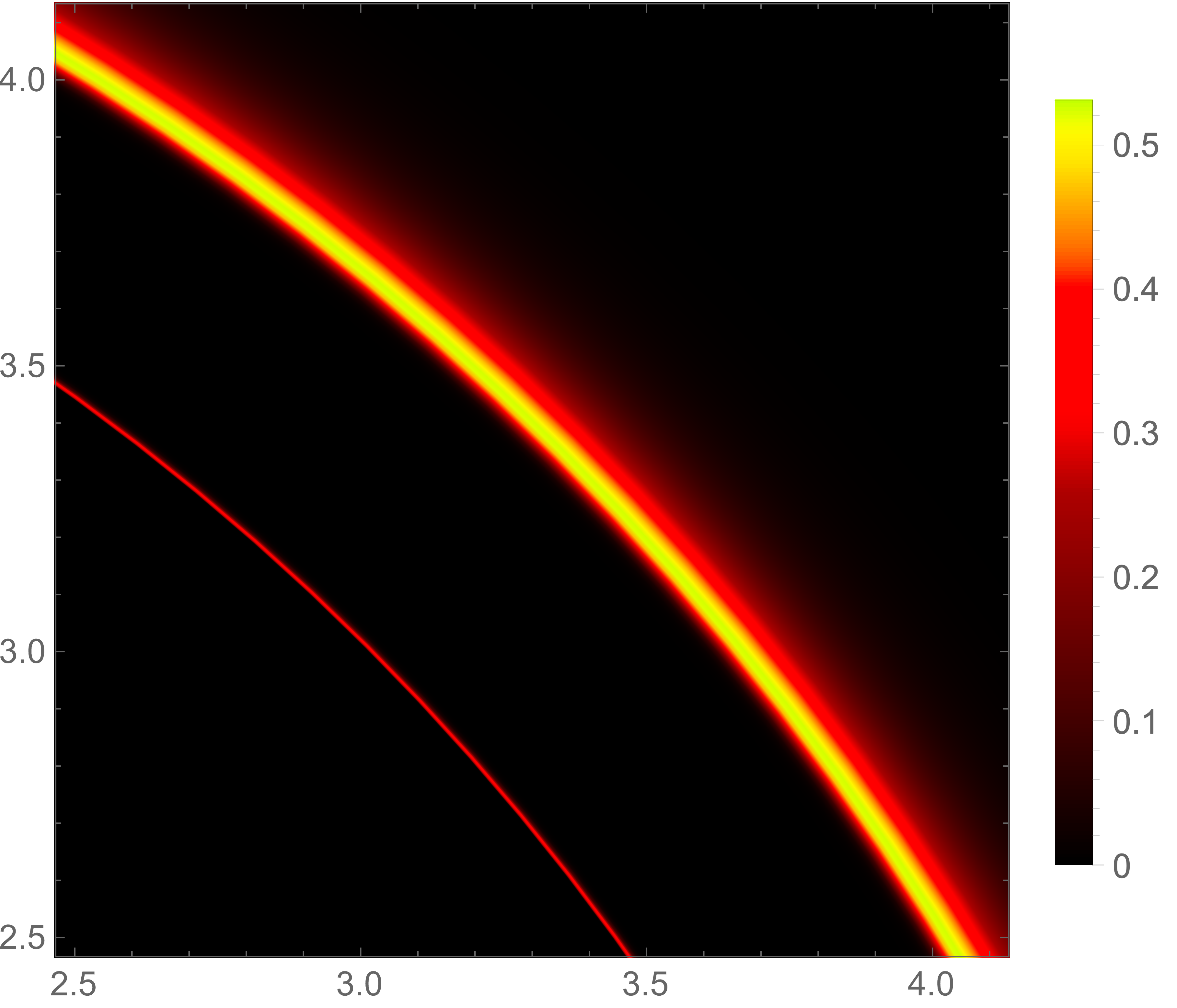}
\includegraphics[width=4.25cm,height=3.75cm]{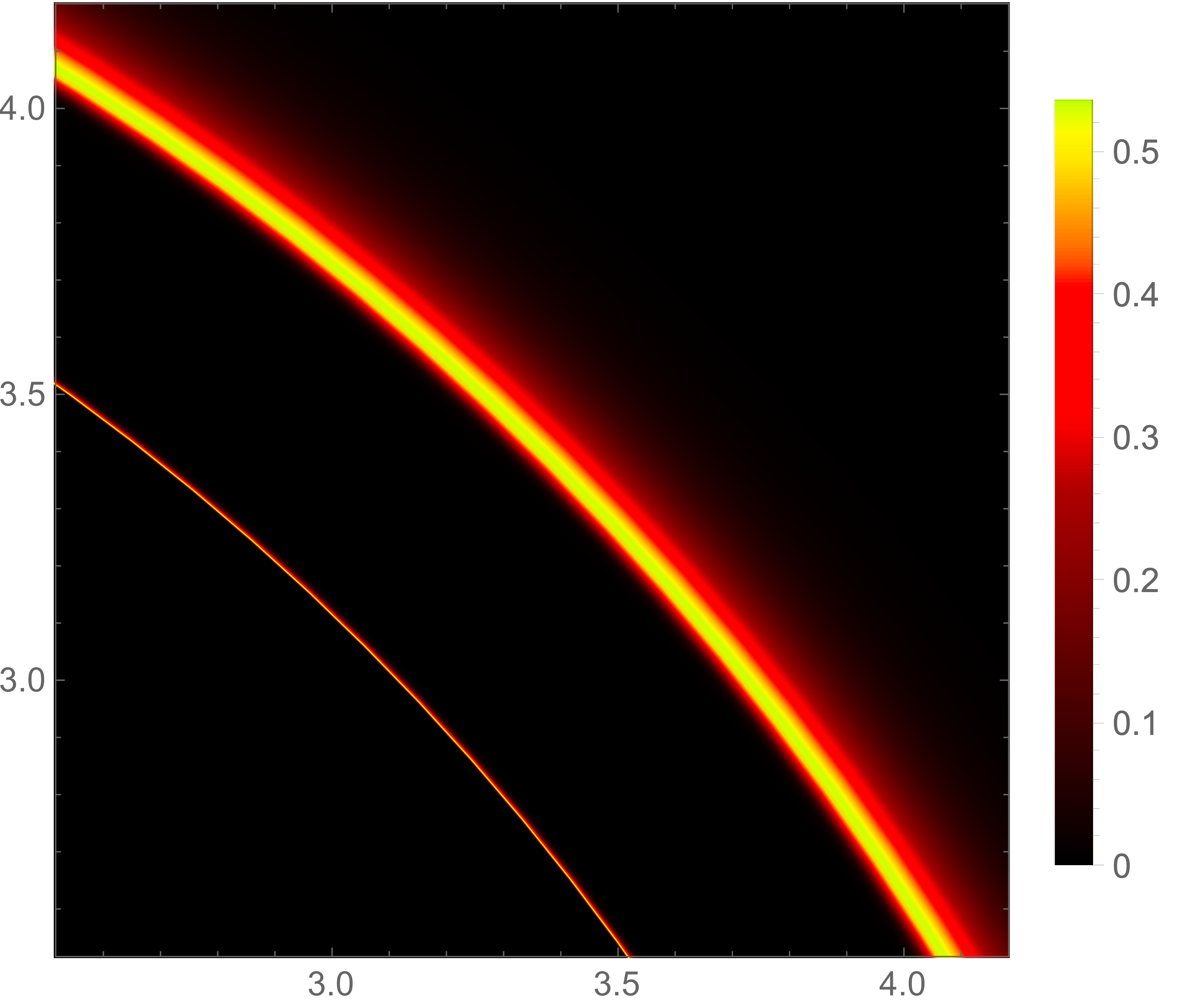}
\includegraphics[width=4.25cm,height=3.75cm]{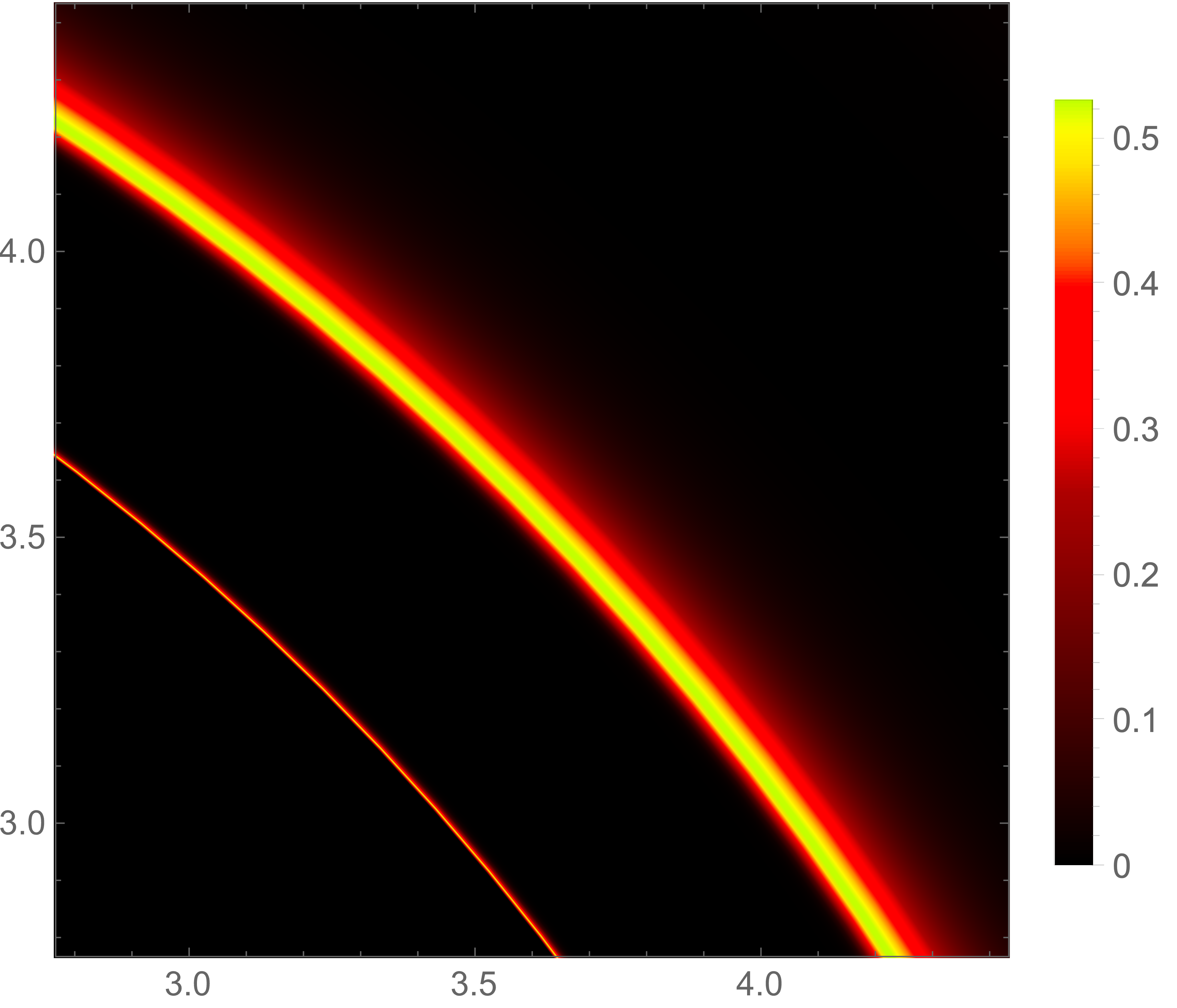}
\includegraphics[width=4.25cm,height=3.75cm]{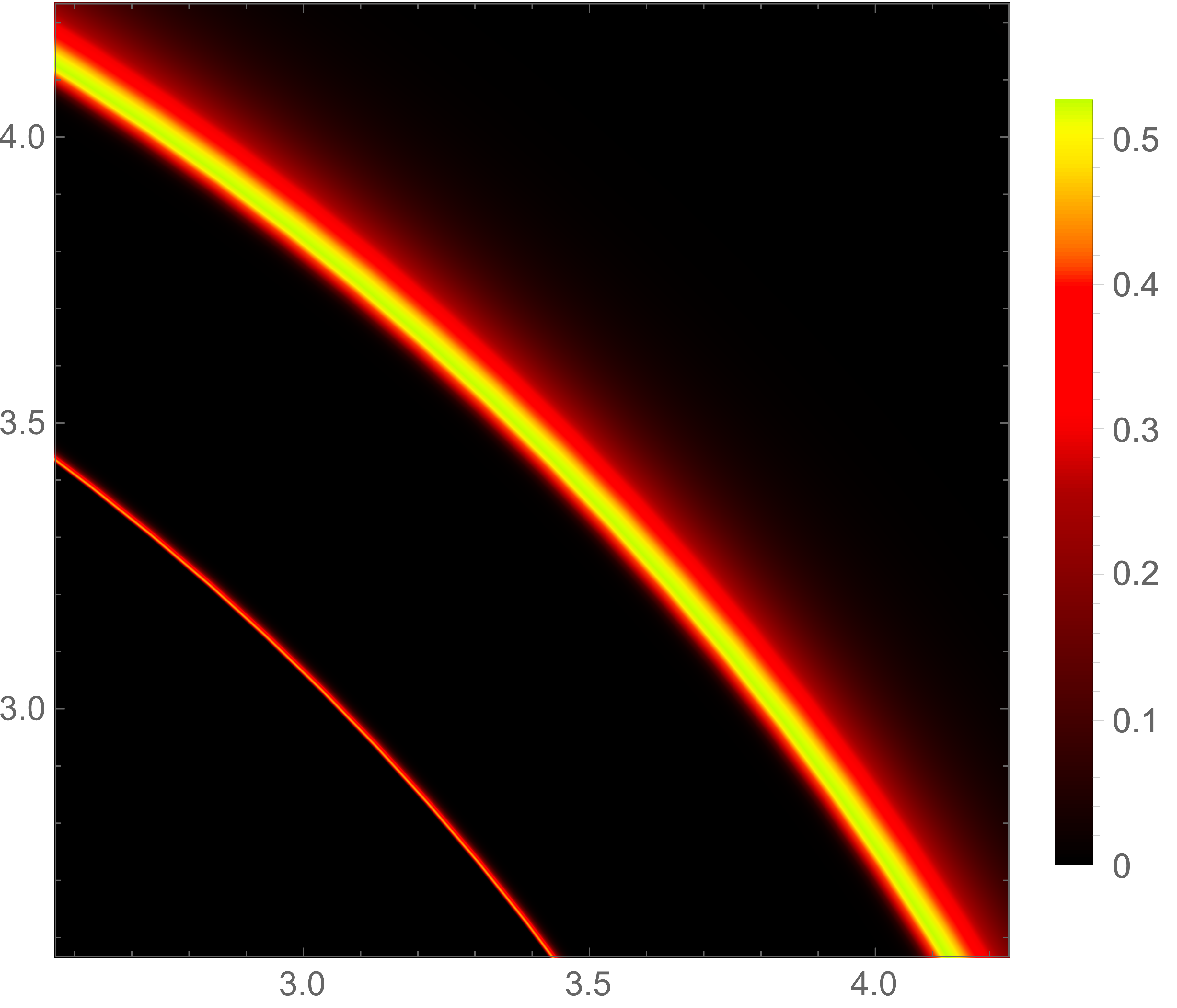}
\caption{Zoom in of the $n=1$ (brighter) and $n=2$ (dimmer) photon rings in the impact parameter space for (from left to right and top to bottom) LQG, KS, ConfSca, DM, SV, JNW, Sen, GCSV (e), EMD, Bronnikov (e), Hayward, RN, EH (e), Frolov, Bardeen, and GK, ordered in decreasing values of their Lyapunov exponent (units of $M=1$) using the emission model GLM3 in Eqs.(\ref{eq:GLM}) and (\ref{eq:GLM3}).}
\label{fig:prs}
\end{figure*}

Keeping with the discussion of the Lyapunov exponent and its relation to the exponential decay of the luminosity of the photon rings, one should note that this is a theoretical expectancy based on the assumption that every photon trajectory will cross emission regions with similar properties. This is certainly not the case since the profile is sensitive to the radius, and hence to the impact parameter (furthermore we are not taking into account any source variabilities on the typical timescale of an orbit). In other words, the Lyapunov exponent is not a direct observable but one would expect deviations in the actual  intensity fluxes fed by the pick of the emission profile; to what extent such observable deviates from the theoretical Lyapunov number is a question of great interest in connecting theoretical properties with actual observables. Indeed, this is what we find when computing the (inverse) flux ratio between the $n=1$ and $n=2$ (which shall be referred to as the {\it extinction rate}) rings for the GLM models, as reported in Table \ref{table}, and whose images (disregarding the Schwarzschild black hole itself) we discuss next. We point out that in generating such images the observed luminosity is normalized to its total value for every geometry and GLM model via Eq.(\ref{eq:Io}), in order to consider as similar settings for each case as possible.

\begin{figure*}[t!]
\includegraphics[width=4.25cm,height=3.75cm]{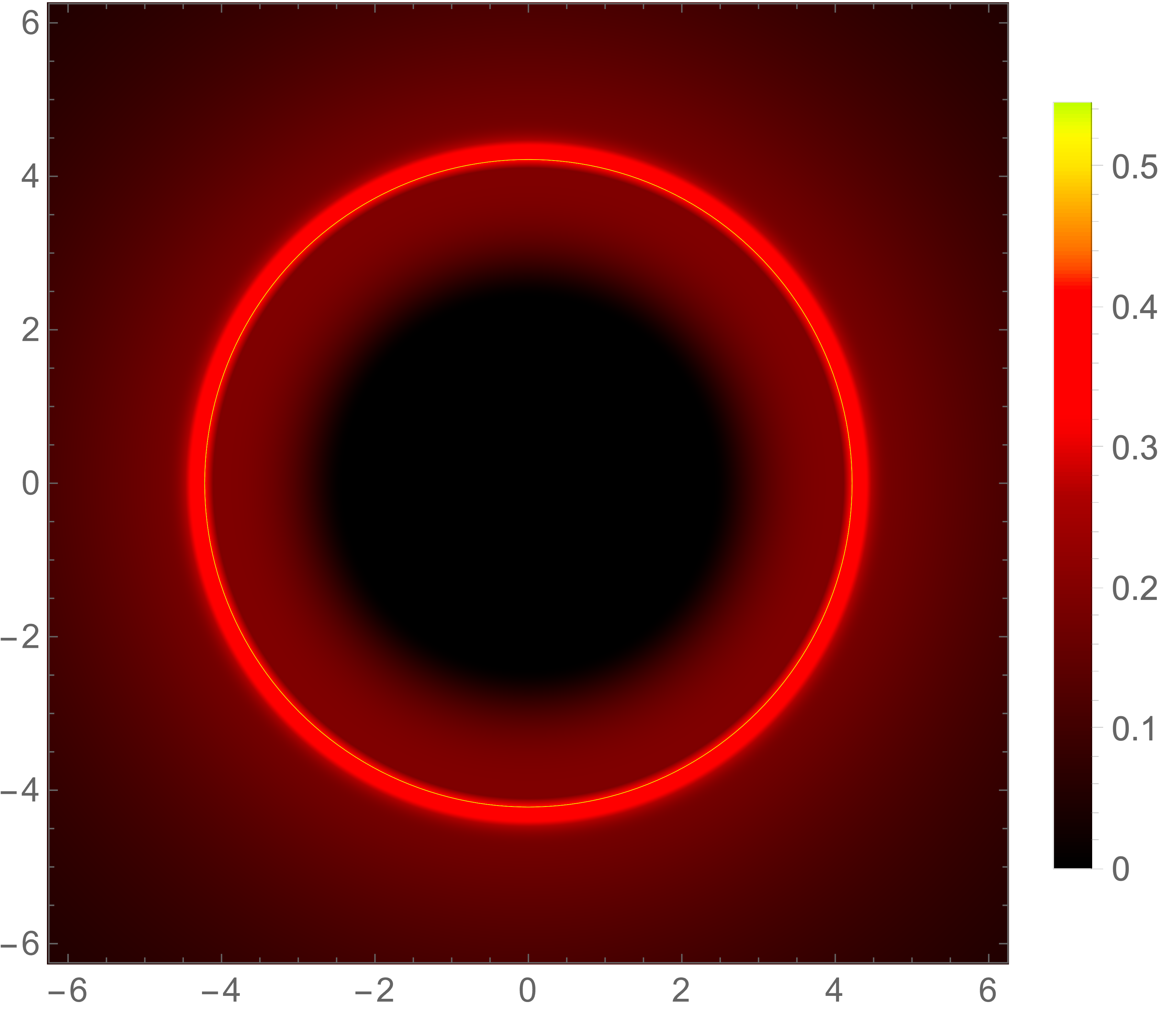}
\includegraphics[width=4.25cm,height=3.75cm]{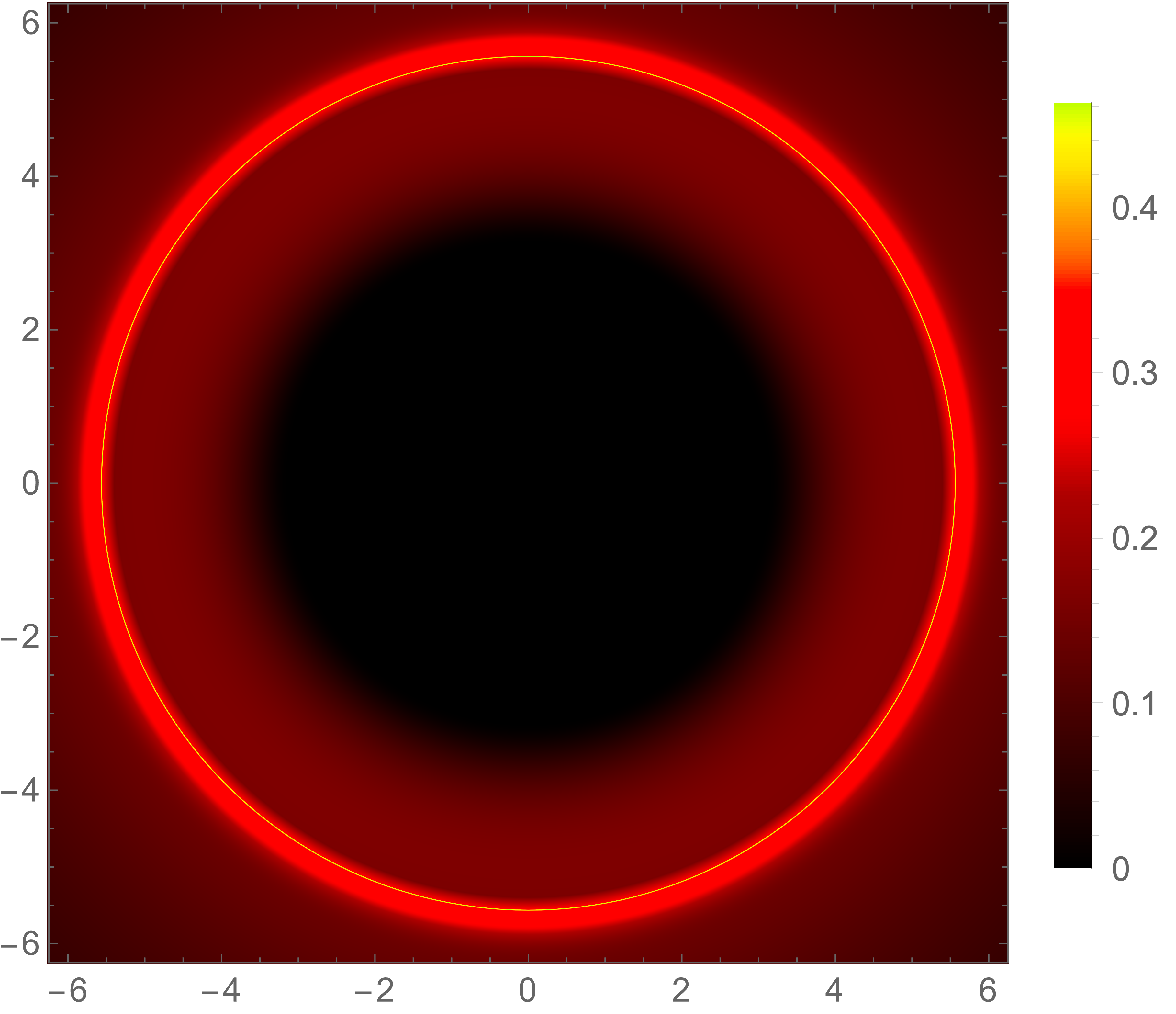}
\includegraphics[width=4.25cm,height=3.75cm]{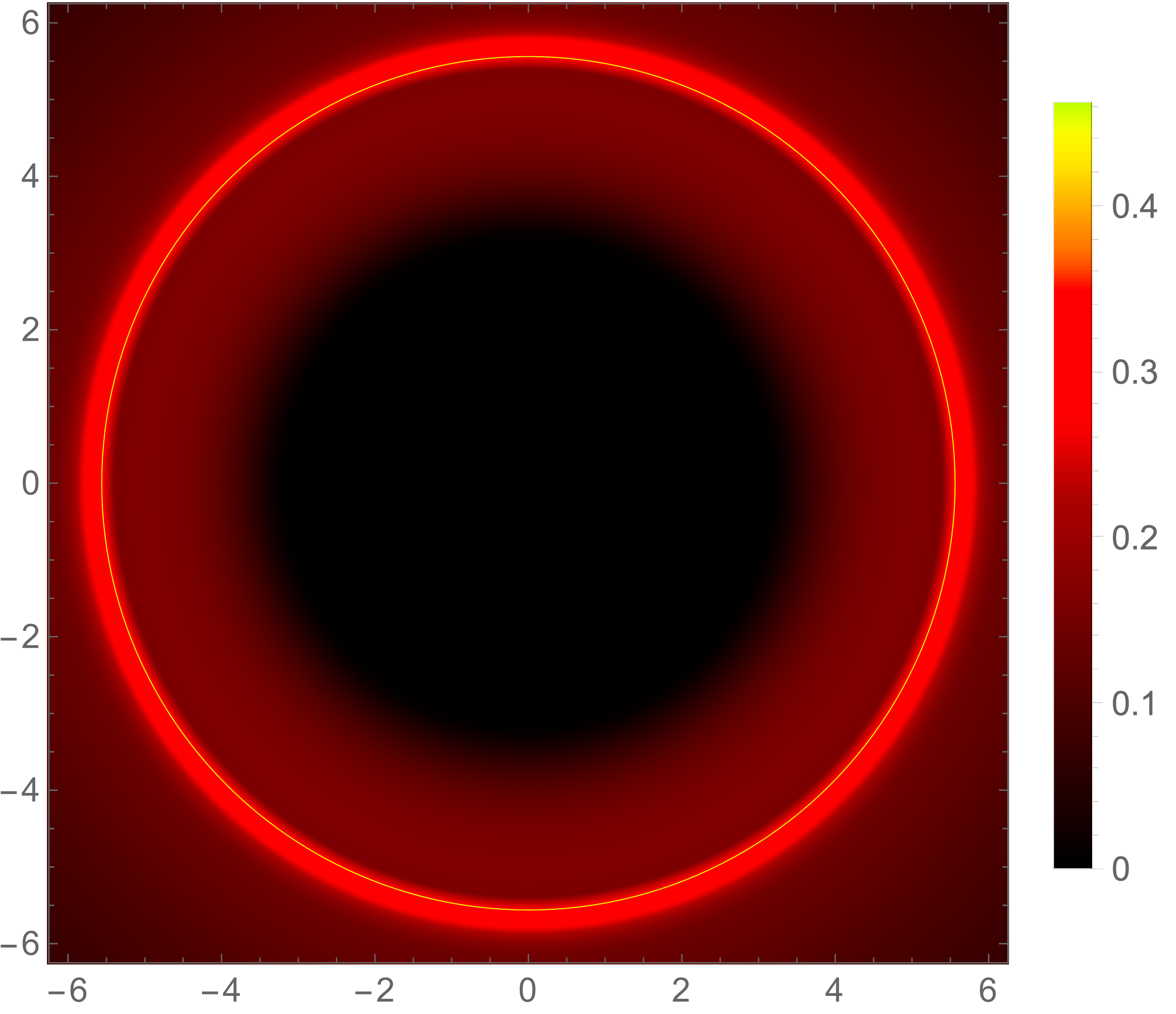}
\includegraphics[width=4.25cm,height=3.75cm]{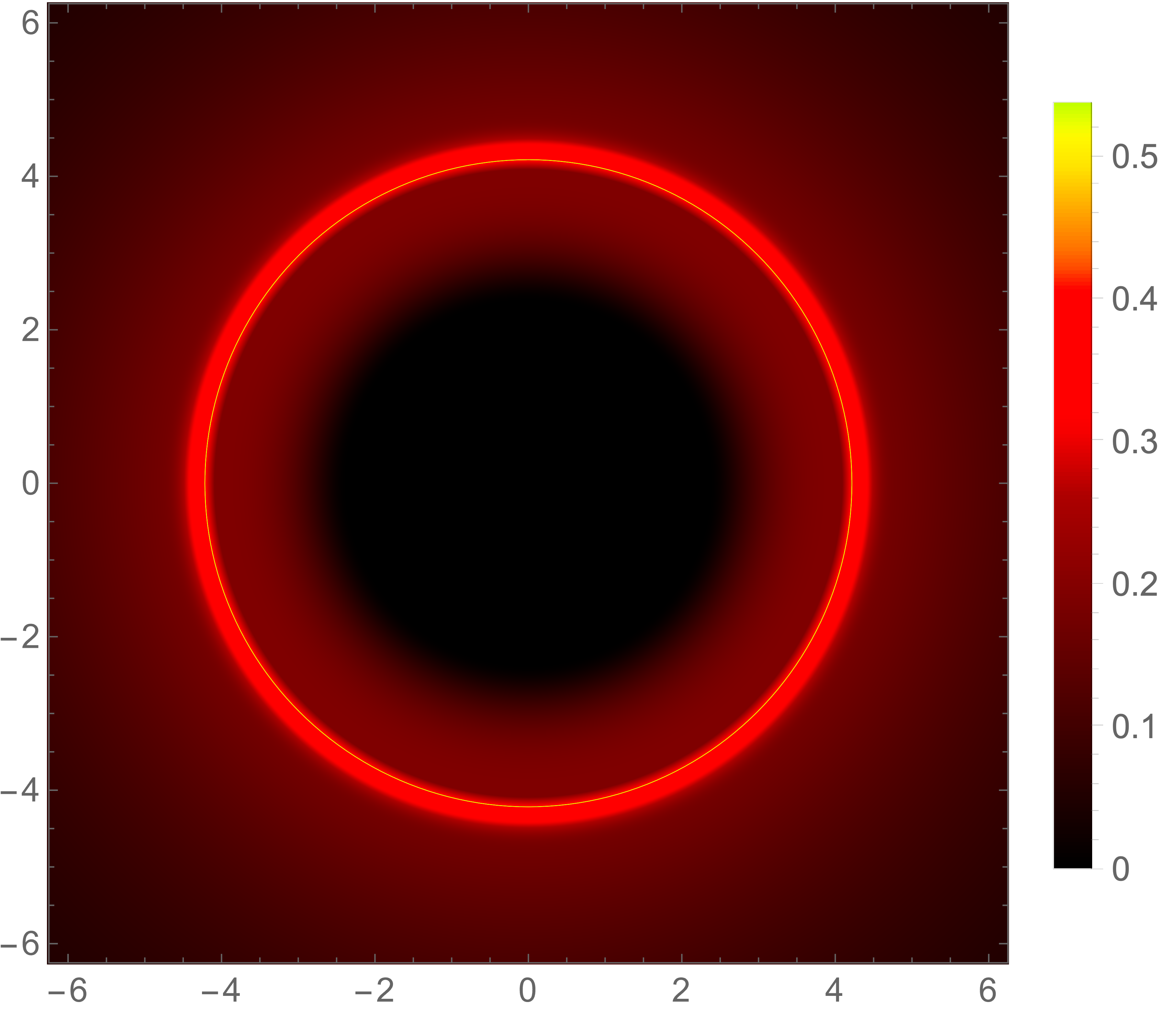}
\includegraphics[width=4.25cm,height=3.75cm]{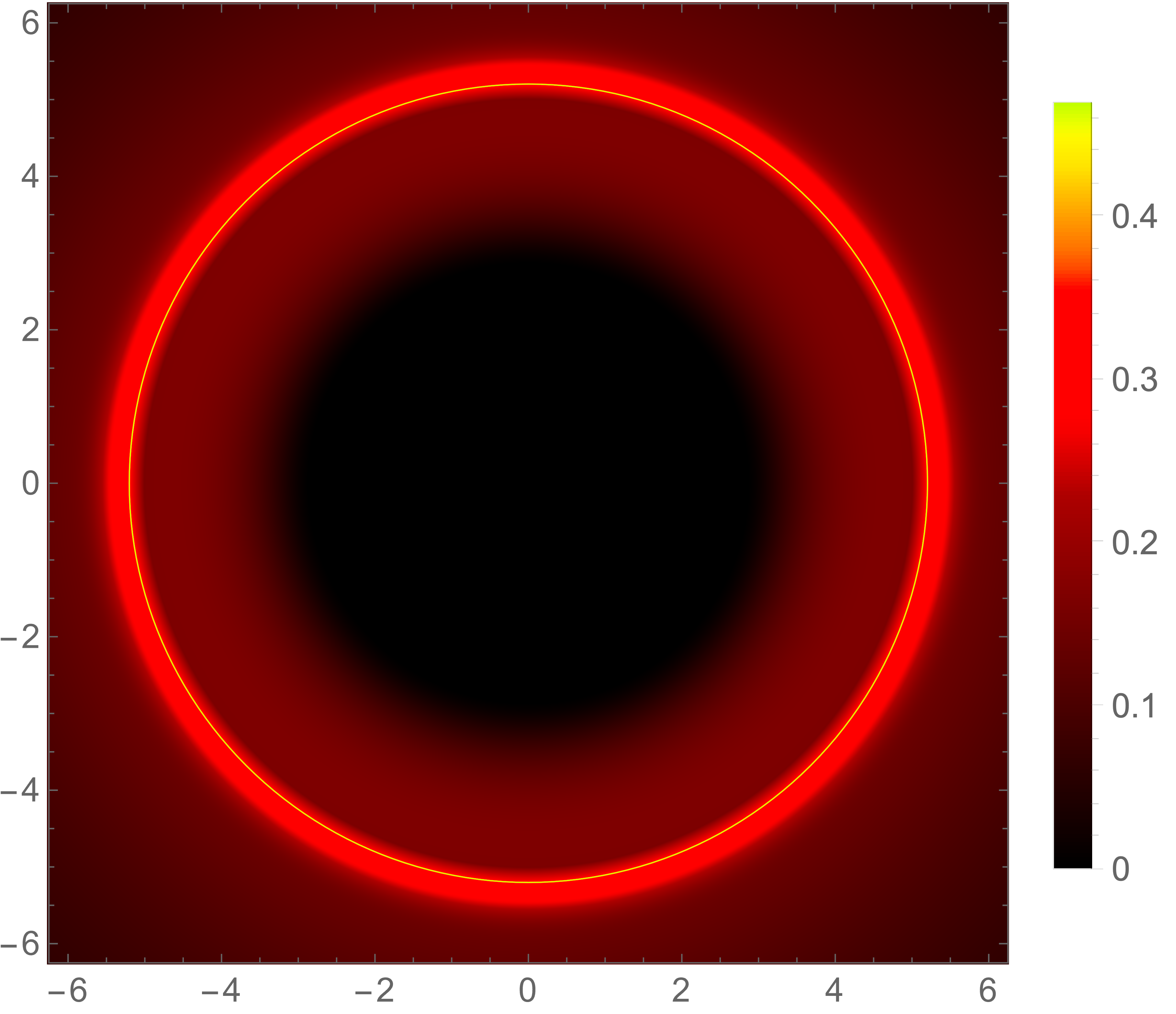}
\includegraphics[width=4.25cm,height=3.75cm]{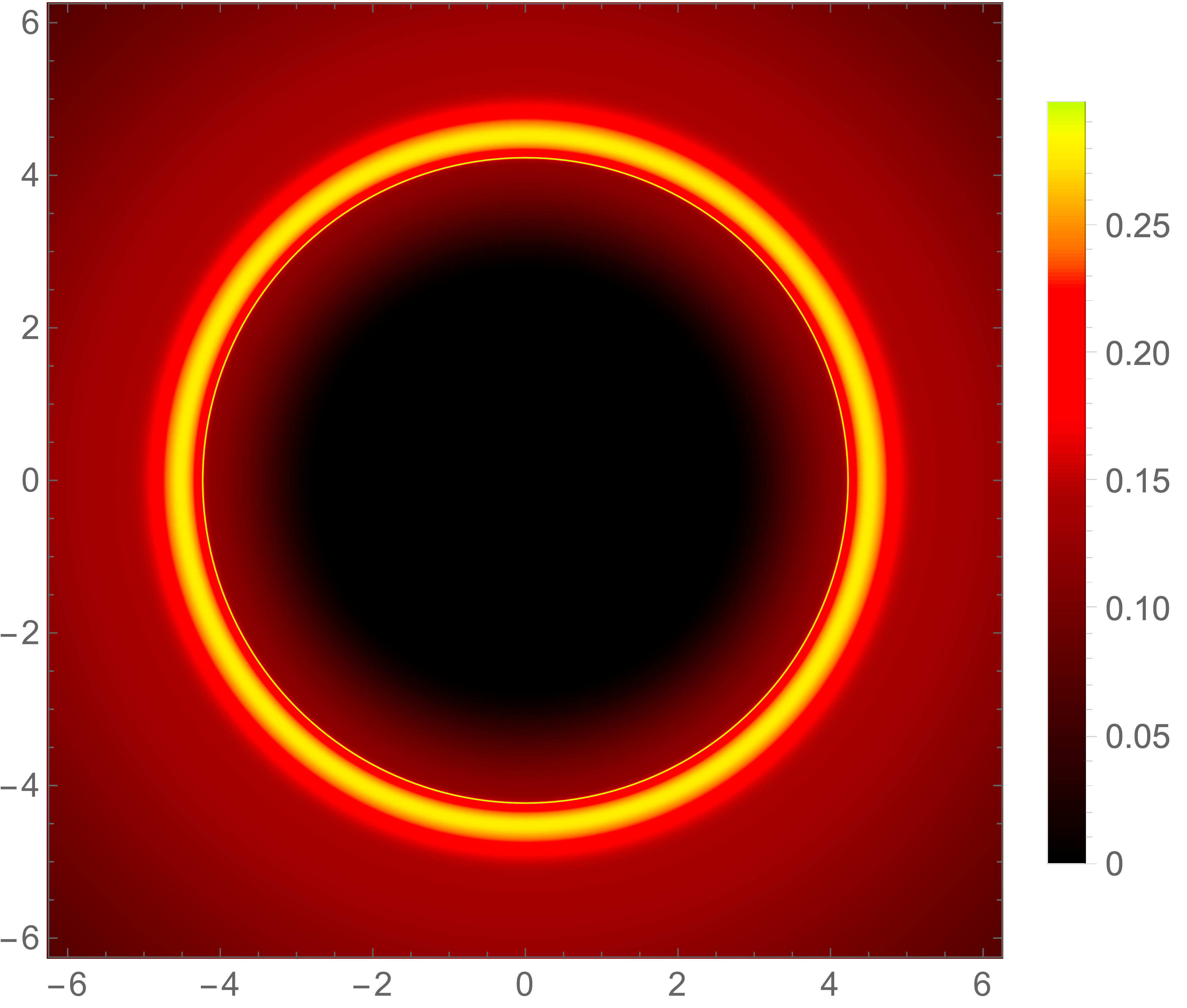}
\includegraphics[width=4.25cm,height=3.75cm]{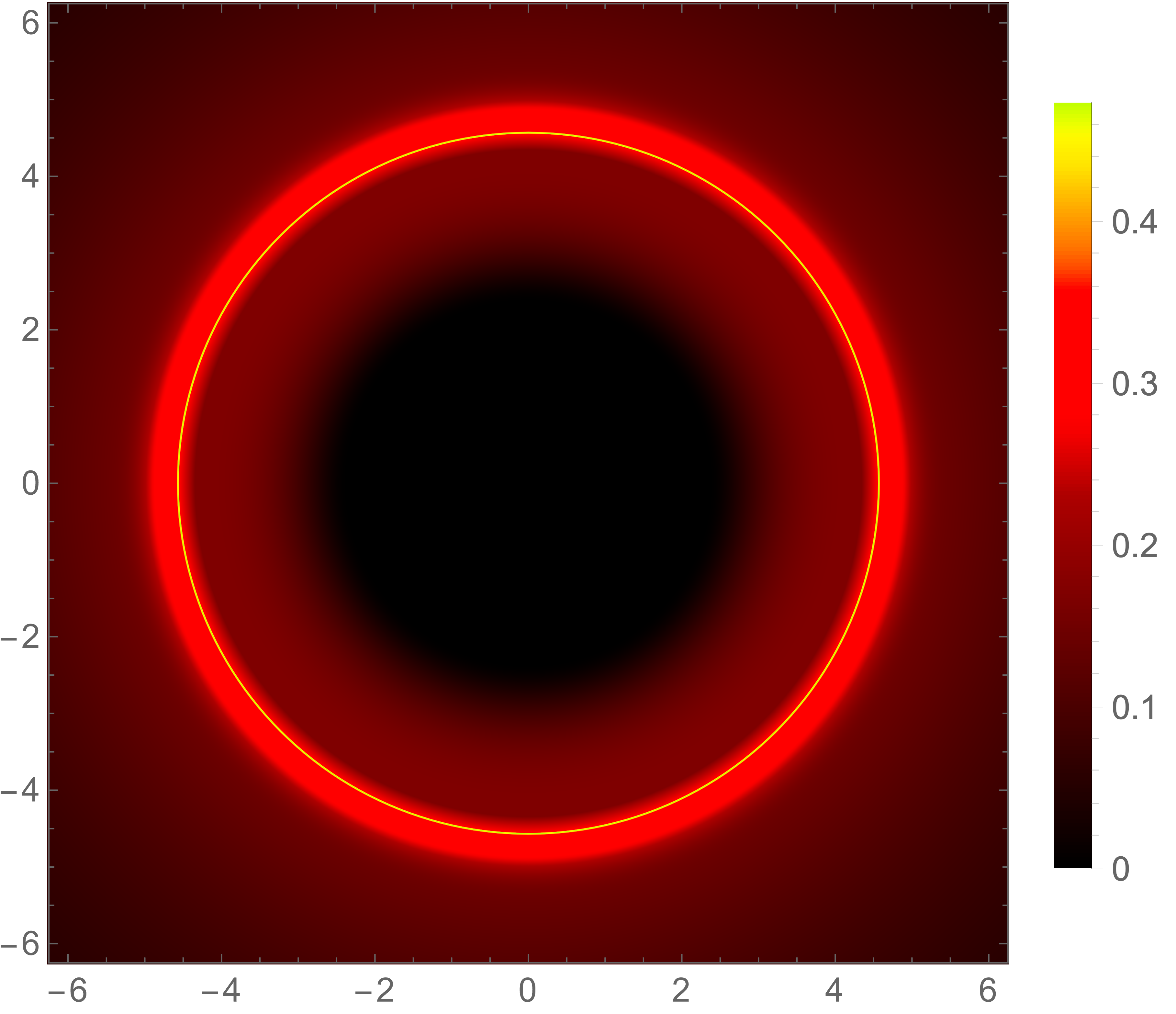}
\includegraphics[width=4.25cm,height=3.75cm]{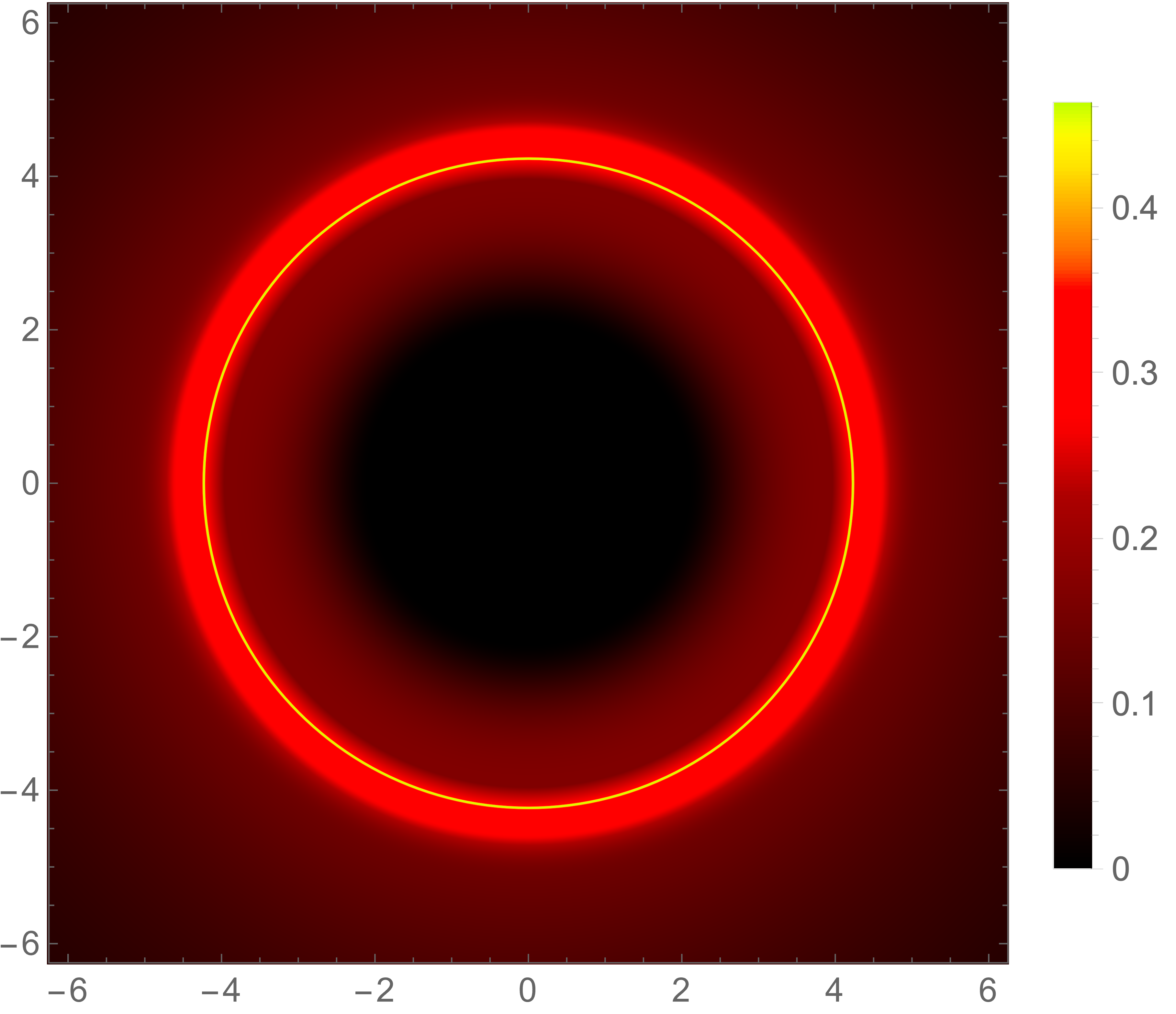}
\includegraphics[width=4.25cm,height=3.75cm]{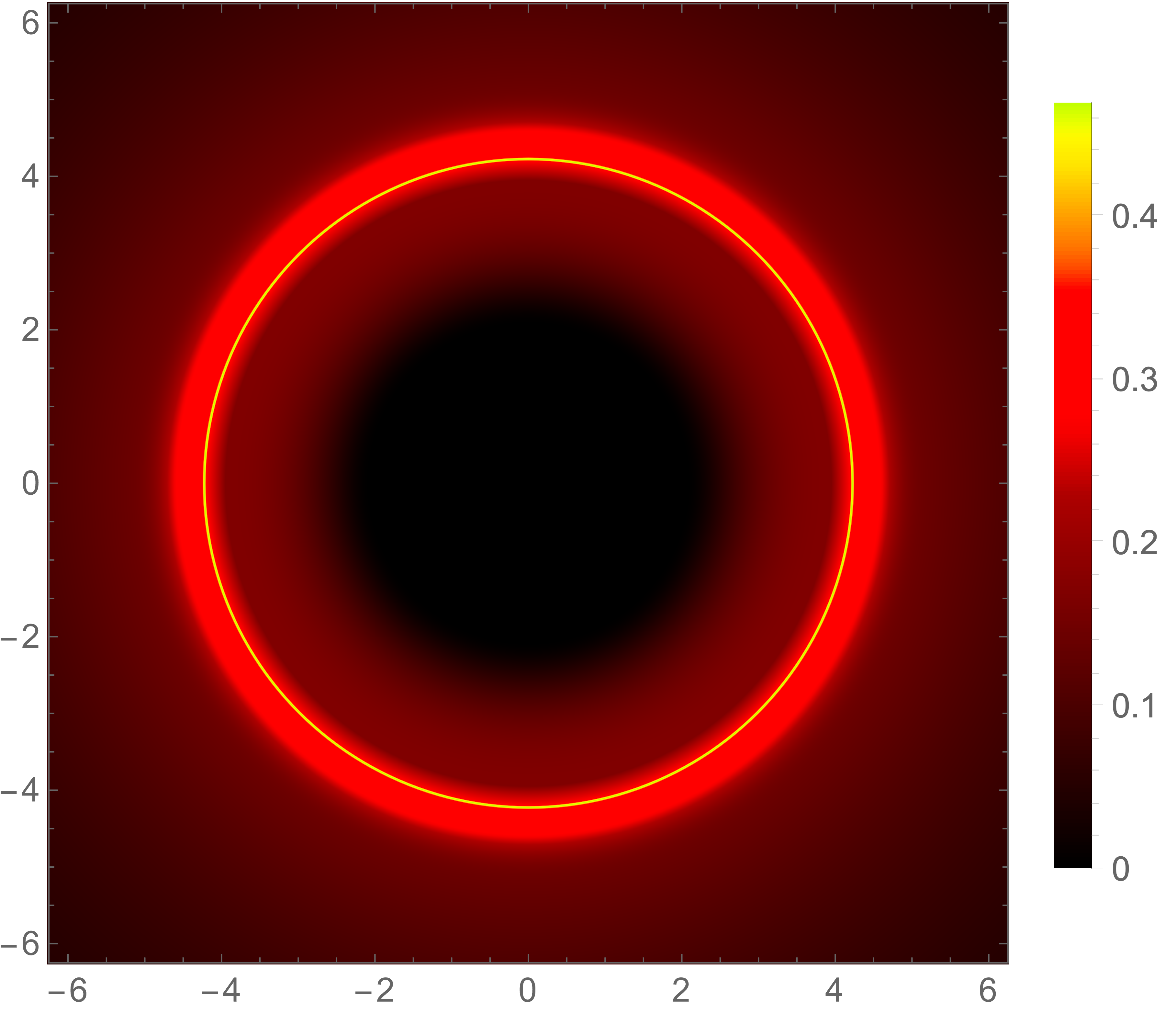}
\includegraphics[width=4.25cm,height=3.75cm]{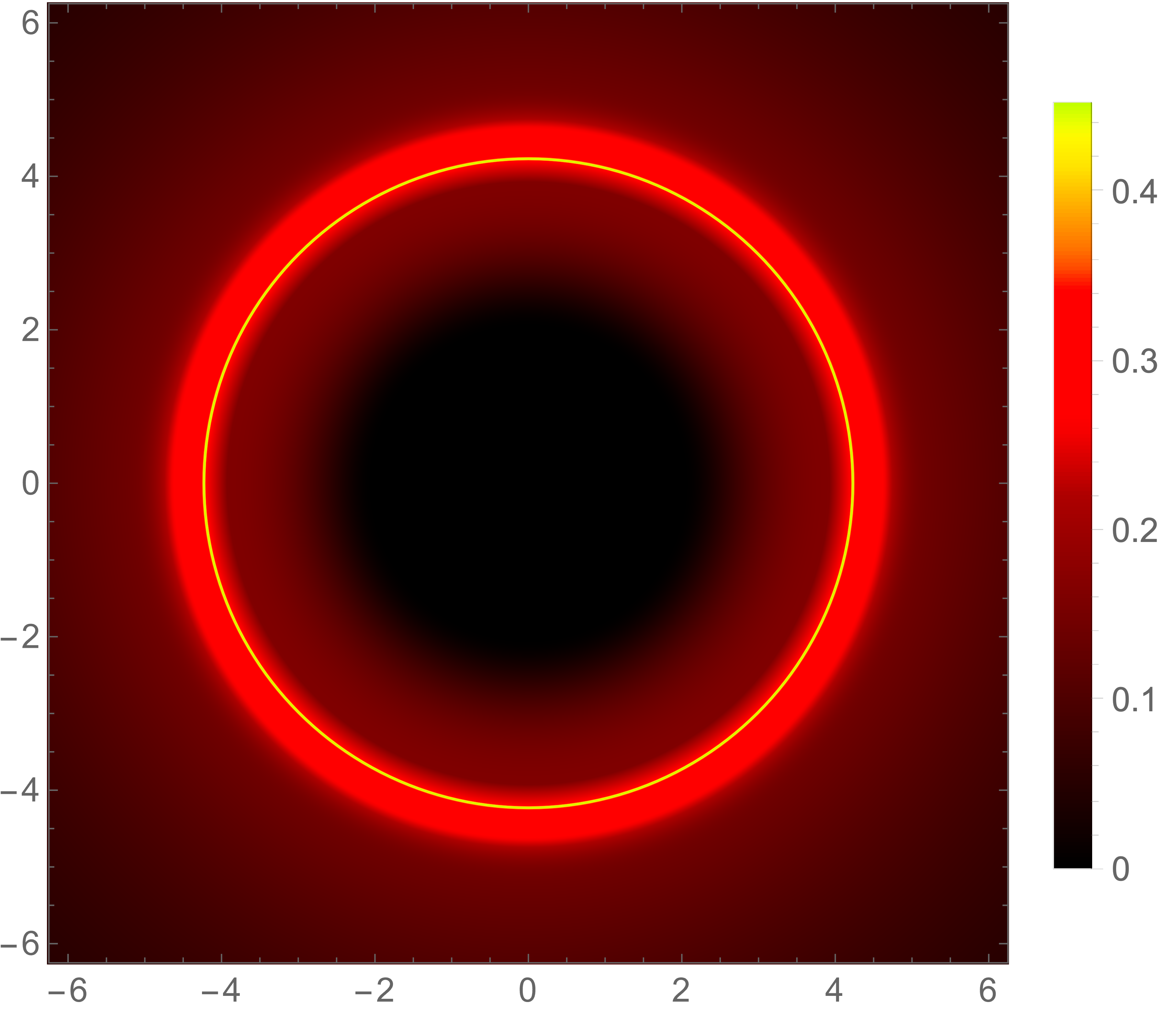}
\includegraphics[width=4.25cm,height=3.75cm]{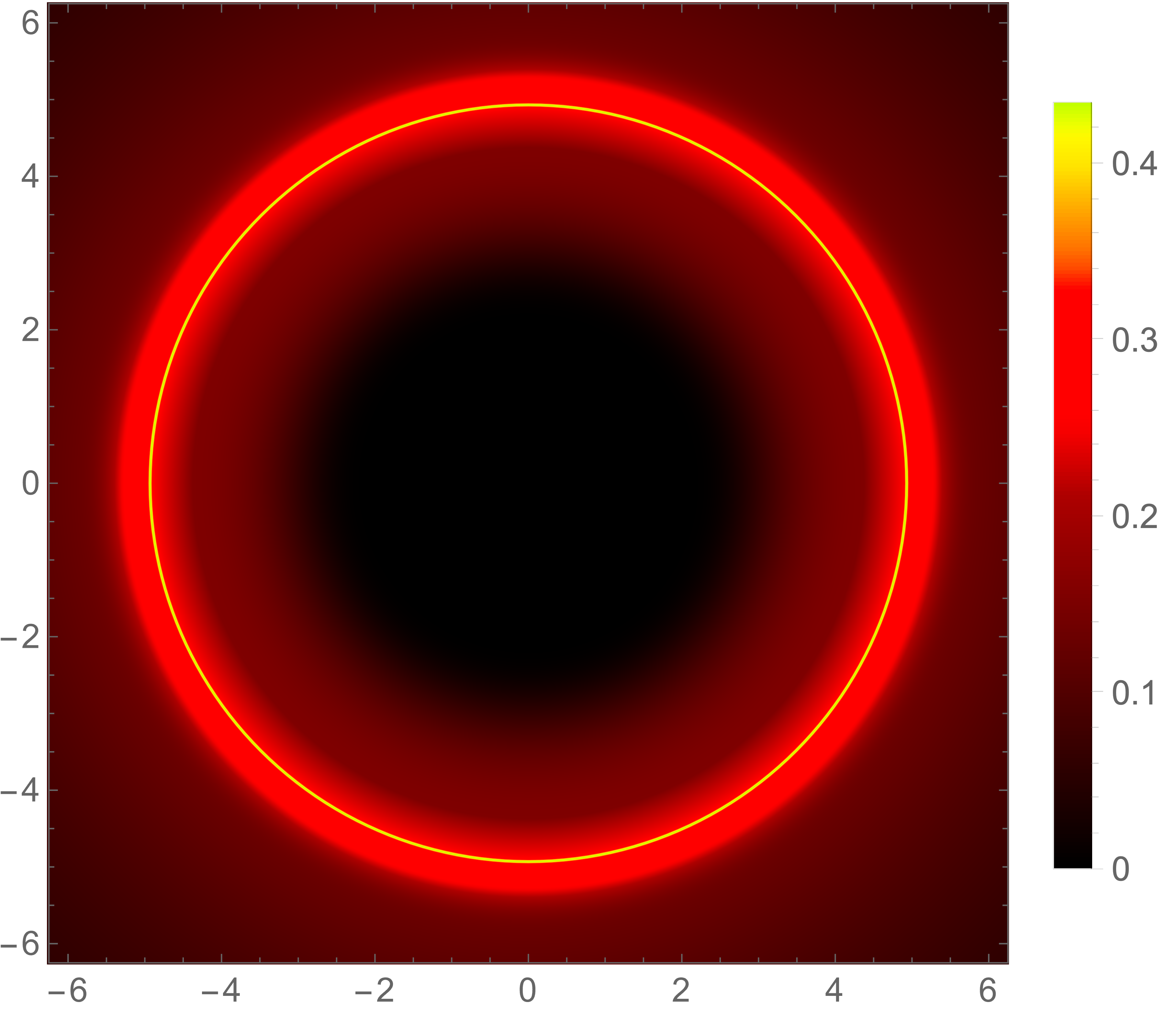}
\includegraphics[width=4.25cm,height=3.75cm]{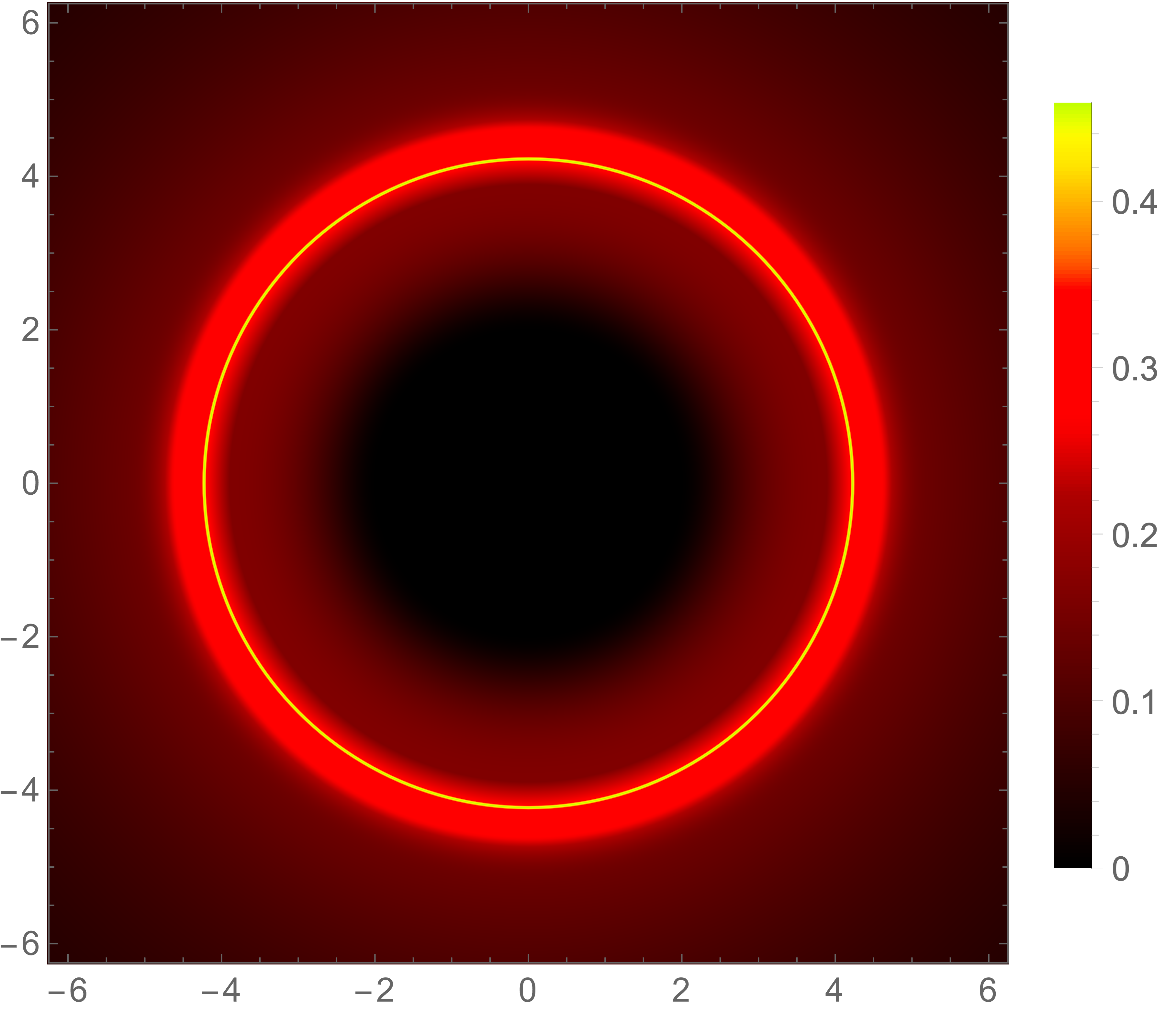}
\includegraphics[width=4.25cm,height=3.75cm]{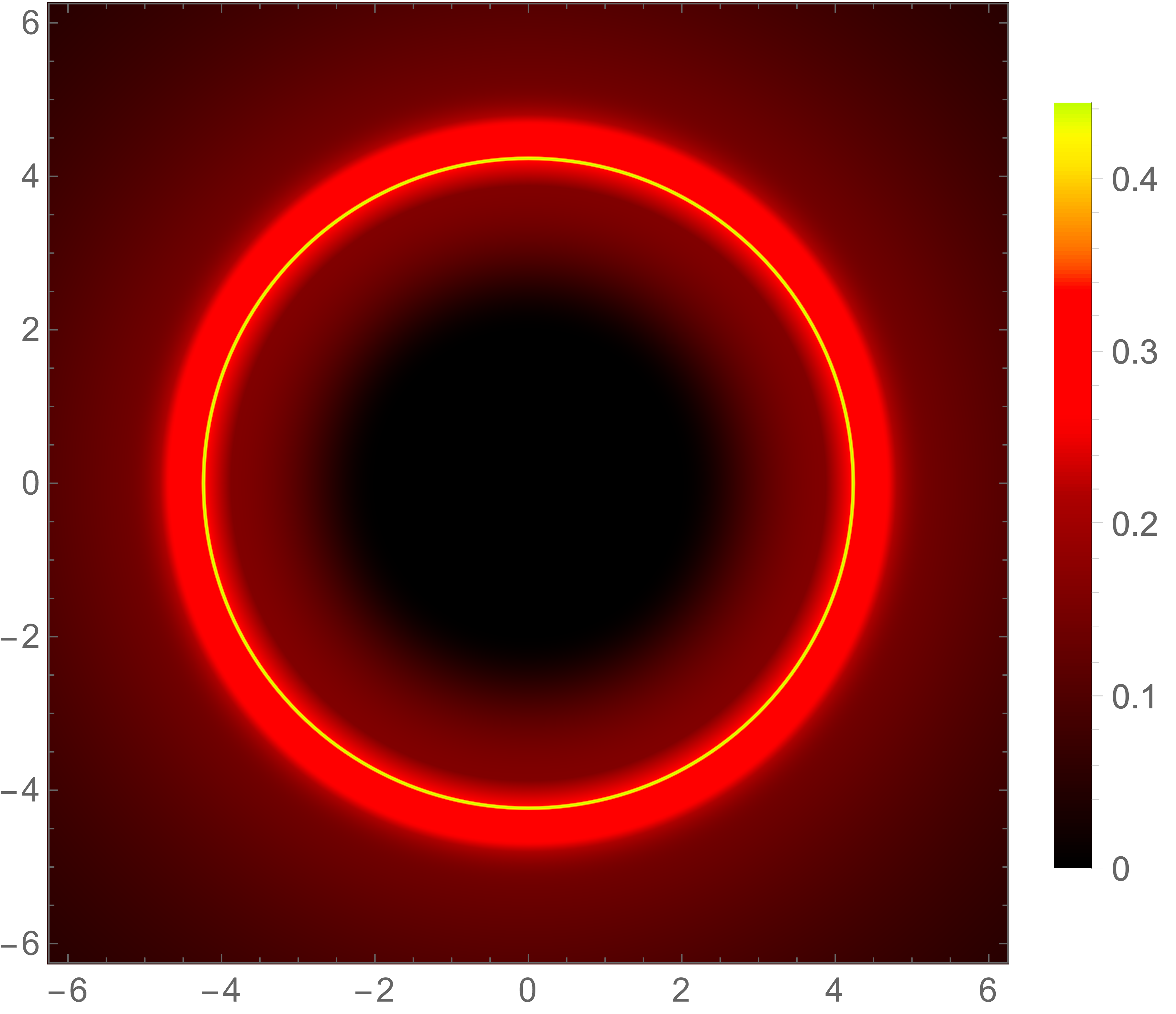}
\includegraphics[width=4.25cm,height=3.75cm]{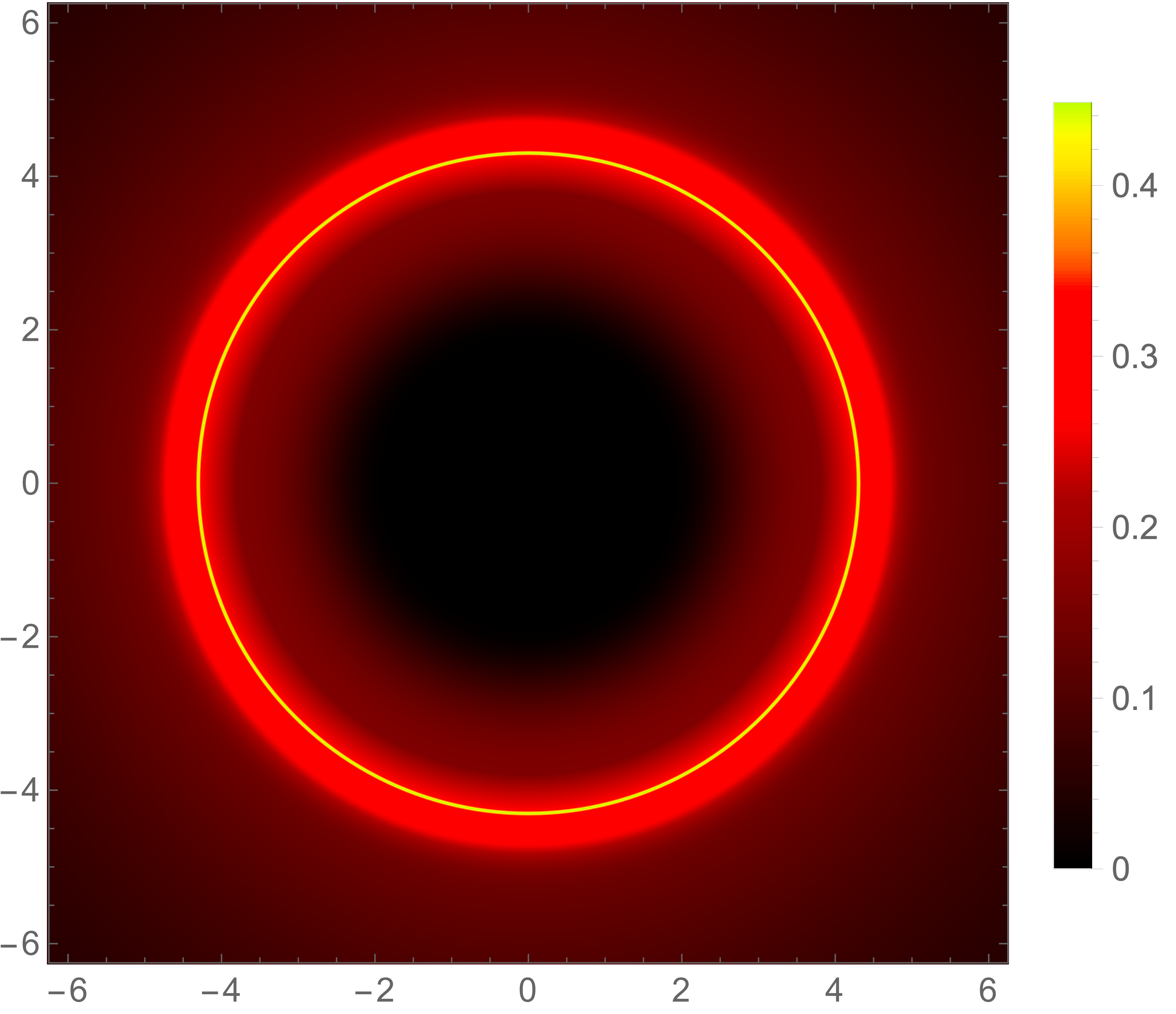}
\includegraphics[width=4.25cm,height=3.75cm]{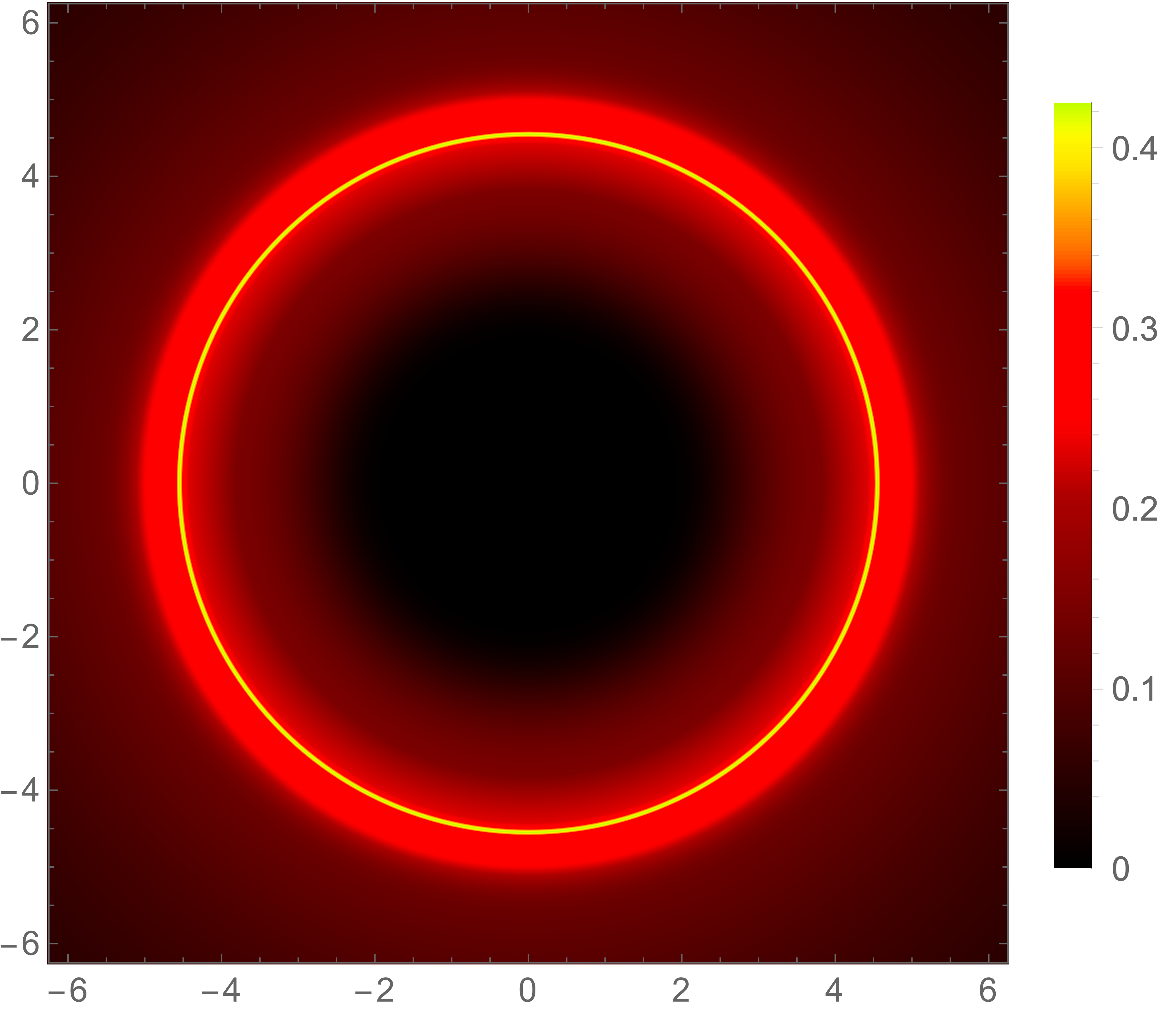}
\includegraphics[width=4.25cm,height=3.75cm]{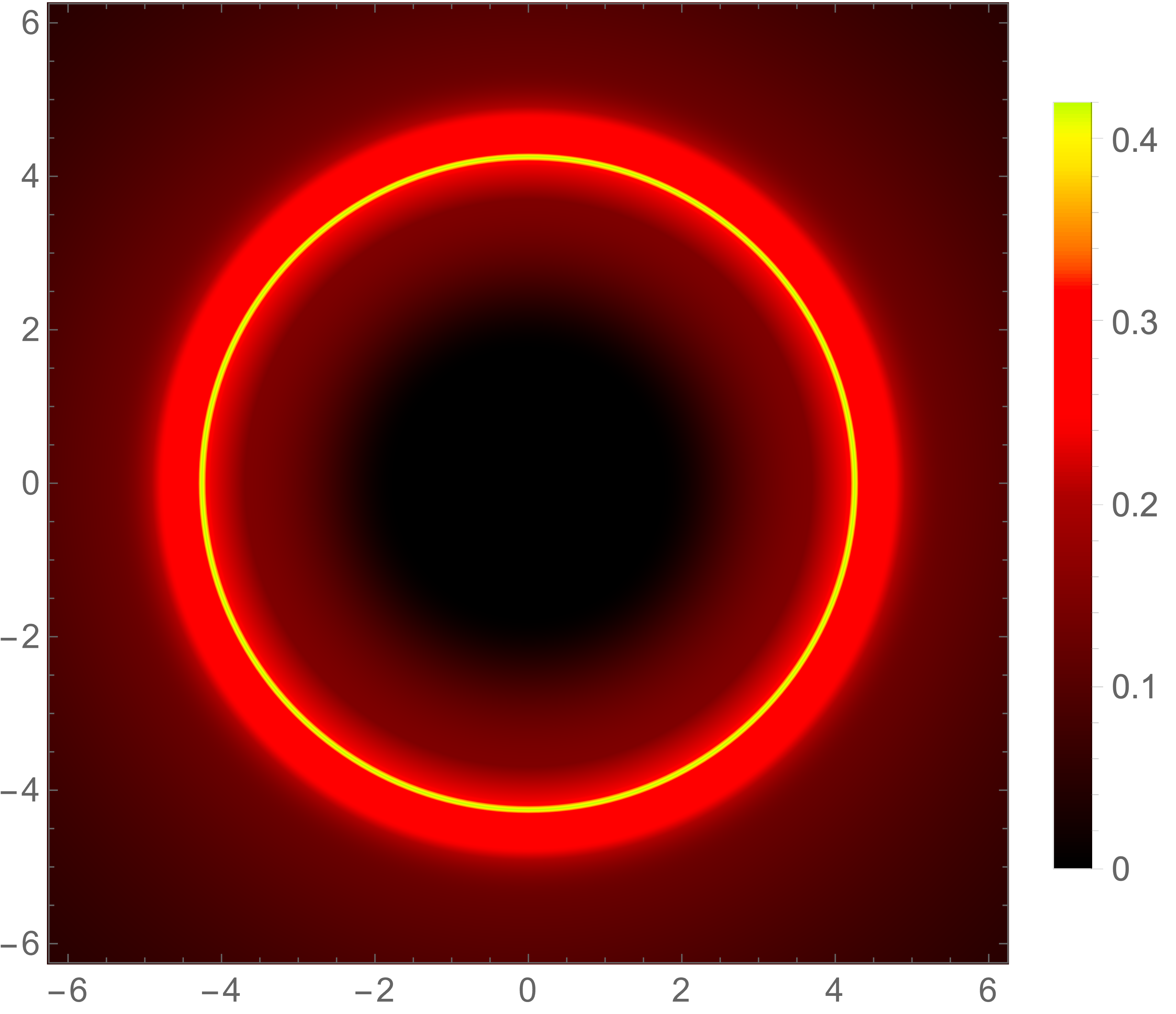}
\caption{Images (from left to right and top to bottom) for LQG, KS, ConfSca, DM, SV, JNW, Sen, GCSV (e), EMD, Bronnikov (e), Hayward, RN, EH (e), Frolov, Bardeen, and GK, ordered in decreasing values of their Lyapunov exponent (units of $M=1$) using the emission model GLM1 in Eqs.(\ref{eq:GLM}) and (\ref{eq:GLM1}).}
\label{fig:shadow1}
\end{figure*}

\subsection{GLM3 model}

We first consider the GLM3 model, which allows us to clearly isolate the $n=1$ and $n=2$ rings, as depicted in Fig. \ref{fig:prs}. There we provide a zoom in of the image around the photon rings for each spherically symmetric geometry ordered according to the decreasing values of their Lyapunov exponent, and restricting the relevant impact parameter space to (mostly) remove the direct emission from the figures. It is transparent that there are significant differences regarding several aspects of these rings: their locations in the impact parameter space, their widths, luminosities, and finally in the distance to each other. Furthermore, despite the fact that both the effective NED geometries and the naked JNW solution trouble the comparison, we can appreciate a trend in the evolution of these photon rings, most notably in the width separating them, which tends to increase as the Lyapunov exponent decreases. In particular, the $n=1$ photon ring has a non-negligible thickness preventing the identification of a well-defined diameter, while the $n=2$ one does appear as a sharp feature in all images. All these aspects allow to distinguish spherically symmetric geometries from each other at equal emission model, something in agreement with our initial expectations regarding the features of photon rings to depend less on the emission properties as we get to larger values of $n$. In this case, the extinction rate clearly correlates with the Lyapunov index: save by a few exceptions lower values of the latter leads to lower extinction rates; indeed the Lyapunov exponent systematically underestimates the extinction rate as compared to the GLM3 one. Another comment is related to the special features of the naked JNW geometry, as given by a much wider $n=1$ ring and a closer distance to the direct emission, clearly appearing in the top right end of its figure. This goes along our previous warning on the fact that horizonless compact objects have special features regarding the contributions to the luminosity of its rings depending on the shape of its effective potential,  which in some cases may (even if partially) neglect the assumption on the exponential suppression of the luminosity of successive rings, troubling its comparison with usual black hole space-times.

\begin{figure*}[t!]
\includegraphics[width=4.25cm,height=3.75cm]{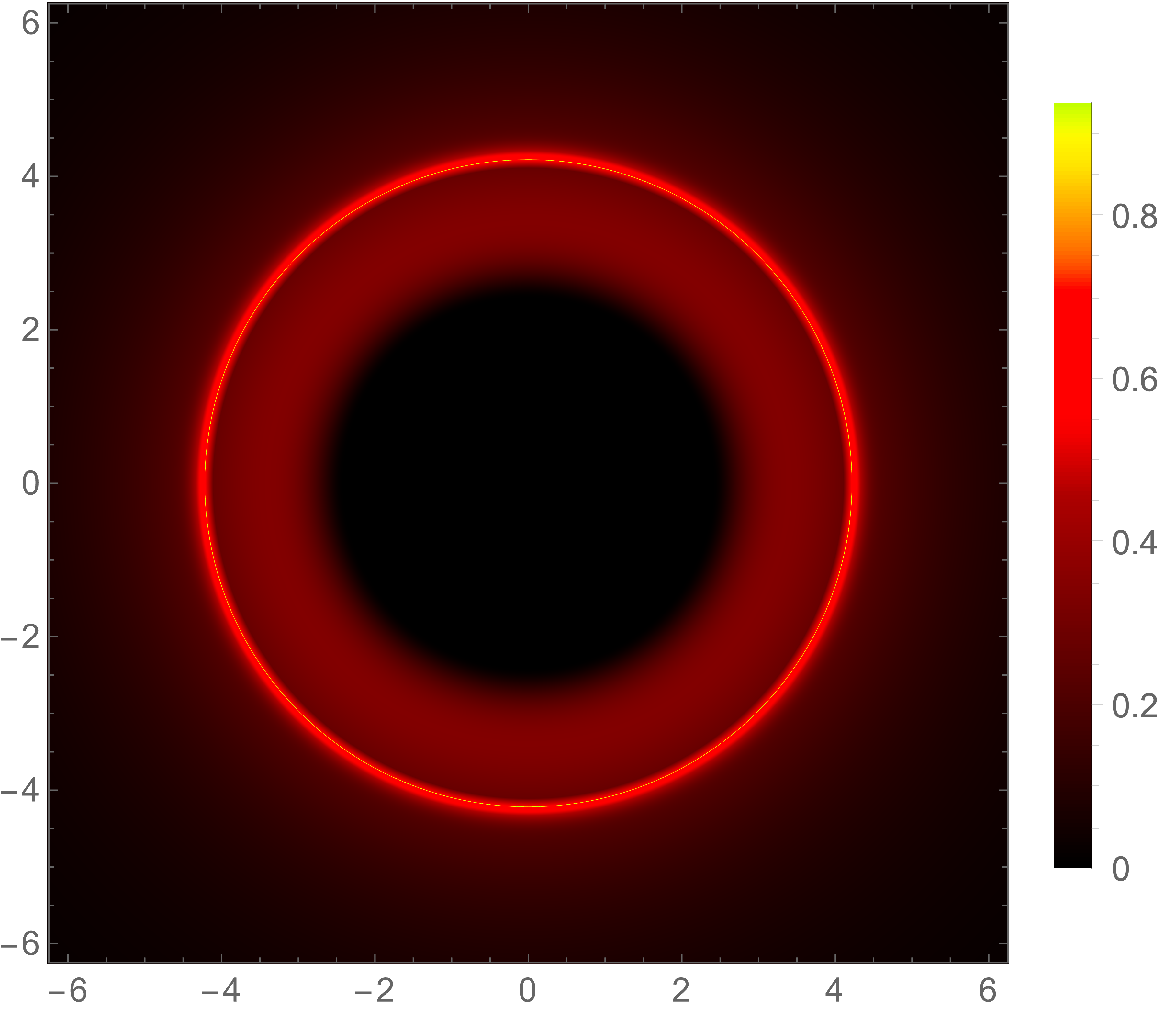}
\includegraphics[width=4.25cm,height=3.75cm]{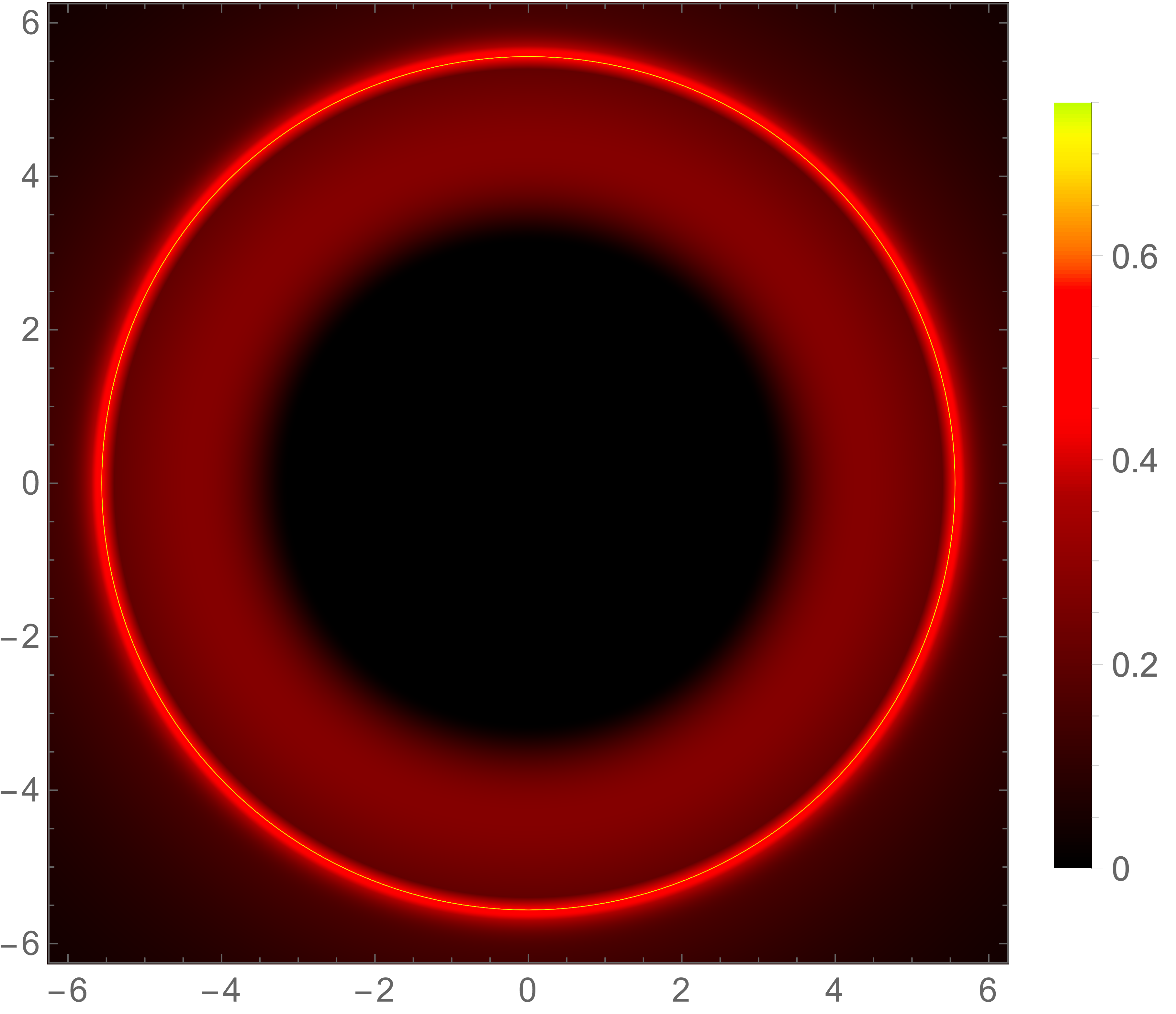}
\includegraphics[width=4.25cm,height=3.75cm]{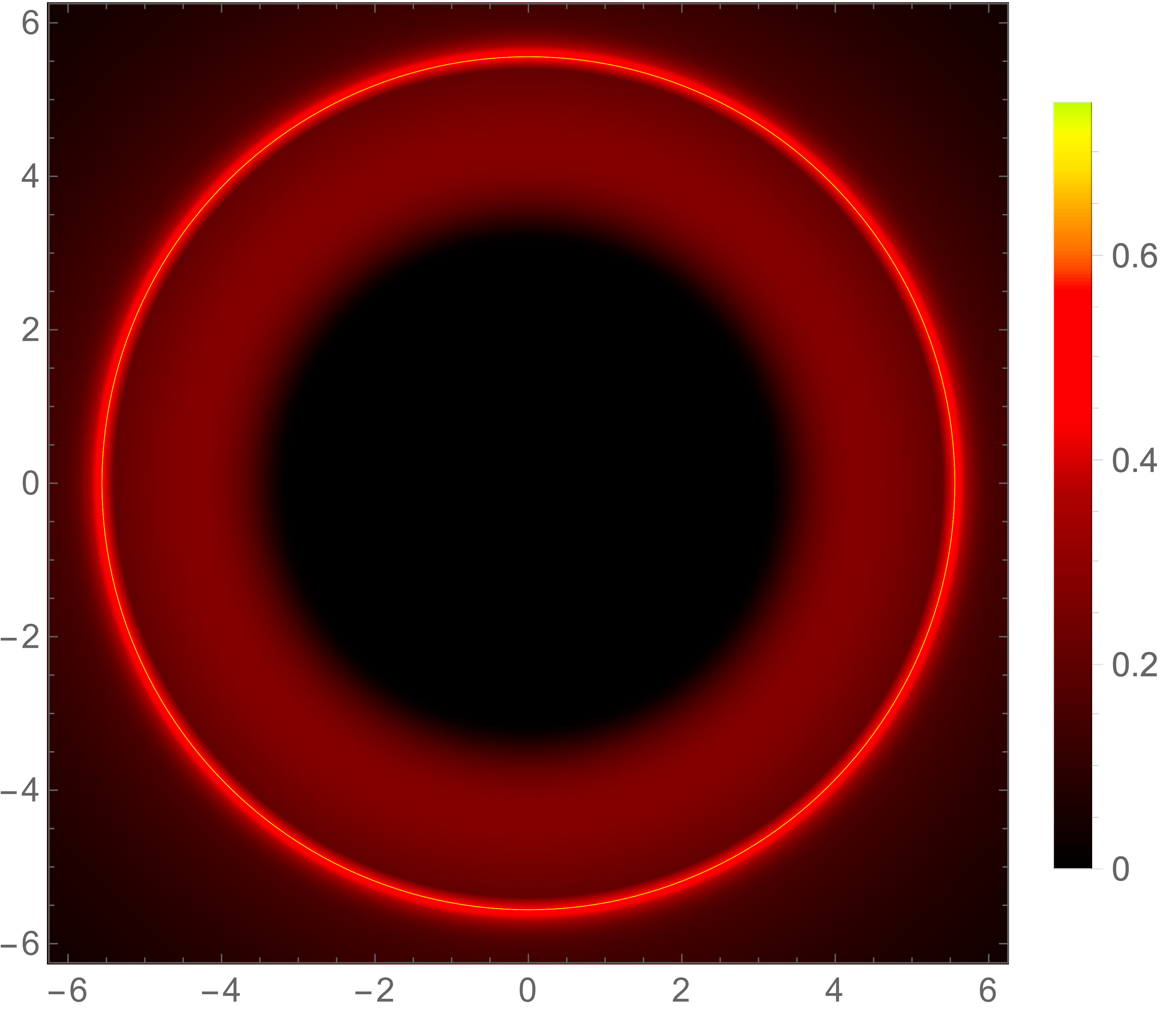}
\includegraphics[width=4.25cm,height=3.75cm]{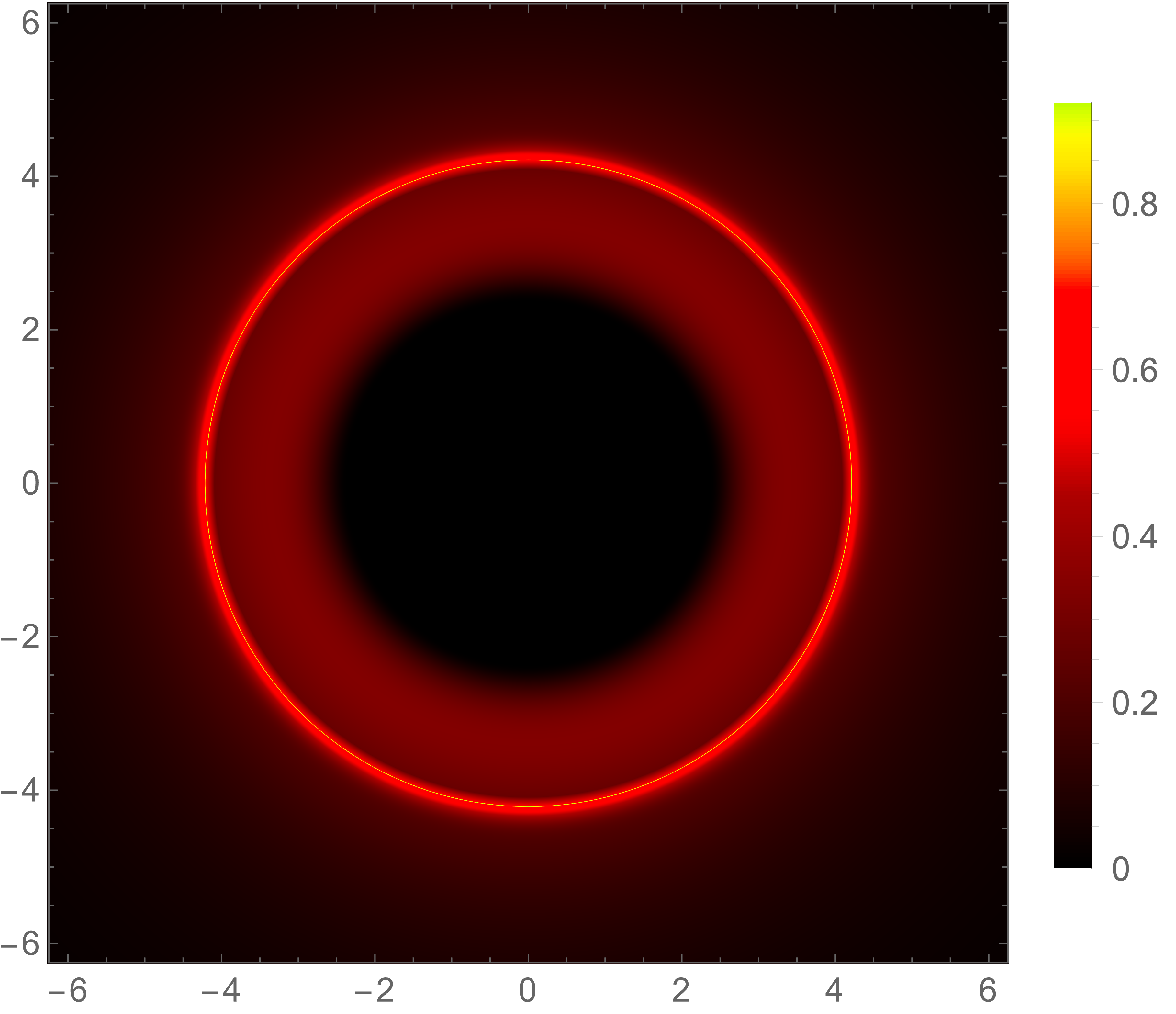}
\includegraphics[width=4.25cm,height=3.75cm]{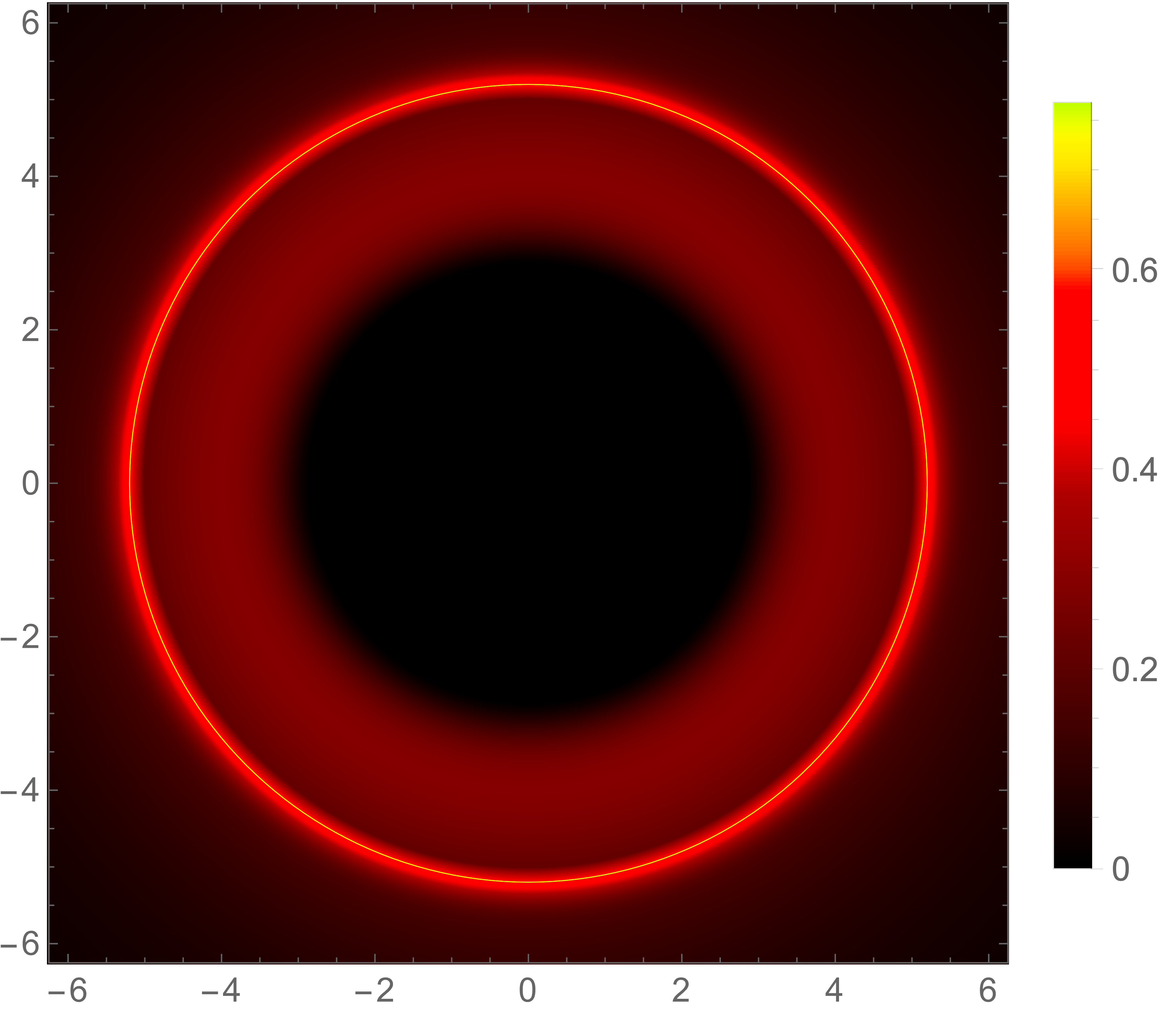}
\includegraphics[width=4.25cm,height=3.75cm]{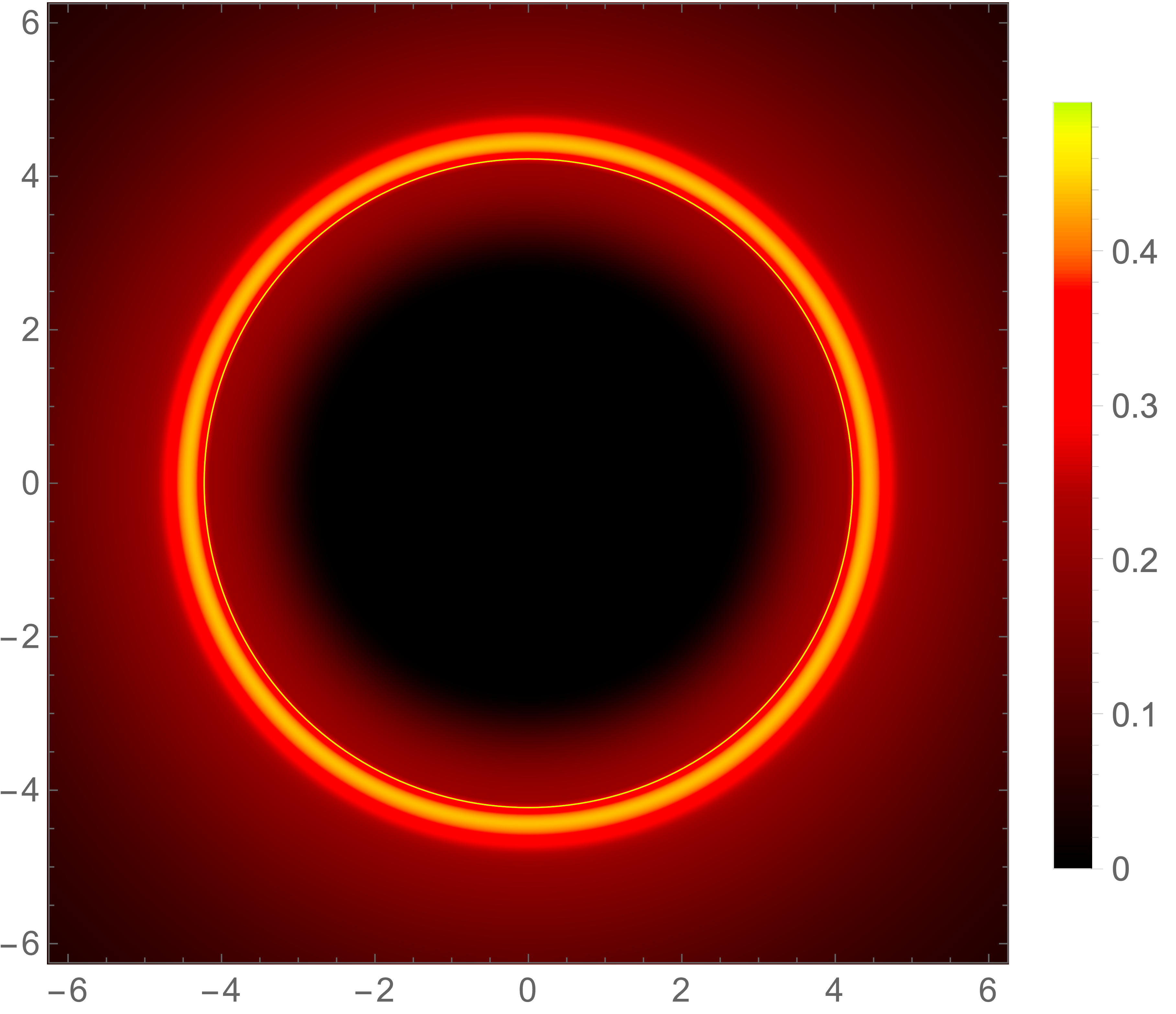}
\includegraphics[width=4.25cm,height=3.75cm]{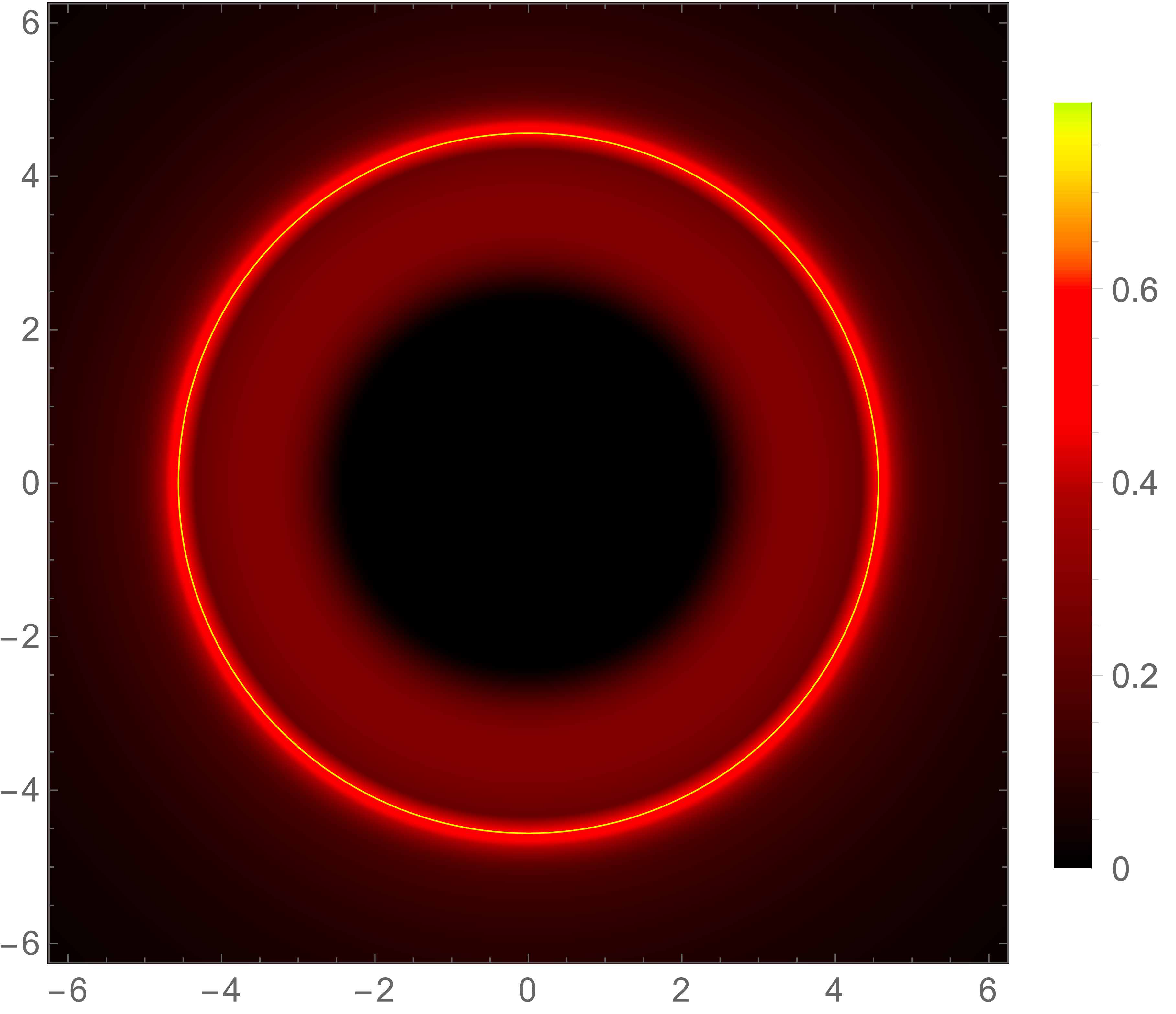}
\includegraphics[width=4.25cm,height=3.75cm]{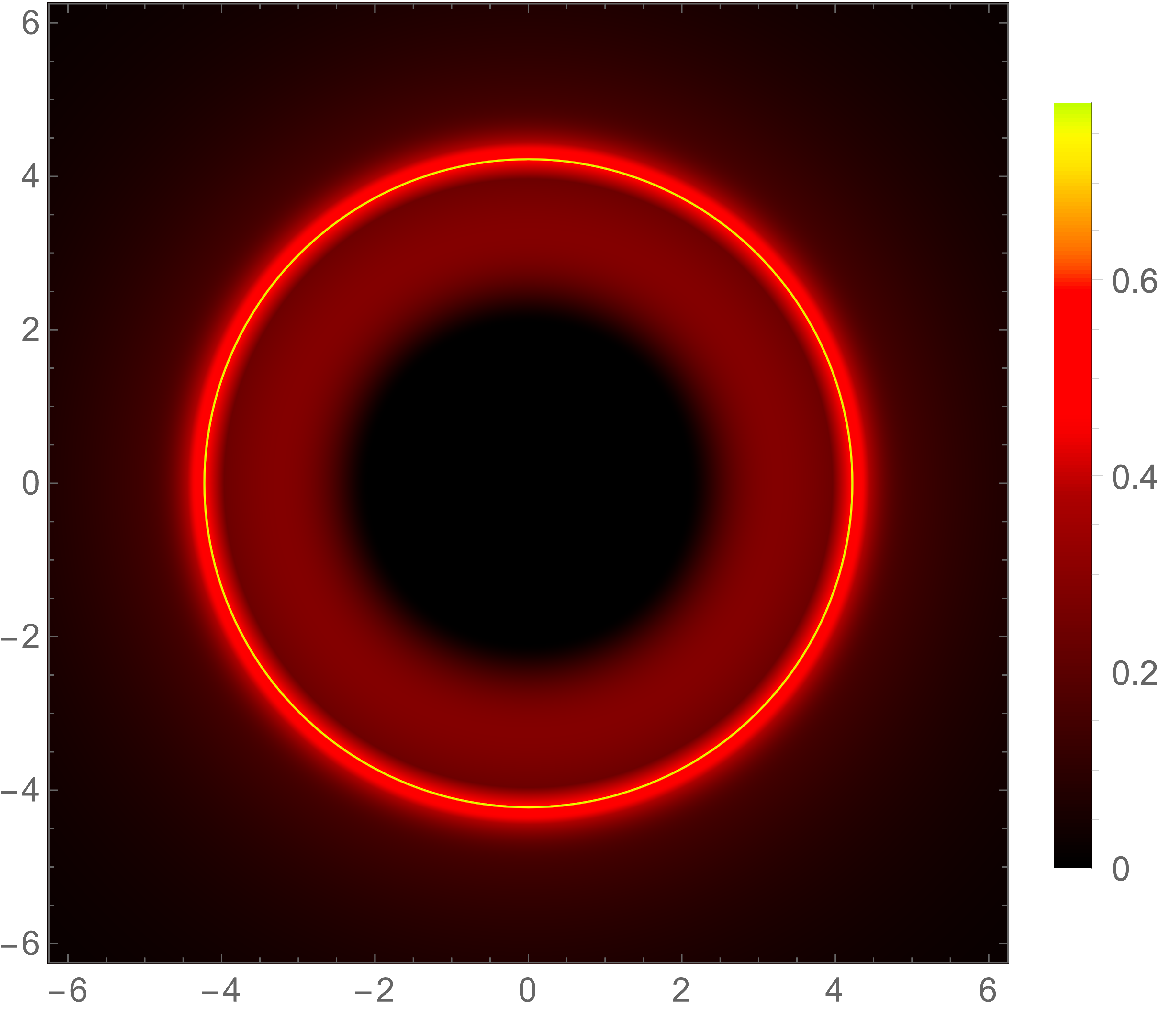}
\includegraphics[width=4.25cm,height=3.75cm]{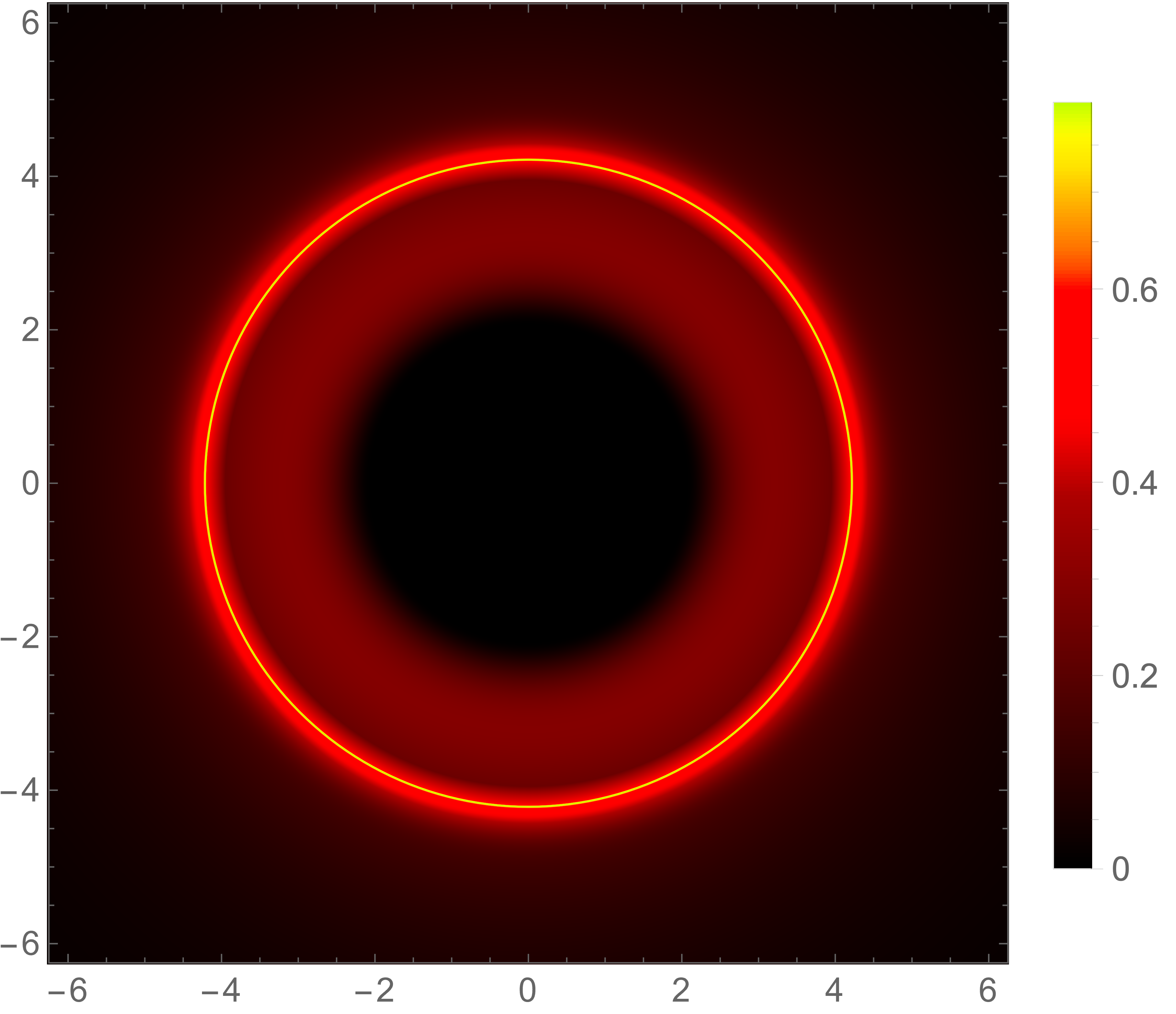}
\includegraphics[width=4.25cm,height=3.75cm]{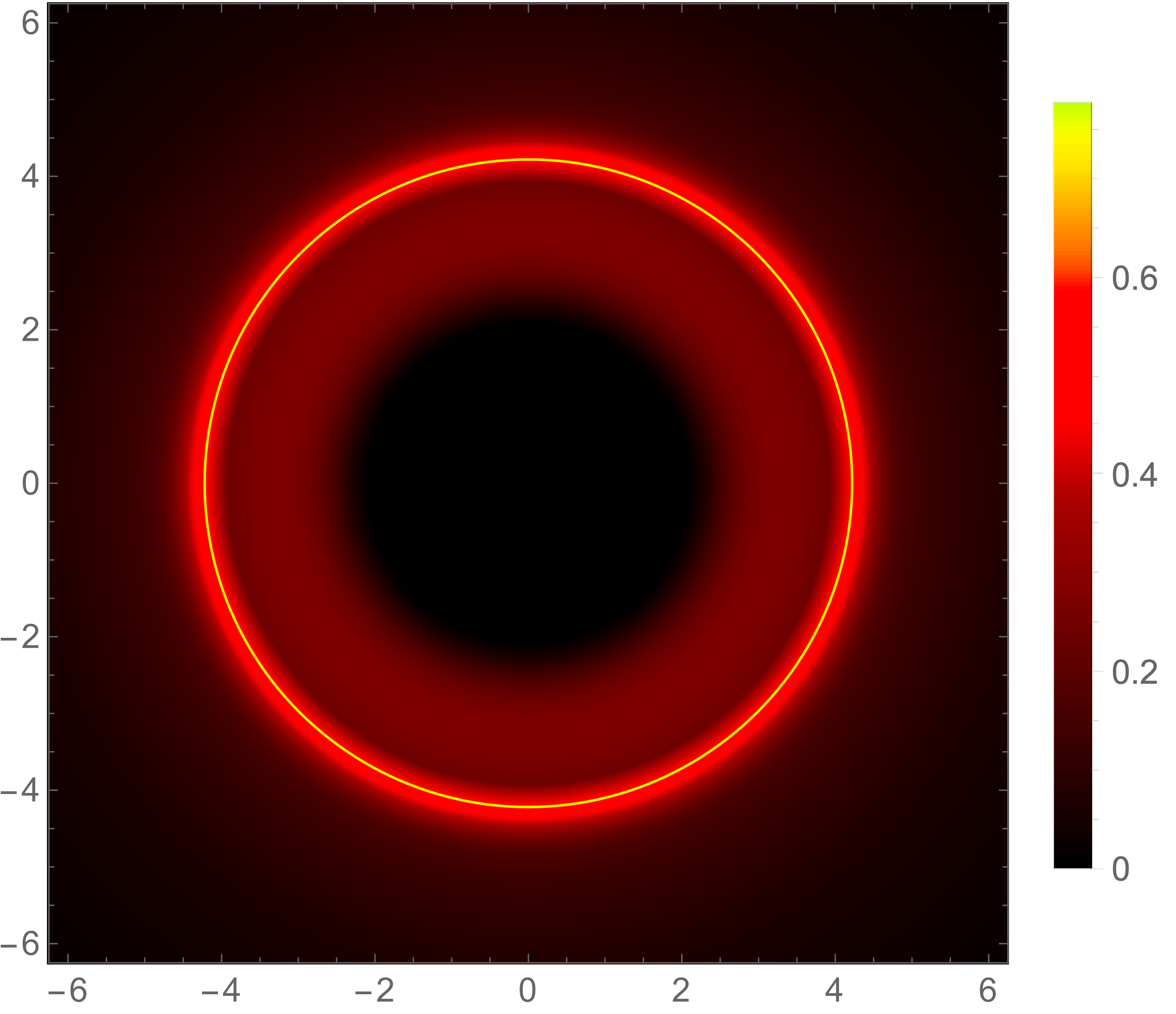}
\includegraphics[width=4.25cm,height=3.75cm]{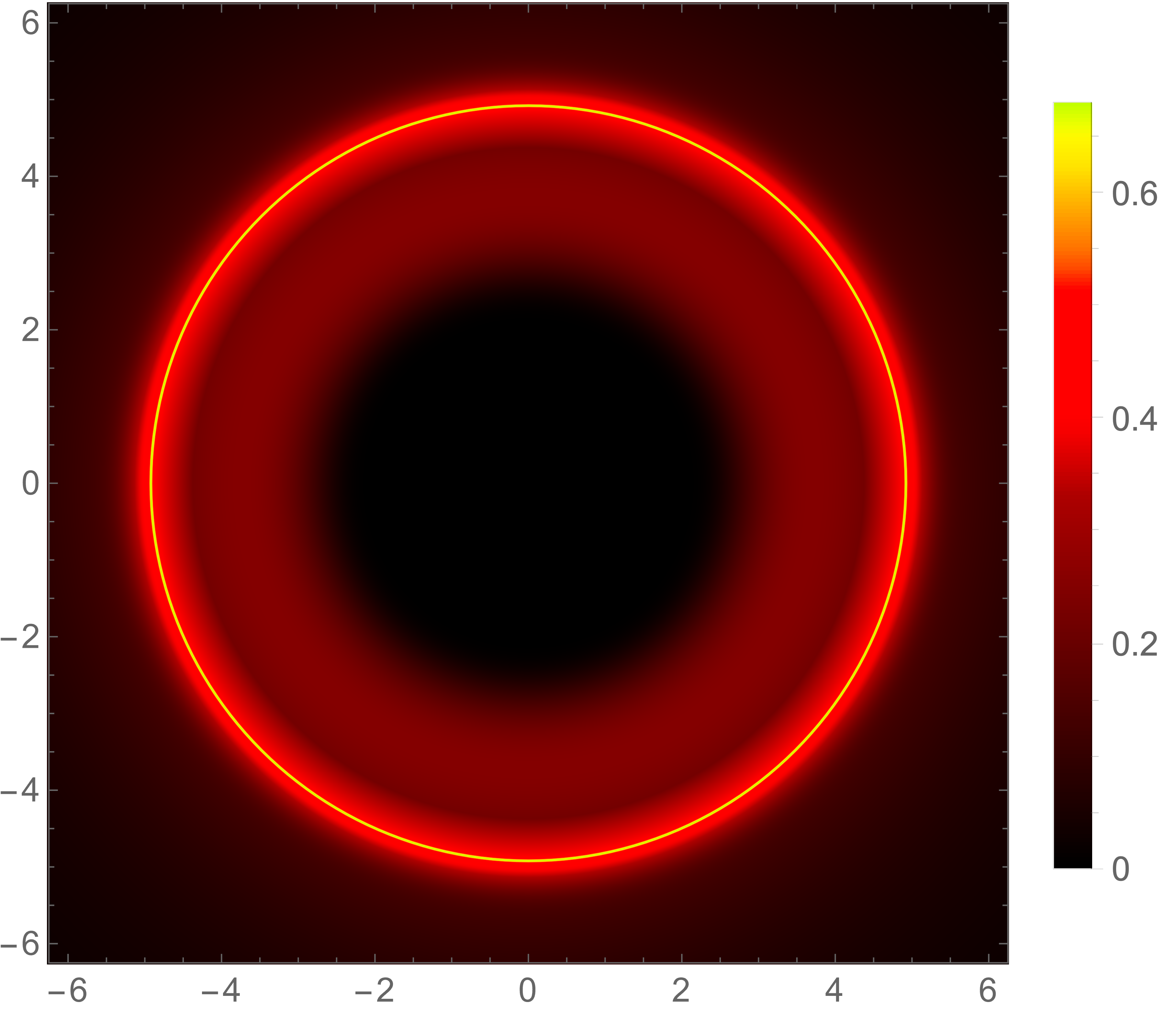}
\includegraphics[width=4.25cm,height=3.75cm]{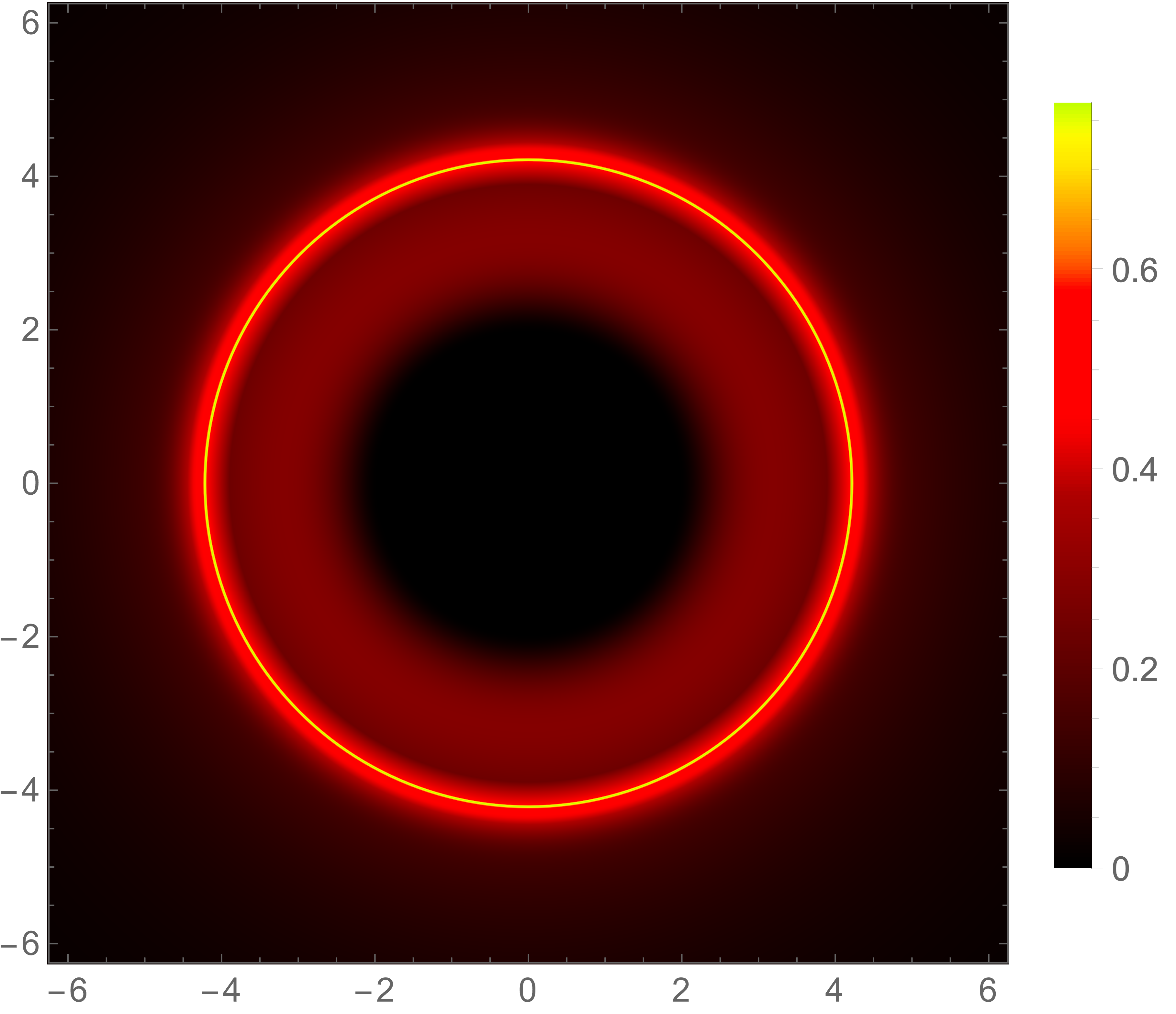}
\includegraphics[width=4.25cm,height=3.75cm]{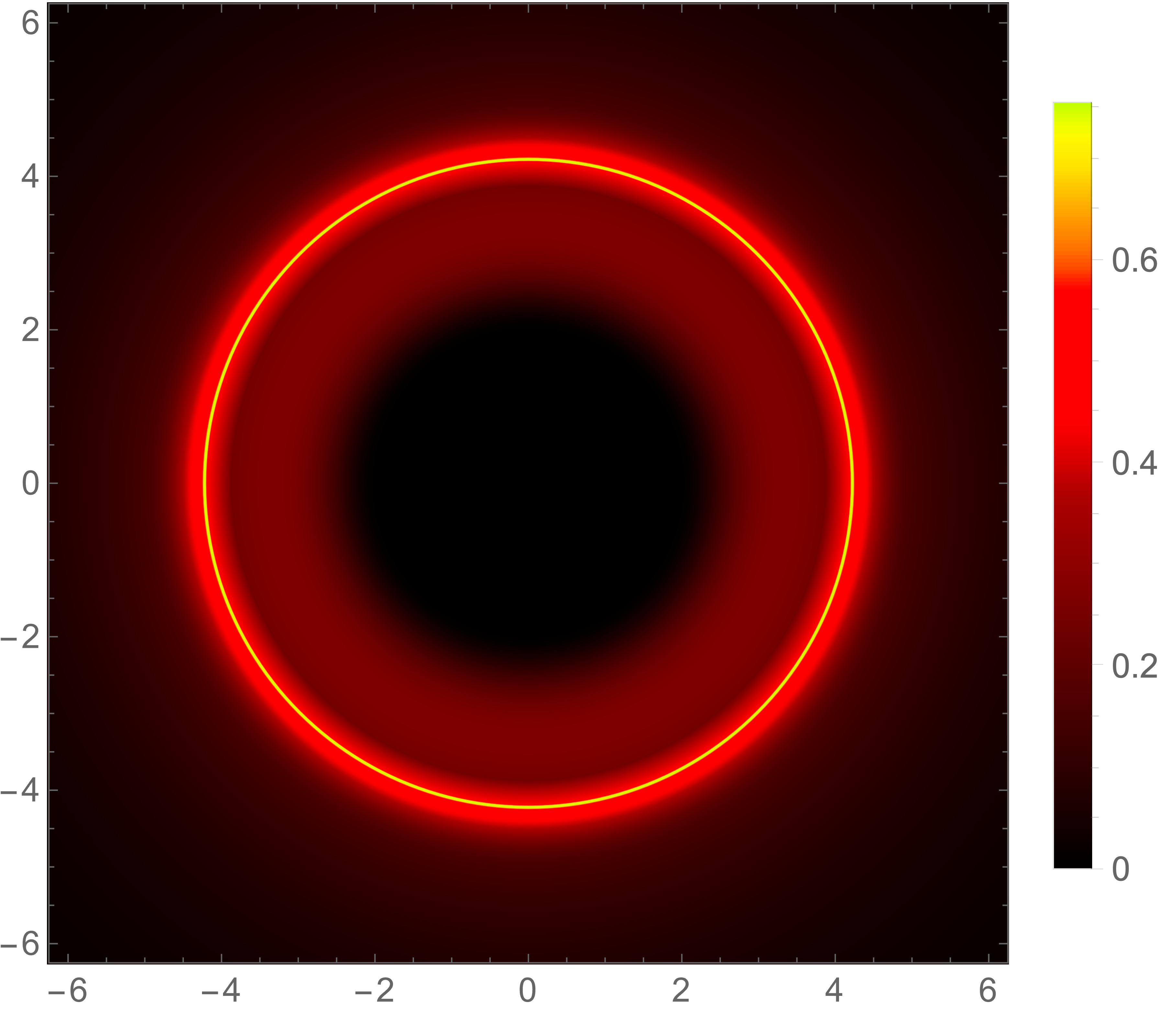}
\includegraphics[width=4.25cm,height=3.75cm]{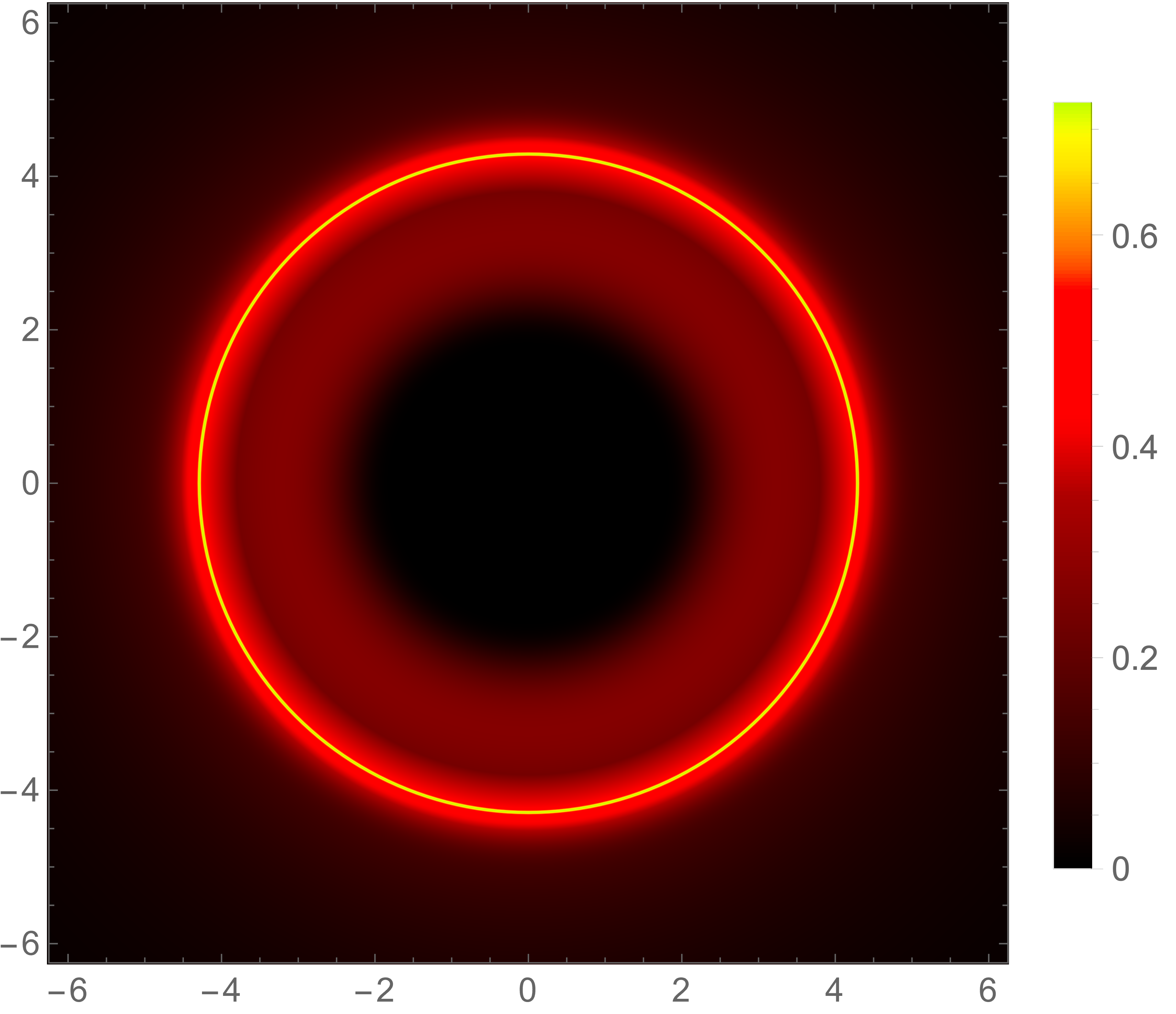}
\includegraphics[width=4.25cm,height=3.75cm]{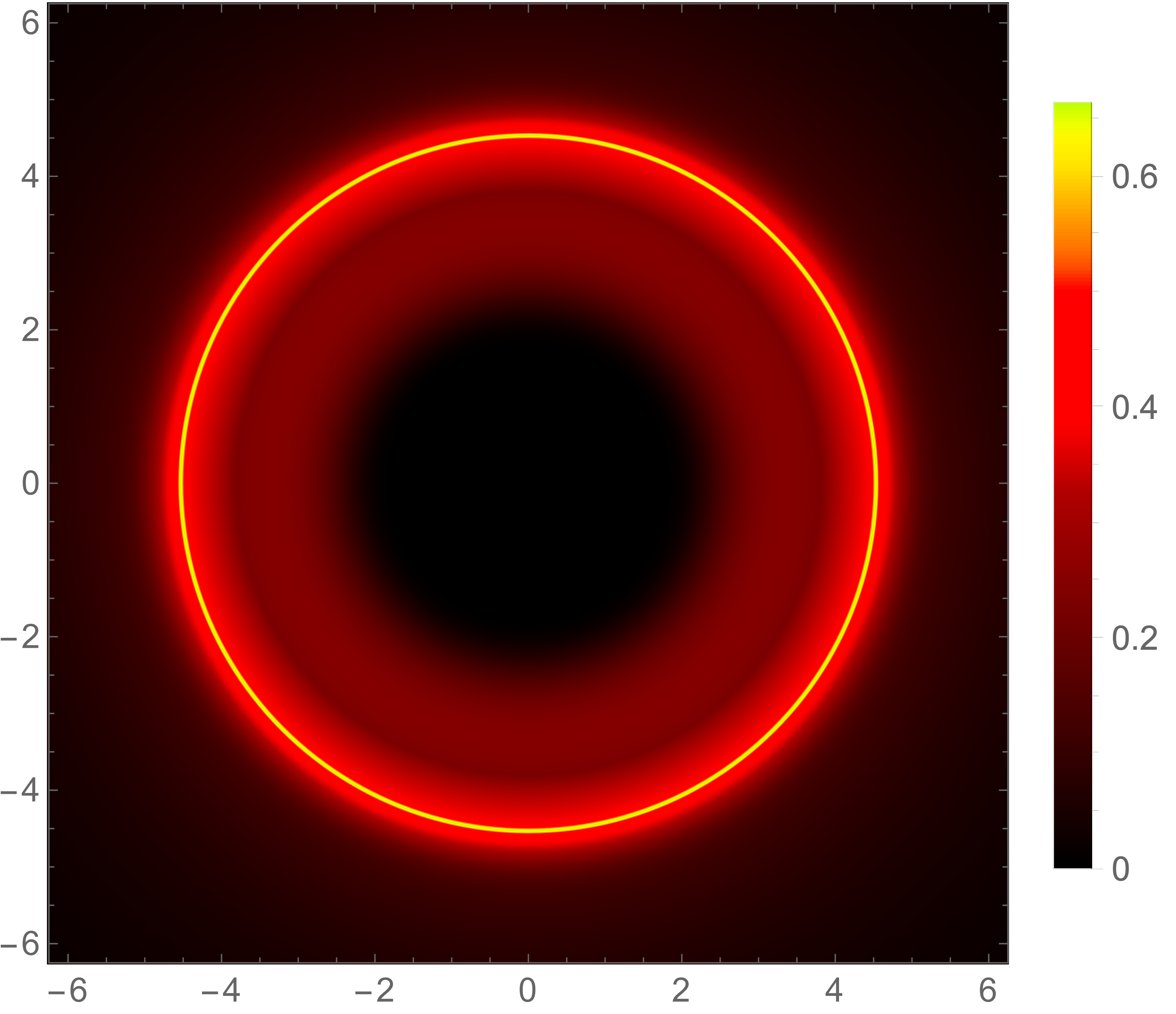}
\includegraphics[width=4.25cm,height=3.75cm]{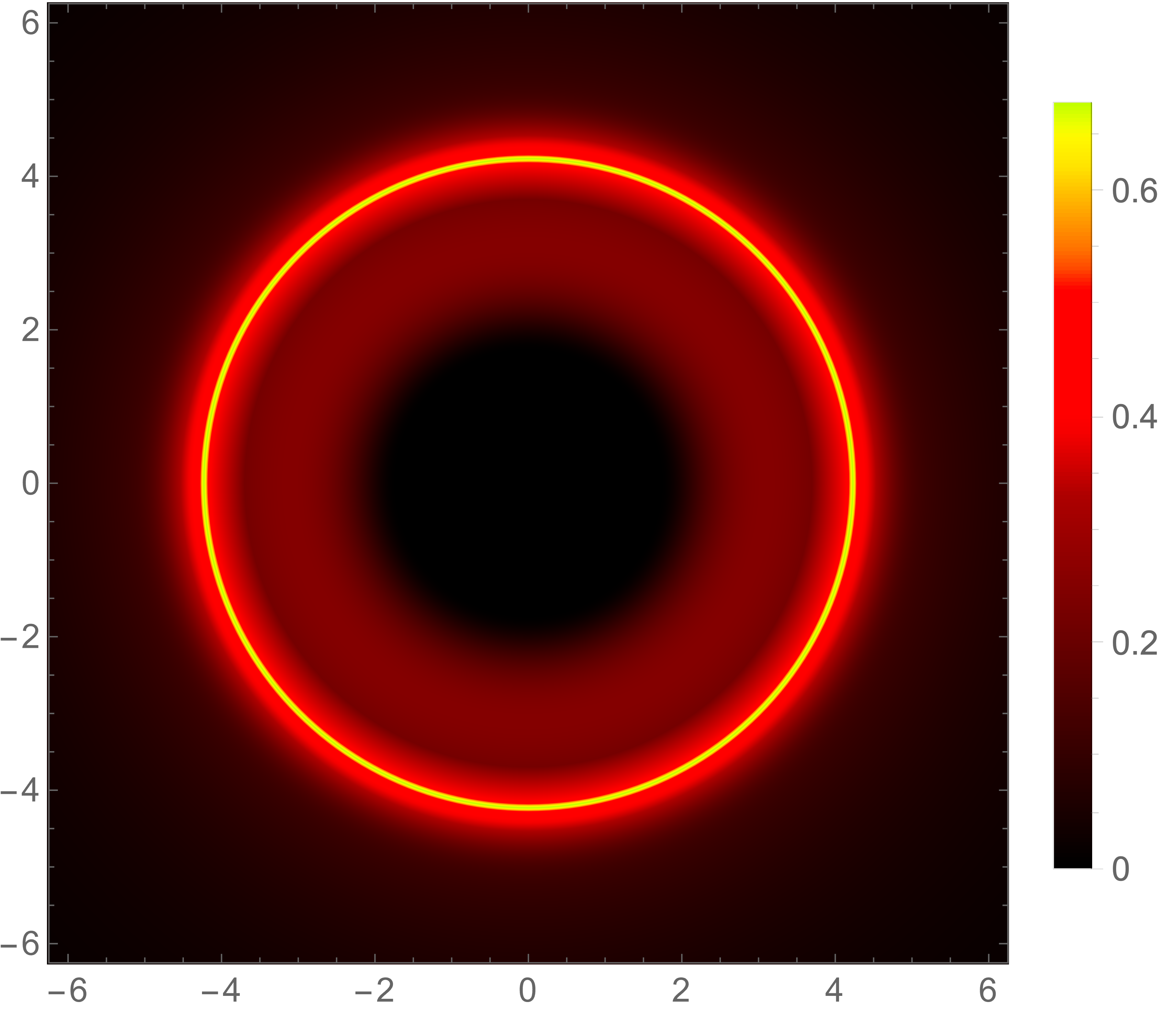}
\caption{Images (from left to right and top to bottom) for LQG, KS, ConfSca, DM, SV, JNW, Sen, GCSV (e), EMD, Bronnikov (e), Hayward, RN, EH (e), Frolov, Bardeen, and GK, ordered in decreasing values of their Lyapunov exponent (units of $M=1$) using the emission model GLM2 in Eqs.(\ref{eq:GLM}) and (\ref{eq:GLM2})}
\label{fig:shadow2}
\end{figure*}

\subsection{GLM1/GLM2 models}

The GLM3 model is rather unnatural given the fact that we place the emission region at a truncated and arbitrary region (but the same) for every spherically symmetric geometry, and it is designed only to probe the structure of the photon rings without the {``contamination"} of the direct emission. As opposed to this, in the GLM1/GLM2 models the accretion flow goes all the way down to the event horizon (whenever present), and thus are better aligned with astrophysical expectations. The imaging of the 16 alternative configurations according to these two models is presented in Figs. \ref{fig:shadow1} and \ref{fig:shadow2}, respectively. Such images are consistent with what we know about observed (by the EHT) images: they are largely dominated by the bright ring of radiation caused by the direct emission of the disk, and have the typical central brightness depression on their center; superimposed on the direct emission we find the slight boost of luminosity caused by the $n=1$ and $n=2$ photon rings, though only the former is neatly visible. This is a trivial consequence of the extinction rate between the photon rings, as reported in the corresponding columns of Table \ref{table}. Indeed, such a rate closely tracks the Lyapunov index, with deviations between the latter (theoretical) and the former (observational) being $\lesssim 15\%$ between each other for every geometry in the GLM1 model, and $ \lesssim  5\%$ in the GLM2 one, and typically underestimated in the theoretical prediction (note in this sense that effective geodesics slightly decrease such rates as compared to what one would find should it use the background geodesics instead). This implies that the (theoretical) Lyapunov index is not a bad guidance in the actual (observable) luminosity of the photon rings after all.

We also observe neat differences in the location and width of the photon rings as well as in the depth of the brightness depression among background geometries, which is just a reflection of the data displayed in Table \ref{table}.  Similarly as in the GLM3 model, the naked JNW solution distorts the trend of images, since in such a case the distribution of luminosities of the photon rings inserted in the direct emission is significantly changed as compared to black hole space-times, and so the depth is of the central brightness depression. A comment related to this is that in these thin-disk models the depth of the central brightness depression can be much smaller than the inferred EHT shadow's size: in the GLM3 model this is translated into a $n=2$ ring that can penetrate well inside the corresponding critical curve in the observer's plane image, while in the GLM1/GLM2 it is the direct emission itself which clearly lies inside it. This is expected on the grounds of previous studies in the field with thick but not fully spherical disk \cite{Vincent:2022}, where the size of the central blackness depression is tied to the apparent (lensed) location of the (equatorial) horizon, leading to the inner shadow.

The bottom line of our results is that even when the shadow's boundary is assumed to be degenerate between different spherically symmetric geometries \cite{Lima:2021las} (assuming EHT data, hypothesis, modelling and interpretation), it turns out that their corresponding first and second photon rings contain sufficiently sharp differences to allow to distinguish between such geometries in a thin-disk context, something in agreement with other findings in the field \cite{Eichhorn:2022oma}. In practical terms, however, the properties of the disk are comparatively poorly known, and this may significantly alter the results to the point of mistaking alternative geometries from each another (and from Schwarzschild's). Here we resorted to the simplified GLM-type models of the disk, yet there is still plenty of room for improvement in the comparison with observed images.

\section{Conclusion and prospects} \label{C:V}

In this work we have generated images of a selected pool of alternative spherically symmetric geometries, extracted from the work of Vea in \cite{Vagnozzi:2022moj}. To do so, we applied (and refined) the constraints derived there in the space of parameters of each model from the inferred correlation between the size of the bright ring and the shadow's size itself by the EHT Collaboration on Sgr A$^\star$ \cite{EventHorizonTelescope:2022xqj} (subject to the caveats pointed out there), and generated such images when each geometry is surrounded by an infinitesimally-thin accretion disk with three samples of analytical profiles for the emission provided by the GLM ones. We thus computed the Lyapunov exponent of nearly-bound orbits and seek for any correlation with actual extinction rates of the luminosity between the $n=1$ and $n=2$ photon rings.

Our results show that, when pushed to the extreme of its parameter's space by the calibrated shadow's size in a thick disk geometry, in the opposite end of an infinitely-thin geometry different alternative spherically symmetric geometries significantly deviate in the physical features relevant for such images (horizon and photon sphere radius), and dramatically in their extinction rates, up to a factor three from one end of the (upward) modifications to the shadow's size to the other (downward). Furthermore, such rates strongly correlate with the theoretical (Lyapunov) prediction, particularly in the GLM2 model (and to a lesser extent in the GLM1), thus rendering a usefulness to such theoretical quantities in connecting them to observations. Indeed, significant visual differences exist between the photon rings of each GLM model, as seen when isolated from each other and from the direct emission in the GLM3 model, as well as in the features of the full images of the (more realistic) GLM1/GLM2 models. This suggests that, in this scenario, it would be possible to distinguish between this pool of alternative spherically symmetric geometries in their optical appearance (at fixed emission profile). 

\begin{figure*}[t!]
\includegraphics[width=5.6cm,height=4.6cm]{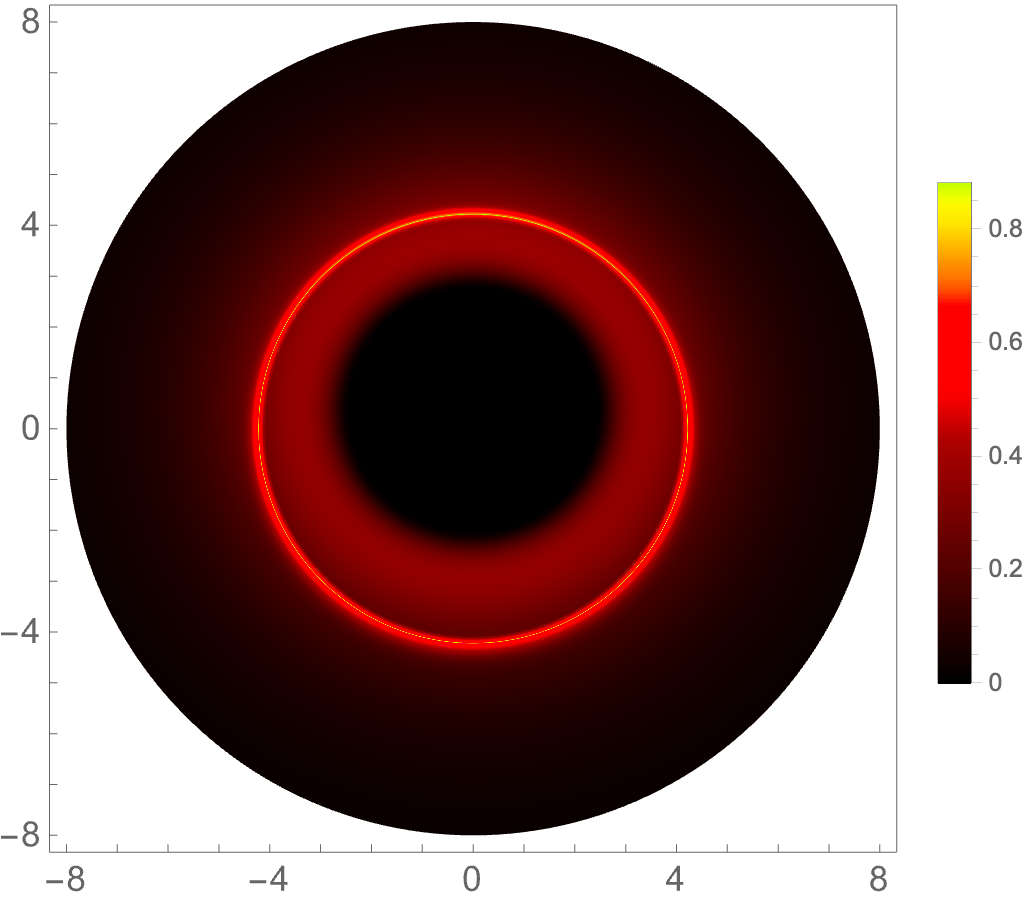}
\includegraphics[width=5.6cm,height=4.6cm]{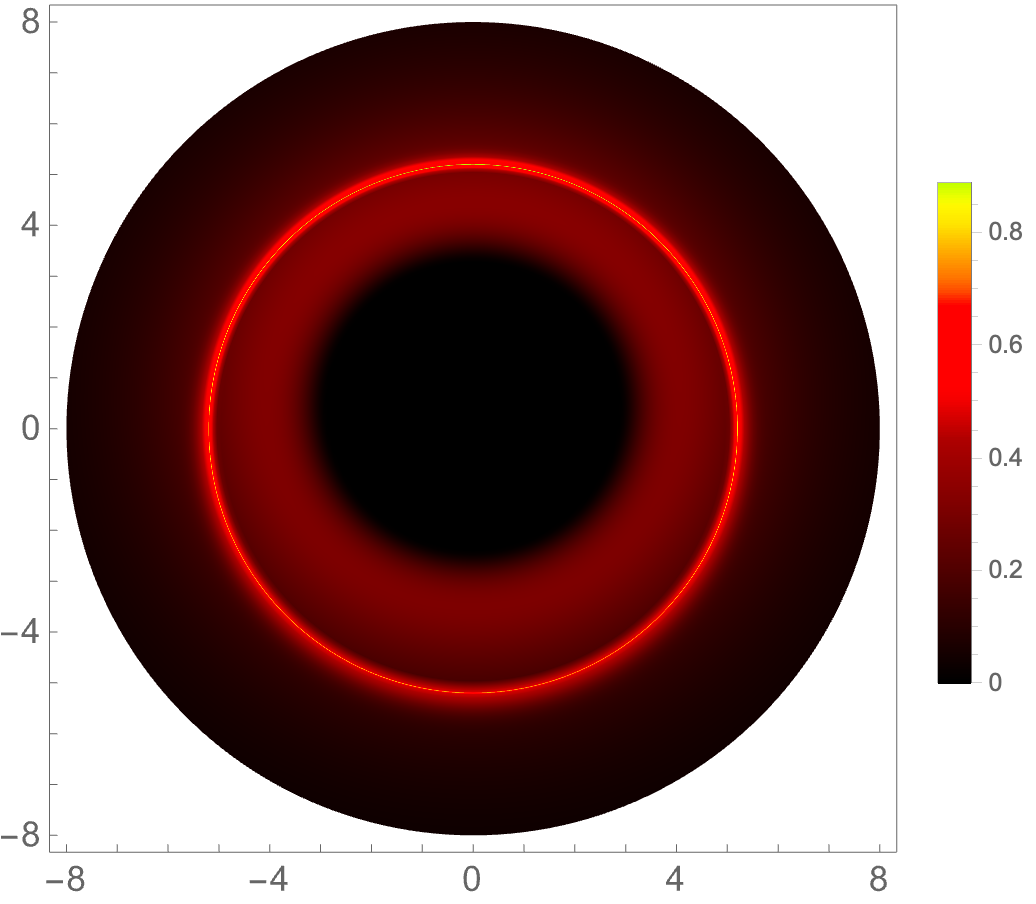}
\includegraphics[width=5.6cm,height=4.6cm]{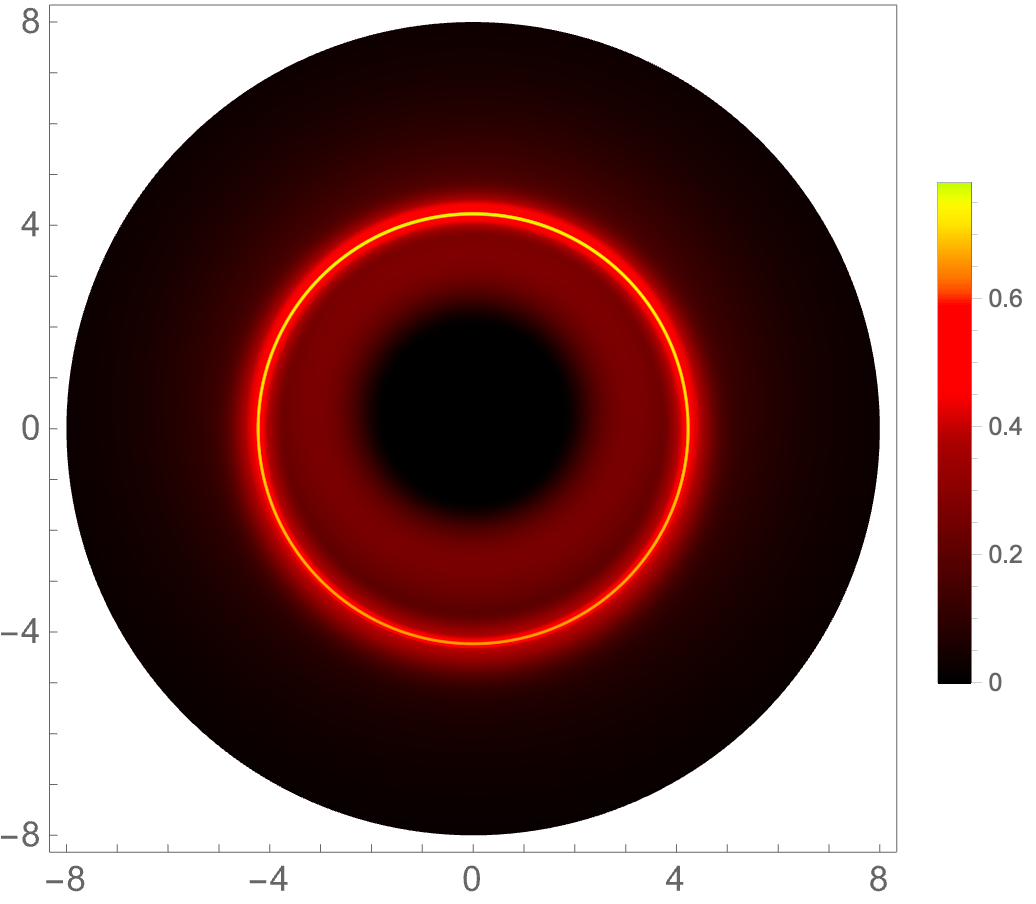}
\includegraphics[width=5.6cm,height=4.6cm]{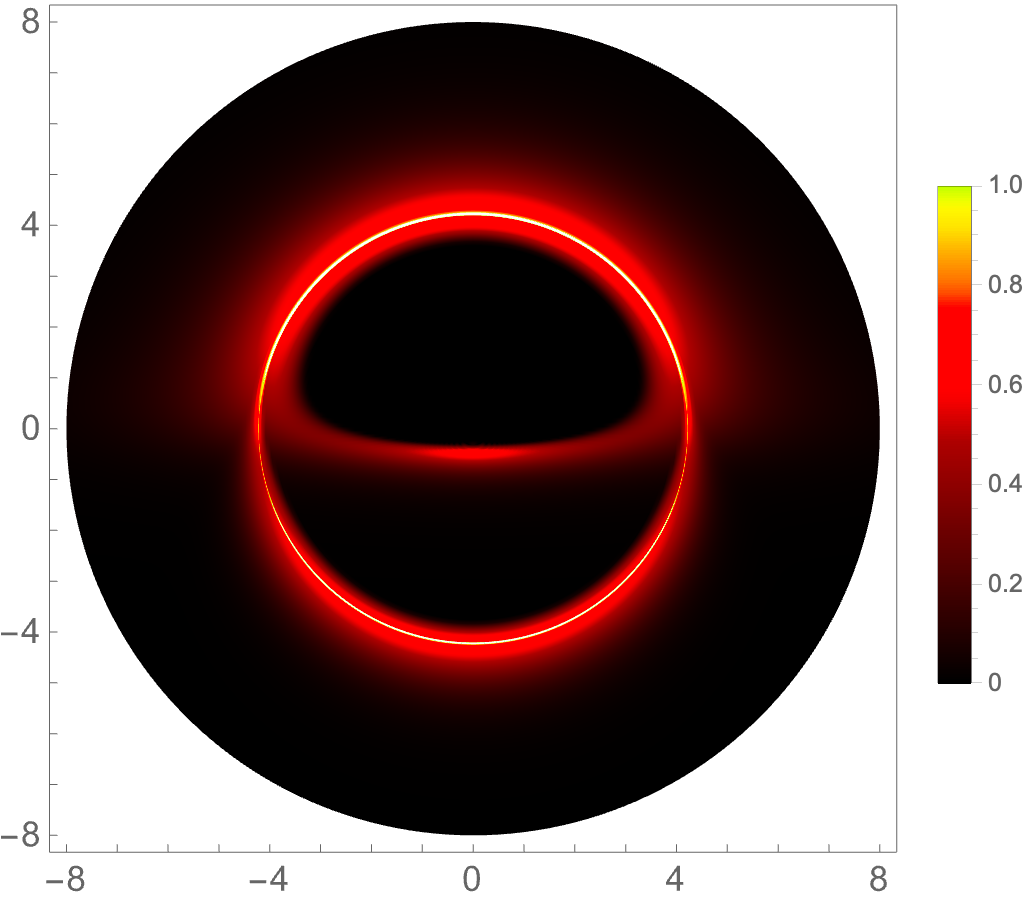}
\includegraphics[width=5.6cm,height=4.6cm]{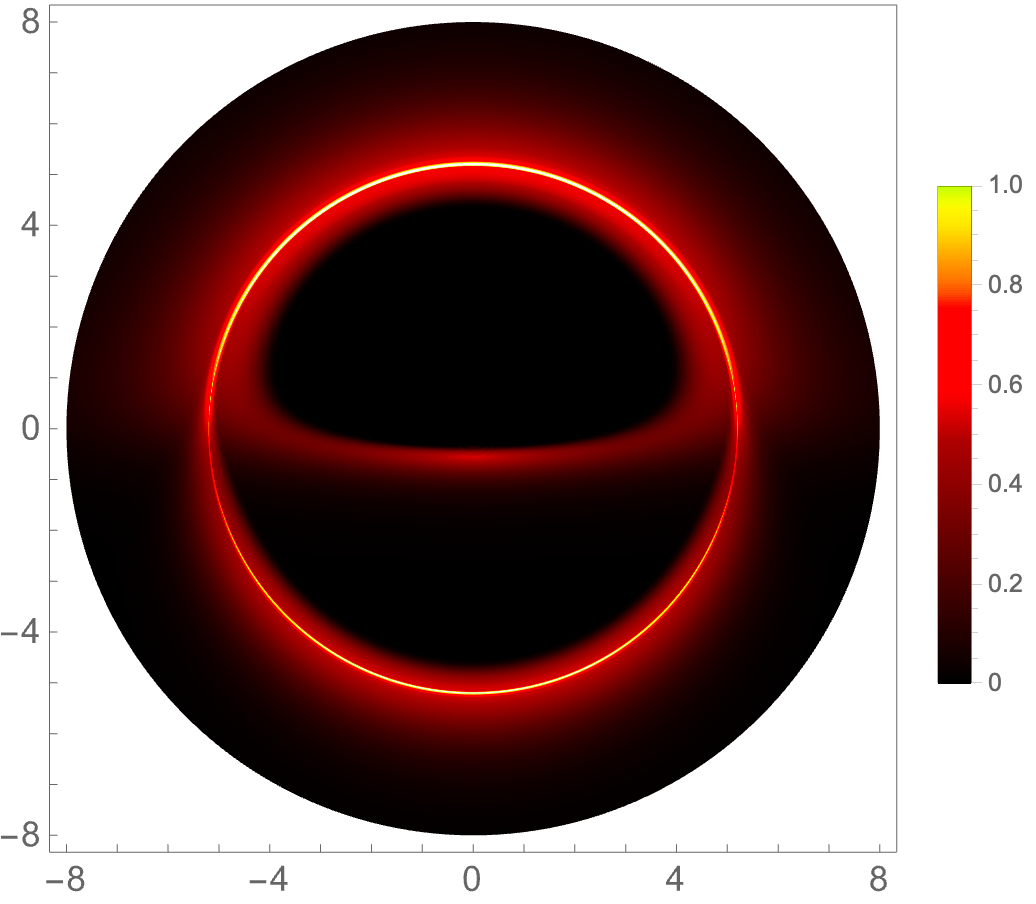}
\includegraphics[width=5.6cm,height=4.6cm]{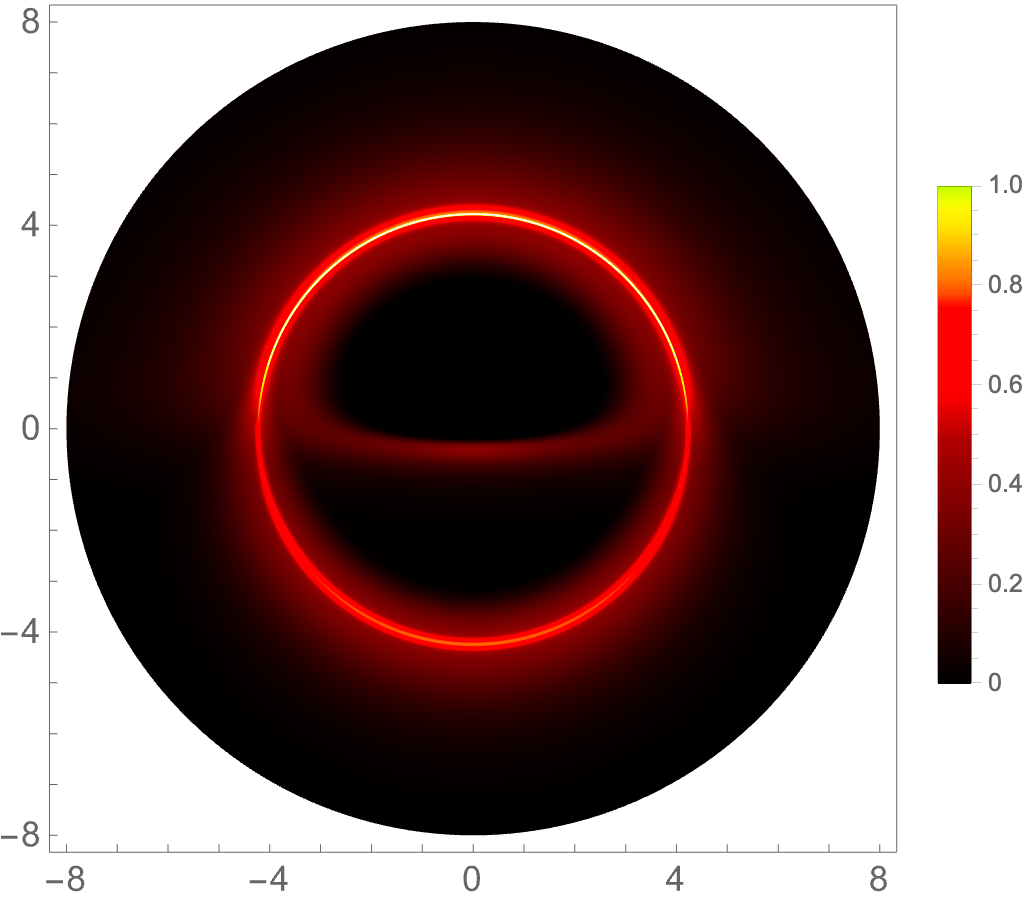}
\caption{Inclined images at $17^o$ (top) and $80^o$ (bottom) degrees of inclination for LQG (left), Schwarzschild (middle) and GK (right) for the GLM2 model.}
\label{fig:shadowinc}
\end{figure*}

There are, however, many known caveats that render the above conclusion premature. First of all, there are the assumptions on the optical, geometrical, and emission properties of the disk. In addition to the previously discussed assumption of an infinitesimally-thin accretion disk, our analysis also assumes that emission is completely monochromatic and optically thin, in the accretion disk's frame. In reality, accretion disks possess complex emission profiles that are not expected to be optically thin at all frequencies. Thus there is room for modelling improvement in this area. In this sense, we note that the EHT collaboration operates at a constant $230$ GHz frequency (i.e. in the observer's frame) \cite{EventHorizonTelescope:2022xqj}; furthermore at such a frequency opacity tends to suppress the $n=2$ ring, though the signal is expected to reappear at higher frequencies, such as the planned 345GHz of future upgrades of VLBI \cite{Vincent:2022}. On the emission profiles, the fact that GLM models are analytical approximations to GRMHD simulations for Kerr black holes, means that we have no solid reason to expect that other black holes will have the same exact intensity profile, since the relevant geometrical features for the generation of images (e.g. horizon and photon sphere radius) may vary significantly from one geometry to another. This way, should we be able to look at a compact object and retrieve data on photon ring intensities, this would not immediately translate into reliable constraints for the background geometry without priors on the properties of the disk. This difficulty could be circumvented by appealing to universal polarimetric signatures \cite{Himwich:2020msm}, or to the more recent concept of photon ring {\it autocorrelations}, a two-point correlation of fluctuations in the intensity along a given photon ring \cite{Chesler:2020gtw,Wong:2020ziu,Hadar:2020fda}.

Finally, the inclusion of rotation and inclination is expected to moderately modify the extinction rate numbers. Rotation actually turns the photon sphere into a photon shell of unstable geodesics, and adds two more critical exponents in the characterization of photon rings \cite{Gralla:2019drh} (besides altering the depth of the central brightness depression), which have a non-negligible impact in the theoretical luminosity. As for inclination, the critical curve also depends on it, so one should also expect significant deviations in the features of the associated rings \cite{Gralla:2020srx}: for instance, for a Kerr black hole at full speed and at the $\theta_0=17^{\text{o}}$ inclination of M87 the factor $e^{-\gamma}$ gets a $\sim 13\%$ modification (on the observer's spin-oriented part of the ring \cite{CraigWalker:2018vam}). In Fig. \ref{fig:shadowinc} we provide a quick glance to the inclined images at the M87 angle $\theta_0=17^o$ (top), and at a much more extreme angle of $\theta_0=80^o$ (bottom), of a Schwarzschild black hole and the two alternative black hole geometries with the largest modifications (upward and downward) to its Lyapunov index considered in this work, namely, LQG and GK,  for the seemingly favored GLM2 model. Even in this spherical symmetry setting, there are apparent visual differences among each model. The incorporation of both rotation and inclination would render the problem of characterizing photon rings in detail significantly more complicated than the simplified analysis made here (see e.g. the analysis of \cite{Paugnat:2022qzy} on this problem), and goes well beyond the scope of this work.

This problem would be further exacerbated in misaligned (tilted) accretion disk scenarios, which is consistent with low-luminosity active galaxy nuclei \cite{Fragile:2007dk}. This scenario introduces another dimension to the problem in the form of an additional angle between the disk orientation and the black hole spin vector.  Such a disk tilt would affect the emission geometry and brightness because it breaks the axisymmetric nature of the accretion flow and results in increased flux variability, thus also significantly altering the emission profile, and supposedly the extinction rates too (see \cite{White:2020cos}, and also \cite{EventHorizonTelescope:2022urf} for a discussion on this problem). 

To conclude, while photon rings may contain valuable information of putative non-Kerr geometries, much work and far better modelling is still necessary in order to hope to disentangle the contribution of background geometries and accretion disk features in black hole images for photon rings to become useful as tests for the presence of new gravitational Physics \cite{Johnson:2023ynn}.

\section*{Acknowledgements}

FSNL acknowledges support from the Funda\c{c}\~{a}o  para a Ci\^{e}ncia e a Tecnologia (FCT) Scientific Employment Stimulus contract with reference CEECINST/00032/2018, and funding through the research grants CERN/FIS-PAR/0037/2019 and PTDC/FIS-AST/0054/2021. FSNL and LS also acknowledge support from the research grants UIDB/04434/2020 and UIDP/04434/2020. This work is also supported by the Spanish Agencia Estatal de Investigaci\'on (grants PID2020-116567GB-C21 and PID2022-138607NB-I00) funded by MCIN/AEI/10.13039/501100011033 and ERDF A way of making Europe), the project PROMETEO/2020/079 (Generalitat Valenciana), the EU's Horizon 2020 research and innovation (RISE) programme H2020-MSCA-RISE-2017 (FunFiCO-777740), and by the European Horizon Europe staff exchange (SE) programme HORIZON-MSCA-2021-SE-01 (NewFunFiCO10108625). This article is based upon work from COST Action CA21136.


\begin{thebibliography}{00}

\bibitem{Kerr:1963ud}
R.~P.~Kerr,
%``Gravitational field of a spinning mass as an example of algebraically special metrics,''
Phys. Rev. Lett. \textbf{11} (1963) 237.

\bibitem{Herdeiro:2015waa}
C.~A.~R.~Herdeiro and E.~Radu,
%``Asymptotically flat black holes with scalar hair: a review,''
Int. J. Mod. Phys. D \textbf{24} (2015) 1542014.

\bibitem{Bambi:2016sac}
C.~Bambi, A.~Cardenas-Avendano, T.~Dauser, J.~A.~Garcia and S.~Nampalliwar,
%``Testing the Kerr black hole hypothesis using X-ray reflection spectroscopy,''
Astrophys. J. \textbf{842} (2017) 76.

\bibitem{Cardoso:2019rvt}
V.~Cardoso and P.~Pani,
%``Testing the nature of dark compact objects: a status report,''
Living Rev. Rel. \textbf{22} (2019) 4.

\bibitem{EventHorizonTelescope:2019dse}
K.~Akiyama \textit{et al.} [Event Horizon Telescope],
%``First M87 Event Horizon Telescope Results. I. The Shadow of the Supermassive Black Hole,''
Astrophys. J. Lett. \textbf{875} (2019) L1.

\bibitem{EventHorizonTelescope:2022wkp}
K.~Akiyama \textit{et al.} [Event Horizon Telescope],
%``First Sagittarius A* Event Horizon Telescope Results. I. The Shadow of the Supermassive Black Hole in the Center of the Milky Way,''
Astrophys. J. Lett. \textbf{930} (2022) L12.

\bibitem{Hadar:2022xag}
S.~Hadar, D.~Kapec, A.~Lupsasca and A.~Strominger,
%``Holography of the photon ring,''
Class. Quant. Grav. \textbf{39} (2022) 215001.

\bibitem{Falcke:1999pj}
H.~Falcke, F.~Melia and E.~Agol,
%``Viewing the shadow of the black hole at the galactic center,''
Astrophys. J. Lett. \textbf{528} (2000) L13.

\bibitem{Narayan:2019imo}
R.~Narayan, M.~D.~Johnson and C.~F.~Gammie,
%``The Shadow of a Spherically Accreting Black Hole,''
Astrophys. J. Lett. \textbf{885} (2019) L33.

\bibitem{Gralla:2019xty}
S.~E.~Gralla, D.~E.~Holz and R.~M.~Wald,
%``Black Hole Shadows, Photon Rings, and Lensing Rings,''
Phys. Rev. D \textbf{100} (2019) 024018.

\bibitem{Chael:2021rjo}
A.~Chael, M.~D.~Johnson and A.~Lupsasca,
%``Observing the Inner Shadow of a Black Hole: A Direct View of the Event Horizon,''
Astrophys. J. \textbf{918} (2021)  6.

\bibitem{Gralla:2020srx}
S.~E.~Gralla, A.~Lupsasca and D.~P.~Marrone,
%``The shape of the black hole photon ring: A precise test of strong-field general relativity,''
Phys. Rev. D \textbf{102} (2020) 124004.

\bibitem{Staelens:2023jgr}
S.~Staelens, D.~R.~Mayerson, F.~Bacchini, B.~Ripperda and L.~K\"uchler,
%``Black hole photon rings beyond general relativity,''
Phys. Rev. D \textbf{107} (2023) 124026.

\bibitem{Eichhorn:2022oma}
A.~Eichhorn, A.~Held and P.~V.~Johannsen,
%``Universal signatures of singularity-resolving physics in photon rings of black holes and horizonless objects,''
JCAP \textbf{01} (2023) 043.

\bibitem{Tiede:2022grp}
P.~Tiede, {\it et. al.}
%``Measuring Photon Rings with the ngEHT,''
Galaxies \textbf{10} (2022) no.6, 111.

\bibitem{Johnson:2020}
M.~D.~Johnson, {\it et. al.}
%``Universal interferometric signatures of a black hole's photon ring,´´
Science Advances \textbf{6} (2020) no.12, eaaz1310.

\bibitem{Gralla:2020nwp}
S.~E.~Gralla,
%``Measuring the shape of a black hole photon ring,''
Phys. Rev. D \textbf{102} (2020) 044017.

\bibitem{Gralla:2020yvo}
S.~E.~Gralla and A.~Lupsasca,
%``Observable shape of black hole photon rings,''
Phys. Rev. D \textbf{102} (2020) 124003.

\bibitem{Vincent:2022}
F.~H.~Vincent, S.~E.~Gralla, A.~Lupsasca, M.~Wielgus
%``Images and photon ring signatures of thick disks around black holes´´
Astronomy \& Astrophysics \textbf{667} (2022) A170.

\bibitem{Gurvits:2022wgm}
L.~I.~Gurvits,  \textit{et al.}
%``The science case and challenges of space-borne sub-millimeter interferometry,''
Acta Astronaut. \textbf{196} (2022) 314.



\bibitem{Paugnat:2022qzy}
H.~Paugnat, A.~Lupsasca, F.~Vincent and M.~Wielgus,
%``Photon ring test of the Kerr hypothesis: Variation in the ring shape,''
Astron. Astrophys. \textbf{668} (2022) A11.



\bibitem{Cardenas-Avendano:2023dzo}
A.~C\'ardenas-Avenda\~no and A.~Lupsasca,
%``Prediction for the interferometric shape of the first black hole photon ring,''
Phys. Rev. D \textbf{108} (2023) 064043.

\bibitem{Wielgus:2021peu}
M.~Wielgus,
%``Photon rings of spherically symmetric black holes and robust tests of non-Kerr metrics,''
Phys. Rev. D \textbf{104} (2021) 124058.

\bibitem{EventHorizonTelescope:2022xqj}
K.~Akiyama \textit{et al.} [Event Horizon Telescope],
%``First Sagittarius A* Event Horizon Telescope Results. VI. Testing the Black Hole Metric,''
Astrophys. J. Lett. \textbf{930} (2022) L17.

\bibitem{Vagnozzi:2022moj}
S.~Vagnozzi, \textit{et al.}
%``Horizon-scale tests of gravity theories and fundamental physics from the Event Horizon Telescope image of Sagittarius A$^*$,''
Class. Quant. Grav. \textbf{40} (2023)  165007.



\bibitem{Lima:2021las}
H.~C.~D.~Lima, Junior., L.~C.~B.~Crispino, P.~V.~P.~Cunha and C.~A.~R.~Herdeiro,
%``Can different black holes cast the same shadow?,''
Phys. Rev. D \textbf{103} (2021) 084040.


\bibitem{Gold:2020iql}
R.~Gold,  \textit{et al.}
%``Verification of Radiative Transfer Schemes for the EHT,''
Astrophys. J. \textbf{897} (2020) 148.

\bibitem{Dymnikova:2015hka}
I.~Dymnikova and E.~Galaktionov,
%``Regular rotating electrically charged black holes and solitons in non-linear electrodynamics minimally coupled to gravity,''
Class. Quant. Grav. \textbf{32} (2015) 165015.

\bibitem{MarotoBook}
A. Dobado, A. Gomez-Nicola, A. L. Maroto, J. R, Pelaez, {\it ``Effective Lagrangians for the Standard Model"} (Springer Science and Business Media, 2012).

\bibitem{Rasheed:1997ns}
D.~A.~Rasheed,
%``Nonlinear electrodynamics: Zeroth and first laws of black hole mechanics,''
[arXiv:hep-th/9702087].

\bibitem{Bronnikov:2021uta}
K.~A.~Bronnikov and R.~K.~Walia,
%``Field sources for Simpson-Visser spacetimes,''
Phys. Rev. D \textbf{105} (2022) 044039.


\bibitem{Novello:1999pg}
M.~Novello, V.~A.~De Lorenci, J.~M.~Salim and R.~Klippert,
%``Geometrical aspects of light propagation in nonlinear electrodynamics,''
Phys. Rev. D \textbf{61} (2000) 045001.

\bibitem{DeLorenci:2000yh}
V.~A.~De Lorenci, R.~Klippert, M.~Novello and J.~M.~Salim,
%``Light propagation in nonlinear electrodynamics,''
Phys. Lett. B \textbf{482} (2000) 134.

\bibitem{GRAVITY:2018ofz}
R.~Abuter \textit{et al.} [GRAVITY],
%``Detection of the gravitational redshift in the orbit of the star S2 near the Galactic centre massive black hole,''
Astron. Astrophys. \textbf{615} (2018) L15.

\bibitem{Psaltis:2018xkc}
D.~Psaltis,
%``Testing General Relativity with the Event Horizon Telescope,''
Gen. Rel. Grav. \textbf{51} (2019) 137.

\bibitem{EventHorizonTelescope:2022urf}
K.~Akiyama \textit{et al.} [Event Horizon Telescope],
%``First Sagittarius A* Event Horizon Telescope Results. V. Testing Astrophysical Models of the Galactic Center Black Hole,''
Astrophys. J. Lett. \textbf{930} (2022) L16.

\bibitem{Blandford:1977ds}
R.~D.~Blandford and R.~L.~Znajek,
%``Electromagnetic extractions of energy from Kerr black holes,''
Mon. Not. Roy. Astron. Soc. \textbf{179} (1977) 433.

\bibitem{Zajacek:2018ycb}
M.~Zaja\v{c}ek, A.~Tursunov, A.~Eckart and S.~Britzen,
%``On the charge of the Galactic centre black hole,''
Mon. Not. Roy. Astron. Soc. \textbf{480} (2018) 4408.

\bibitem{Yajima:2000kw}
H.~Yajima and T.~Tamaki,
%``Black hole solutions in Euler-Heisenberg theory,''
Phys. Rev. D \textbf{63} (2001) 064007.

\bibitem{Allahyari:2019jqz}
A.~Allahyari, M.~Khodadi, S.~Vagnozzi and D.~F.~Mota,
%``Magnetically charged black holes from non-linear electrodynamics and the Event Horizon Telescope,''
JCAP \textbf{02} (2020) 003.

\bibitem{Wen:2022hkv}
S.~Wen, W.~Hong and J.~Tao,
%``Observational Appearances of Magnetically Charged Black Holes in Born-Infeld Electrodynamics,''
Eur. Phys. J. C \textbf{83} (2023) 277.



\bibitem{DobadoBook}
A. Dobado, A. G\'omez-Nicola, A. L. Maroto, J. R. Pel\'aez, ``Effective Lagrangians for the Standard Model", Springer-Verlag, Berlin, Heidelberg, 1997.

\bibitem{Bardeen}
J.M. Bardeen, ``Non-singular general-relativistic gravitational collapse", in Proceedings of of International Conference GR5,Tbilisi, USSR (1968), p. 174.

\bibitem{Ansoldi:2008jw}
S.~Ansoldi,
%``Spherical black holes with regular center: A Review of existing models including a recent realization with Gaussian sources,''
[arXiv:0802.0330 [gr-qc]].



\bibitem{Beato:2000yh}
E.~Ayon-Beato and A.~Garcia,
%``The Bardeen Model as a Nonlinear Magnetic Monopole''
Phys. Lett. B \textbf{493} (2000) 149.

\bibitem{Hayward:2005gi}
S.~A.~Hayward,
%``Formation and evaporation of regular black holes,''
Phys. Rev. Lett. \textbf{96} (2006) 031103.

\bibitem{Frolov:2016pav}
V.~P.~Frolov,
%``Notes on nonsingular models of black holes,''
Phys. Rev. D \textbf{94} (2016) 104056.


\bibitem{Wielgus:2020uqz}
M.~Wielgus, J.~Horak, F.~Vincent and M.~Abramowicz,
%``Reflection-asymmetric wormholes and their double shadows,''
Phys. Rev. D \textbf{102} (2020) 084044.

\bibitem{Tsukamoto:2022vkt}
N.~Tsukamoto,
%``Retrolensing by two photon spheres of a black-bounce spacetime,''
Phys. Rev. D \textbf{105} (2022) 084036.

\bibitem{Guerrero:2022msp}
M.~Guerrero, G.~J.~Olmo, D.~Rubiera-Garcia and D.~S\'aez-Chill\'on G\'omez,
%``Multiring images of thin accretion disk of a regular naked compact object,''
Phys. Rev. D \textbf{106} (2022) 044070.


\bibitem{Kazakov:1993ha}
D.~I.~Kazakov and S.~N.~Solodukhin,
%``On Quantum deformation of the Schwarzschild solution,''
Nucl. Phys. B \textbf{429} (1994) 153.

\bibitem{Sen:1992ua}
A.~Sen,
%``Rotating charged black hole solution in heterotic string theory,''
Phys. Rev. Lett. \textbf{69} (1992) 1006.

\bibitem{Tripathi:2021rwb}
A.~Tripathi, B.~Zhou, A.~B.~Abdikamalov, D.~Ayzenberg and C.~Bambi,
%``Constraints on Einstein-Maxwell dilaton-axion gravity from X-ray reflection spectroscopy,''
JCAP \textbf{07} (2021) 002.

\bibitem{Gibbons:1987ps}
G.~W.~Gibbons and K.~i.~Maeda,
%``Black Holes and Membranes in Higher Dimensional Theories with Dilaton Fields,''
Nucl. Phys. B \textbf{298} (1988) 741.

\bibitem{Li:2012zx}
M.~H.~Li and K.~C.~Yang,
%``Galactic Dark Matter in the Phantom Field,''
Phys. Rev. D \textbf{86} (2012) 123015.

\bibitem{Simpson:2018tsi}
A.~Simpson and M.~Visser,
%``Black-bounce to traversable wormhole,''
JCAP \textbf{02} (2019) 042.

\bibitem{Ellis:1973yv}
H.~G.~Ellis,
%``Ether flow through a drainhole - a particle model in general relativity,''
J. Math. Phys. \textbf{14} (1973) 104.

\bibitem{Guerrero:2021ues}
M.~Guerrero, G.~J.~Olmo, D.~Rubiera-Garcia and D.~S.~C.~G\'omez,
%``Shadows and optical appearance of black bounces illuminated by a thin accretion disk,''
JCAP \textbf{08} (2021) 036.

\bibitem{Modesto:2008im}
L.~Modesto,
%``Semiclassical loop quantum black hole,''
Int. J. Theor. Phys. \textbf{49} (2010) 1649.

\bibitem{Astorino:2013sfa}
M.~Astorino,
%``C-metric with a conformally coupled scalar field in a magnetic universe,''
Phys. Rev. D \textbf{88} (2013) 104027.

\bibitem{Janis:1968zz}
A.~I.~Janis, E.~T.~Newman and J.~Winicour,
%``Reality of the Schwarzschild Singularity,''
Phys. Rev. Lett. \textbf{20} (1968) 878.

\bibitem{Bronnikov:2000vy}
K.~A.~Bronnikov,
%``Regular magnetic black holes and monopoles from nonlinear electrodynamics,''
Phys. Rev. D \textbf{63} (2001) 044005.

\bibitem{Ghosh:2021clx}
S.~G.~Ghosh and R.~K.~walia,
%``Rotating black holes in general relativity coupled to nonlinear electrodynamics,''
Annals Phys. \textbf{434} (2021) 168619.

\bibitem{Ghosh:2014pba}
S.~G.~Ghosh,
%``A nonsingular rotating black hole,''
Eur. Phys. J. C \textbf{75} (2015) 532.

\bibitem{Culetu:2014lca}
H.~Culetu,
%``On a regular charged black hole with a nonlinear electric source,''
Int. J. Theor. Phys. \textbf{54} (2015) 2855.

\bibitem{Simpson:2019mud}
A.~Simpson and M.~Visser,
%``Regular black holes with asymptotically Minkowski cores,''
Universe \textbf{6} (2019) no.1, 8.

\bibitem{Kumar:2020ltt}
R.~Kumar and S.~G.~Ghosh,
%``Photon ring structure of rotating regular black holes and no-horizon spacetimes,''
Class. Quant. Grav. \textbf{38} (2021) 8.



\bibitem{Cardoso:2008bp}
V.~Cardoso, A.~S.~Miranda, E.~Berti, H.~Witek and V.~T.~Zanchin,
%``Geodesic stability, Lyapunov exponents and quasinormal modes,''
Phys. Rev. D \textbf{79} (2009) 064016.

\bibitem{Himwich:2020msm}
E.~Himwich, M.~D.~Johnson, A.~Lupsasca and A.~Strominger,
%``Universal polarimetric signatures of the black hole photon ring,''
Phys. Rev. D \textbf{101} (2020) 084020.

\bibitem{Chesler:2020gtw}
P.~M.~Chesler, L.~Blackburn, S.~S.~Doeleman, M.~D.~Johnson, J.~M.~Moran, R.~Narayan and M.~Wielgus,
%``Light echos and coherent autocorrelations in a black hole spacetime,''
Class. Quant. Grav. \textbf{38} (2021) 125006.

\bibitem{Wong:2020ziu}
G.~N.~Wong,
%``Black Hole Glimmer Signatures of Mass, Spin, and Inclination,''
Astrophys. J. \textbf{909} (2021) 217.

\bibitem{Hadar:2020fda}
S.~Hadar, M.~D.~Johnson, A.~Lupsasca and G.~N.~Wong,
%``Photon Ring Autocorrelations,''
Phys. Rev. D \textbf{103} (2021) 104038.



\bibitem{Gralla:2019drh}
S.~E.~Gralla and A.~Lupsasca,
%``Lensing by Kerr Black Holes,''
Phys. Rev. D \textbf{101} (2020) 044031.

\bibitem{CraigWalker:2018vam}
R.~Craig Walker, P.~E.~Hardee, F.~B.~Davies, C.~Ly and W.~Junor,
%``The Structure and Dynamics of the Subparsec Jet in M87 Based on 50 VLBA Observations over 17 Years at 43 GHz,''
Astrophys. J. \textbf{855} (2018) 128.

\bibitem{Fragile:2007dk}
P.~C.~Fragile, O.~M.~Blaes, P.~Anninois and J.~D.~Salmonson,
%``Global General Relativistic MHD Simulation of a Tilted Black-Hole Accretion Disk,''
Astrophys. J. \textbf{668} (2007) 417.


\bibitem{White:2020cos}
C.~J.~White, J.~Dexter, O.~Blaes and E.~Quataert,
%``The Effects of Tilt on the Images of Black Hole Accretion Flows,"
Astrophys. J. \textbf{894} (2020) 14.


%\cite{Johnson:2023ynn}
\bibitem{Johnson:2023ynn}
M.~D.~Johnson \textit{et al.}
%``Key Science Goals for the Next-Generation Event Horizon Telescope,''
Galaxies \textbf{11}, no.3, 61 (2023).
%doi:10.3390/galaxies11030061
%[arXiv:2304.11188 [astro-ph.HE]].






\end{thebibliography}
\end{document}